\documentclass[twocolumn,nopacs,preprintnumbers,amsmath,amssymb,subfig]{revtex4-2}
\usepackage{graphicx}% Include figure files
\usepackage{multirow}
\usepackage[colorlinks=true, allcolors=blue]{hyperref}

\begin{document}

%\title{Competition between magnetic frustration and spin dimensionality in the ground state of the classical antiferromagnetic $n$-vector model with arbitrary $n$}

%\title{Magnetic-field induced classical configurations that break translational symmetry in icosahedral rings}

%\title{Local variation of the classical magnetization of short molecular arrays effected by an external magnetic field}

\title{Local variations of the magnetization effected by an external field in molecular rings}

\author{N. P. Konstantinidis}
\affiliation{Mathematics and Science Department, American University in Bulgaria, Sv. Buchvarova Str. 8, Blagoevgrad 2700, Bulgaria}

\date{\today}

\begin{abstract}
It is shown that an external magnetic field generates local variations of the classical ground-state magnetization in molecular rings of antiferromagnetic icosahedra with isotropic spin interactions. The magnetic response is characterized by a multitude of magnetization and susceptibility discontinuities occurring across the ring. In addition, a parity effect with respect to the evenness or oddness of the number of icosahedra allows for magnetization jumps that occur at different field values for different molecules and produce an even more pronounced local variation of the magnetization. It is also found that for specific field ranges all canting angles of the molecular magnetizations increase with the field. These findings are in sharp contrast with the ones for rings of individual spins.
\end{abstract}


\maketitle

\section{Introduction}
\label{sec:introduction}

%SPIN FLOP.

%SPIN-FLOP TRANSITION-NOT EXACTLY.

%NONUNIFORM CANTING ALONG THE CHAIN.

%"DOUBLING" (?) OF THE UNIT-CELL.

In frustrated magnetic systems competing interactions lead to a wide range of unconventional phenomena at the classical and quantum level \cite{Auerbach98,Fazekas99}. These occur even in the absence of magnetic anisotropy, as is the case with the isotropic nearest-neighbor antiferromagnetic Heisenberg model (AHM) \cite{Manousakis91,Lhuillier01,Misguich03,Ramirez05,Schnack10}. Apart from extended systems, finite units whose connectivity is associated with frustrated interactions are of particular importance. These include the icosahedral-symmetry Platonic solids \cite{Plato,Prinzbach00,Wang01,Iqbal03,Vaknin14,Engelhardt14,Qin17}, the icosahedron and the dodecahedron, which have been found to support nontrivial magnetic states at the classical and quantum level \cite{Coffey92,NPK01,Schmidt03,Schroeder05,NPK05,NPK07,Hucht11,Strecka15,NPK16,Karlova16,Karlova17,Karlova17-1,NPK23-1,NPK23-2,NPK25,NPK26}. Furthermore, the introduction of an external field leads to discontinuous magnetic response in the absence of magnetic anisotropy.

The above properties of the icosahedral Platonic solids are the motivation to consider rings of icosahedra, with their magnetic properties modeled by the classical AHM. Classical spins forming rings and interacting according to the AHM have an increasing external field turn them toward and eventually align with its direction. In this way the so-called umbrella configuration is realized, in which all the spins form the same angle with the field \cite{Landau81,Holtschneider07,NPK17-1}. Here the magnetic units forming a ring are taken to be the total magnetizations $\vec{M}_i$ of the icosahedra. The interactions within an icosahedron and between neighboring ones are taken to be antiferromagnetic. In order to calculate the influence of an external field on the ground-state magnetization, intermolecular interactions that do not introduce further frustration over that of a single isolated icosahedron in zero field are investigated. In this way additional frustration introduced by specific intermolecular couplings is avoided, and the connectivities investigated are expected to provide a smoother transition towards rings with a very large number of molecules. This is achieved when nearest-neighbor spins belonging to the same icosahedron form the same angle as in the zero-field ground state of an isolated molecule, while interacting spins belonging to neighboring icosahedra are antiparallel, and the magnetization $\vec{M}_i$ of each molecule is zero. At least four different types of intermolecular interactions are found to achieve minimum frustration, even though this is not generally true for any number of molecules.
%Four different types of intermolecular interactions are considered, so that in the ground state in the absence of an external field no additional frustration over that of a single molecule is introduced.
%Nearest-neighbor spins belonging to the same icosahedron form the same angle as in the ground state of an isolated icosahedron, while interacting spins belonging to neighboring molecules are antiparallel, and the magnetization $\vec{M}_i$ of each molecule is zero.
As the field is switched on each molecule develops a finite total magnetization which increases with the field, in contrast to a single isolated spin in a ring whose magnitude is independent of the field value.

The classical total magnetization of an isolated icosahedron has a discontinuity in an external field \cite{Schroeder05} and points along the field direction. Here it is found that for rings of up to six icosahedra the isolated-icosahedron discontinuity survives and occurs in two steps for odd-membered rings, with different molecules undergoing the jump at slightly different fields. This breaks translational symmetry and creates local variations of the magnetization. In addition, the magnitudes $M_i$ of the molecular magnetizations can also differ away from this discontinuity, further enhancing the breaking of translational symmetry. The molecular magnetizations are not necessarily directed along the field direction but can form a finite polar angle $\theta_i$ with it, which is also not the same for different molecules. This is in contrast to a ring of spins, where each spin forms the same angle with the magnetic field. The result is different molecular magnetization projections along the field. For an even number of molecules the isolated-icosahedron jump takes place at the same field value for every molecule. Still, the magnetizations are again not the same in magnitude and direction for all the icosahedra, violating translational symmetry in the ground state.

Apart from the isolated-icosahedron discontinuity, the external field generates a rich magnetic response with multiple distinct ground states separated by magnetization or susceptibility jumps. These are absent for an isolated icosahedron and originate from the interaction between neighboring molecules. Many of them also break translational symmetry.
%In addition, the angles between these magnetizations and the external field are not the same.
The polar angles can all increase simultaneously as a function of the field for specific field ranges. This is in contrast with the variation of the polar angles in the umbrella state of a ring of spins, where all decrease with the field. Even in an isolated icosahedron, where polar angles of individual spins increase for a specific field range, this can not happen for all simultaneously \cite{NPK15}.

Due to computational requirements, rings of more than six icosahedra have not been considered. It is an open question how the magnetic response develops going toward the thermodynamic limit, how the lowest-energy configurations evolve, and if more discontinuities emerge. The results in this paper are relevant to icosahedral quasicrystals and their approximant crystals \cite{Shechtman84,Goldman93,Suzuki21,Tsai00,Eto25,Novat26}. Going from one to three-dimensional structures the results may survive in one form or another.

%\cite{Coffey92,NPK05,NPK07,NPK16,NPK23-1,NPK17,Schroeder05,NPK15,NPK23-2,NPK21,Schulenburg02,Richter04,Schnack06,Nakano13,Nakano14,Nakano14-1,Furuchi21,Furuchi23}.

\begin{figure}[h]
\begin{center}
\includegraphics[width=3.2in,height=1.4in]{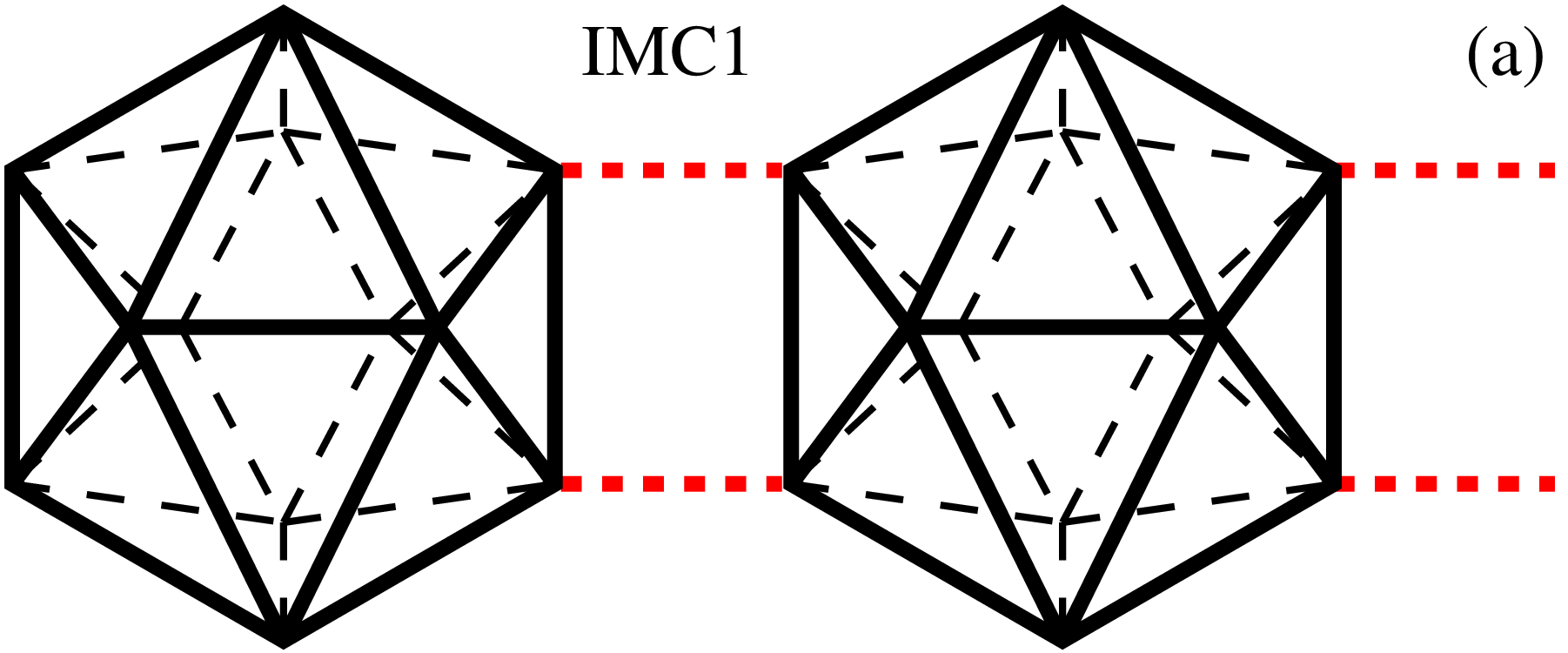}
\end{center}
\begin{center}
\includegraphics[width=3.2in,height=1.4in]{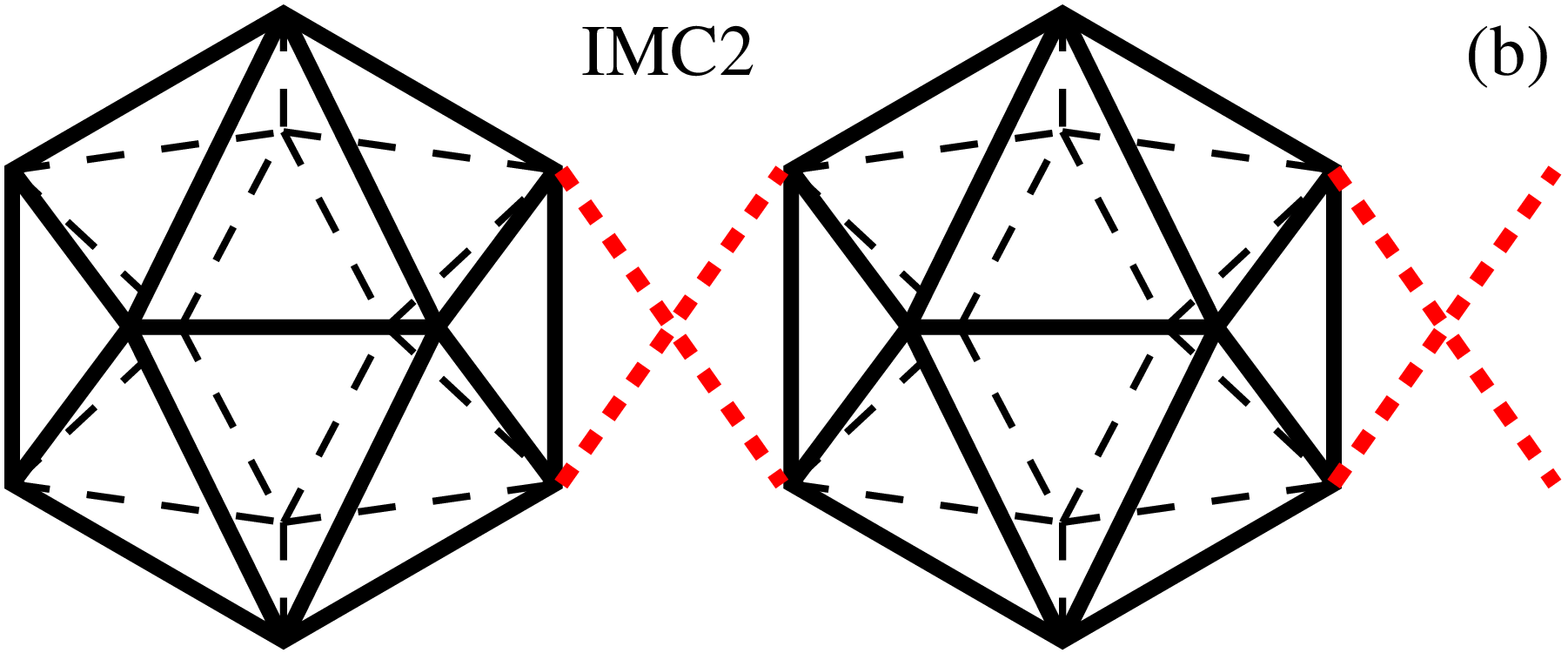}
\end{center}
\begin{center}
\includegraphics[width=3.2in,height=1.4in]{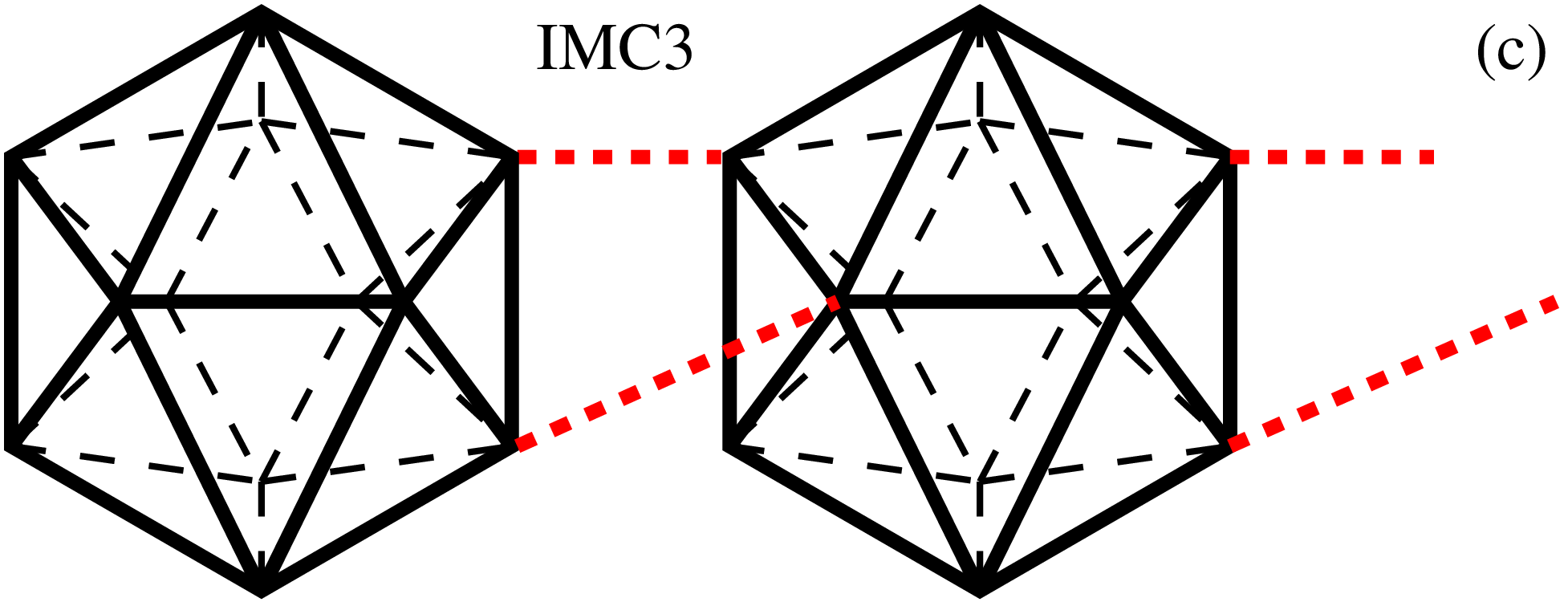}
\end{center}
\begin{center}
\includegraphics[width=3.2in,height=1.4in]{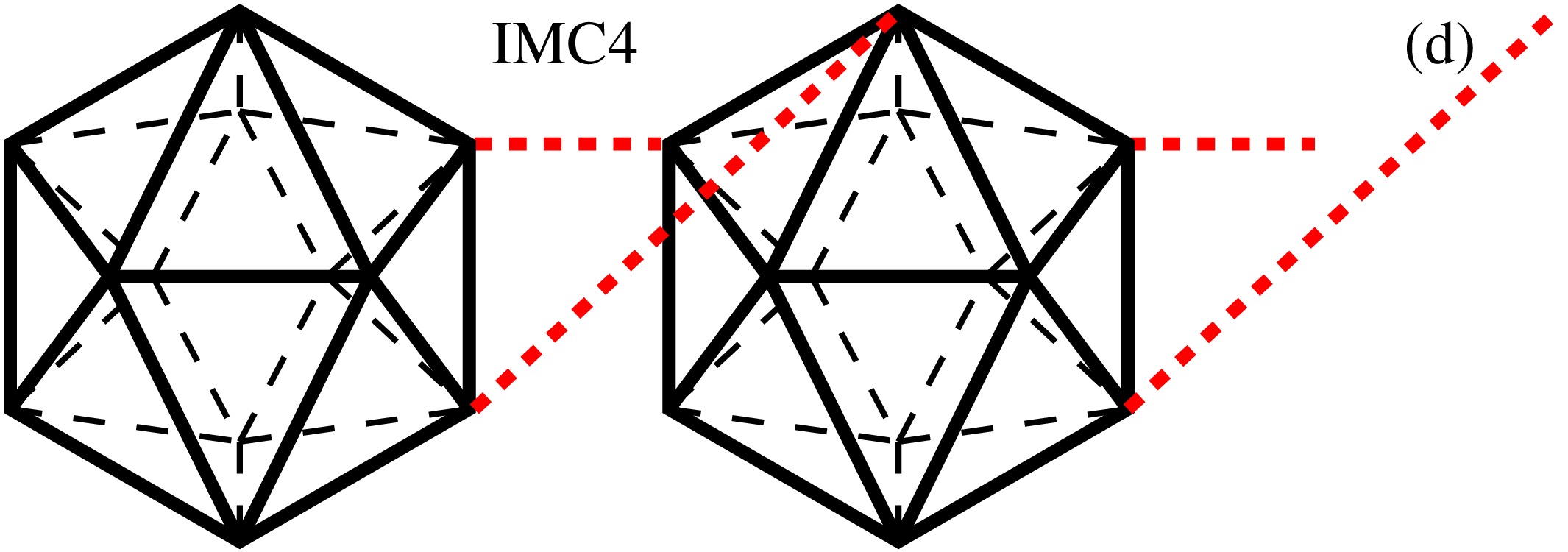}
\end{center}
\caption{Schematic representation of two interacting icosahedra. The (black) solid lines correspond to bonds within the same icosahedron with exchange interaction value $cos\omega$, while the (red) dashed lines to neighboring-icosahedra bonds with exchange interaction value $sin\omega$. The four different types of intermolecular connectivity IMCi, i=1,2,3,4, are shown.
%(~/basic/LATEX/icosahedronarray)
}
\label{fig:truncatedicosahedronconnectivity}
\end{figure}

%The presence of frustrated odd-membered polygons \cite{NPK15-1} leads to classical zero-field ground states with nearest-neighbor spins not pointing in opposite directions as in the case of bipartite structures, which lack any frustration.

%Quantum discontinuities with $\Delta S^z>1$ have also been found for the dodecahedron \cite{NPK05}, the smallest member of the fullerene class \cite{Fowler95}, the pentakis dodecahedron \cite{NPK21}, and for extended systems \cite{Schulenburg02,Richter04,Schnack06,Nakano13,Nakano14,Nakano14-1,Furuchi21,Furuchi23}.

The plan of this paper is as follows: in Sec. \ref{sec:model} the AHM in a magnetic field and the intermolecular connectivities are introduced for a ring of icosahedra. Section \ref{sec:imc1andimc2} calculates the magnetic response for two different types of intermolecular interactions for a dimer of icosahedra. Other types of connectivity are considered in Sec. \ref{sec:imc3} for a triangle and a hexagon of icosahedra, and in Sec. \ref{sec:IMC4} for a pentagon and a hexagon. Finally Sec. \ref{sec:conclusions} presents the conclusions.

\section{Model}
\label{sec:model}

The icosahedron is made of 12 equivalent vertices and 20 identical triangles \cite{Plato}. It has the symmetry of the icosahedral $I_h$ group, the biggest point-symmetry group with 120 symmetry operations \cite{Altmann94}. Here a ring of $N$ icosahedra is considered, and the total number of vertices is $12N$. In each icosahedron a classical spin $\vec{s}_{ij}$, $i=1,2,\dots,N$, $j=1,2,\dots,12$ of unit magnitude is mounted on each of its vertices. The interactions within each icosahedron follow its connectivity. In the cases of the intermolecular couplings investigated the zero-field ground state of each icosahedron has an energy equal to the one when it is isolated \cite{Schmidt03}, while interacting spins of different icosahedra are antiparallel. In this way no additional frustration is introduced by forming the ring. This is true at least for the intermolecular connectivities (IMC) shown in Fig. \ref{fig:truncatedicosahedronconnectivity}, however not for any number of molecules. For IMC1 (Fig. \ref{fig:truncatedicosahedronconnectivity}(a)) they are satisfied for an even number of molecules, while for IMC2 (Fig. \ref{fig:truncatedicosahedronconnectivity}(b)) for any value of $N$. Due to computational requirements, in this paper the magnetic response is considered for a maximum of six molecules in a ring. Connectivity IMC3 (Fig. \ref{fig:truncatedicosahedronconnectivity}(c)) achieves the lowest possible zero-field energy for $N=3$ and 6.  For IMC4 (Fig. \ref{fig:truncatedicosahedronconnectivity}(d)) the corresponding rings have $N=5$ and 6.

The Hamiltonian of the isotropic nearest-neighbor AHM in a magnetic field $\vec{h}$ describes the interactions between the spins. They are parametrized by $\omega$, with $cos\omega$ the interactions between spins in the same icosahedron and $sin\omega$ the ones between spins belonging to different icosahedra. The Hamiltonian is:

\begin{eqnarray}
H & = & cos\omega \sum_{i=1}^N \sum_{<jk>} \vec{s}_{ij} \cdot \vec{s}_{ik} + sin\omega \sum_{i=1}^N \sum_{<jk>'} \vec{s}_{ij} \cdot \vec{s}_{i+1k} \nonumber \\ & & - h \sum_{i=1}^N \sum_{j=1}^{12} s_{ij}^z
\label{eqn:model}
\end{eqnarray}

The brackets indicate that interactions are nonzero only between nearest neighbors. Periodic boundary conditions are assumed. The icosahedral ring has $30N$ bonds of type $cos\omega$ and $2N$ bonds of type $sin\omega$. The magnetic field points along the $z$ axis.

The lowest-energy configuration was calculated numerically as a function of the magnetic field \cite{Coffey92,NPK07,NPK16-1}.
%\cite{NPK07,Machens13,NPK13,NPK15,NPK15-1,NPK16,NPK16-1,NPK17,NPK17-1,NPK18,NPK21}.
%The spins $\vec{s}_i$ are classical unit vectors, and
For a specific field the initial spin configuration is selected by assigning random values to the polar and azimuthal angles of the spins. Then each angle is moved opposite the direction of its gradient, until the energy is minimized with double-precision accuracy. To ensure that the global lowest-energy configuration is found this procedure is repeated for a sufficient number of different random initial configurations. %A theory for the calculation of the ground state in a magnetic field has also been formulated \cite{Schmidt20-1}.

\section{IMC1 and IMC2}
\label{sec:imc1andimc2}

For a ring of icosahedra, frustration is minimum in the zero-field ground state when nearest-neighbor spins have a correlation of $-\frac{\sqrt{5}}{5}$ if they belong to the same icosahedron \cite{Schmidt03}, and are antiparallel if they belong to different icosahedra. For even-membered rings this is the case at least for the intermolecular connectivities IMC1 and IMC2 (Fig. \ref{fig:truncatedicosahedronconnectivity}(a),(b)). For odd-membered rings minimum frustration occurs at least for IMC2. In all these cases the magnetic response is identical and it suffices to consider
%\section{Even-Membered Rings}
%\label{sec:evenmemberedrings}
the smallest assemblage of icosahedra, a dimer ($N=2$).
%For IMC1 and IMC2 zero magnetic field nearest-neighbor spins belonging to the same icosahedron have a correlation equal to$-\frac{\sqrt{5}}{5}$ and are antiparallel if they belong to different icosahedra, achieving minimum frustration.

\subsection{Dimer of Icosahedra ($N=2$)}
\label{subsec:twoicosahedra}

\begin{figure}[h]
\begin{center}
\includegraphics[width=3.5in,height=2.5in]{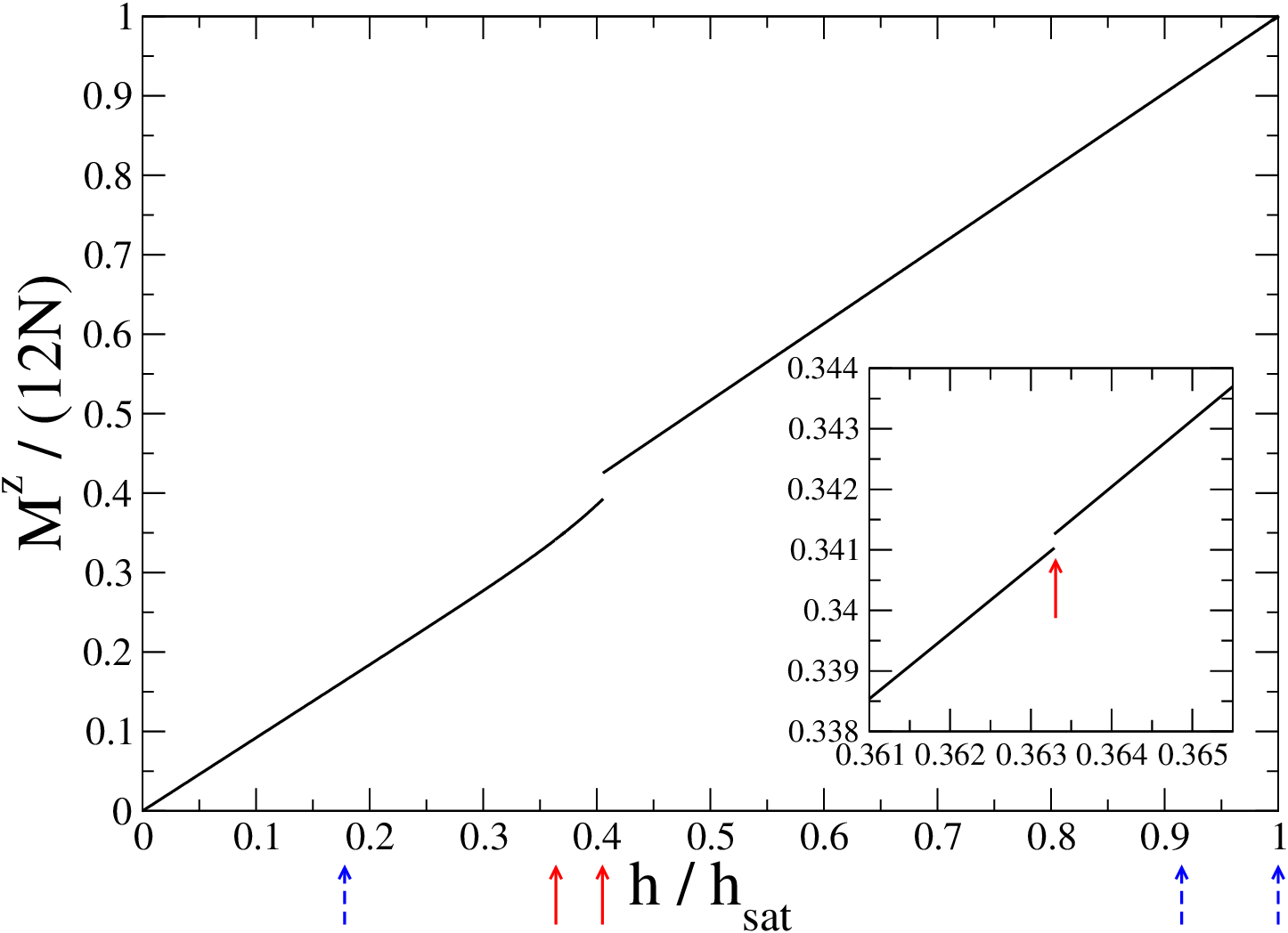}
\end{center}
\caption{Magnetization per spin $\frac{M^z}{12N}$ along the field as a function of the magnetic field over its saturation value $\frac{h}{h_{sat}}$ in the ground state of Hamiltonian (\ref{eqn:model}) for a dimer of icosahedra ($N=2$) and $\omega=\frac{\pi}{1000}$ for connectivities IMC1 and IMC2. The (red) solid arrows point at the locations of the magnetization discontinuities. The (blue) dashed arrows point at the locations of the susceptibility discontinuities.
%(~/basic/classical/icosahedronarraynew/N=2/omega=0.001PI)
}
\label{fig:magnN=2omega=piover1000}
\end{figure}

To establish how the dimer magnetization response emerges from the one of the isolated icosahedron, the case of a very small $\omega=\frac{\pi}{1000}$ is considered. Fig. \ref{fig:magnN=2omega=piover1000} plots the total reduced magnetization $\frac{M^z}{12N}$ of the ground state of Hamiltonian (\ref{eqn:model}) as a function of the field. It is identical to the total reduced magnetization of each one of the two molecules, which have no components perpendicular to the field. By comparison with the isolated icosahedron discontinuity in Ref. \cite{Schroeder05}, which occurs at $\frac{h}{h_{sat}}=0.40603$ between reduced magnetizations $\frac{M^z}{12}=0.39288$ and 0.42565, it is seen to survive the intermolecular coupling and to occur at slightly renormalized values (Table \ref{table:N=2omega=pi/1000discontinuities}). In addition, further magnetization and susceptibility discontinuities develop due to the intermolecular interactions.

\begin{table}
\begin{center}
\caption{Ground-state magnetization discontinuities of Hamiltonian (\ref{eqn:model}) for a dimer of icosahedra ($N=2$) and $\omega=\frac{\pi}{1000}$ for connectivities IMC1 and IMC2. The columns give the magnetic field over the saturation field $\frac{h}{h_{sat}}$ for which the discontinuities occur, the total magnetization along the field just below and just above them over the number of sites $12N$, and the corresponding strength of the magnetization jumps. The three susceptibility discontinuities occur at $\frac{h}{h_{sat}}=0.17698$, 0.91790, and 0.99785, at the corresponding values of the magnetization 0.1629, 0.92090, and 0.99811.}
\begin{tabular}{c|c|c|c}
$\frac{h}{h_{sat}}$ & $(\frac{M^z}{12N})_-$ & $(\frac{M^z}{12N})_+$ & $\frac{\Delta M^z}{12N}$ ($\times 10^{-3}$) \\ % & \\
%\hline
%$\chi$ & 0.17698 & 0.1629 & 0.1629 & 0 \\ % & run3 \\
\hline
 0.363290756 & 0.34103056 & 0.34126146 & 0.23090 \\ % & run10 \\
\hline
 0.40553664 & 0.39265871 & 0.42531783 & 32.65912 \\ % & run11 \\
%\hline
%$\chi$ & 0.91790 & 0.92090 & 0.92090 & 0 \\ % & run5 \\
%\hline
%$\chi$ & 0.99785 & 0.99811 & 0.99811 & 0 \\ % & run7
\end{tabular}
\label{table:N=2omega=pi/1000discontinuities}
\end{center}
\end{table}

\begin{figure}[h]
\begin{center}
\includegraphics[width=3.5in,height=2.5in]{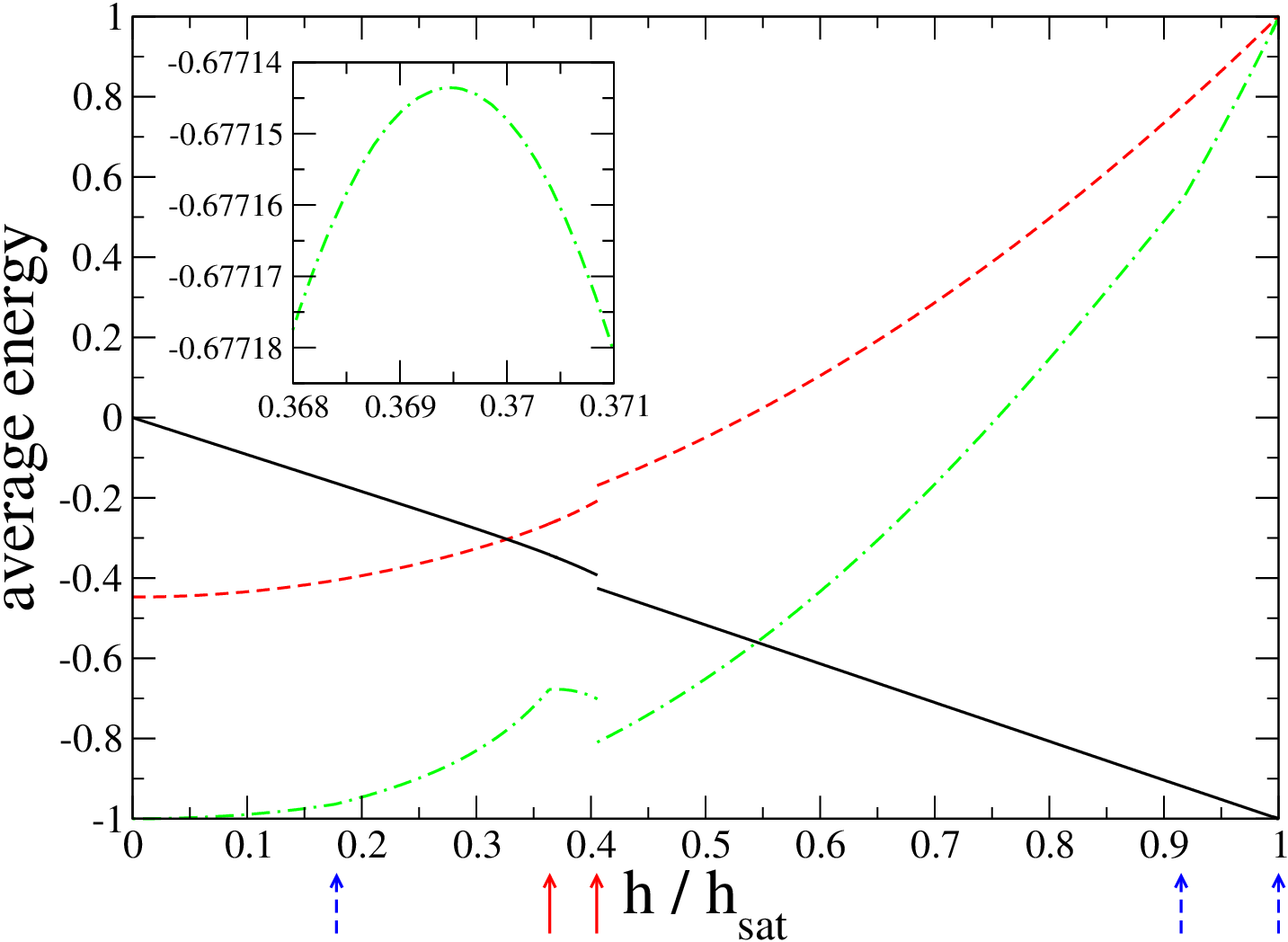}
\end{center}
\caption{Average contribution to the total energy as a function of the magnetic field over its saturation value $\frac{h}{h_{sat}}$ in the ground state of Hamiltonian (\ref{eqn:model}) for a dimer of icosahedra ($N=2$) and $\omega=\frac{\pi}{1000}$ for connectivities IMC1 and IMC2. The (red) dashed line is for bonds in the same icosahedron, the (green) dot-dashed line for bonds in neighboring icosahedra, and the (black) solid line for the spin magnetic energy. The (red) solid arrows point at the locations of the magnetization discontinuities. The (blue) dashed arrows point at the locations of the susceptibility discontinuities.
%(~/basic/classical/icosahedronarraynew/N=2/omega=0.001PI)
}
\label{fig:energycontrN=2omega=piover1000}
\end{figure}

\begin{figure}[h]
\begin{center}
\includegraphics[width=3.5in,height=2.5in]{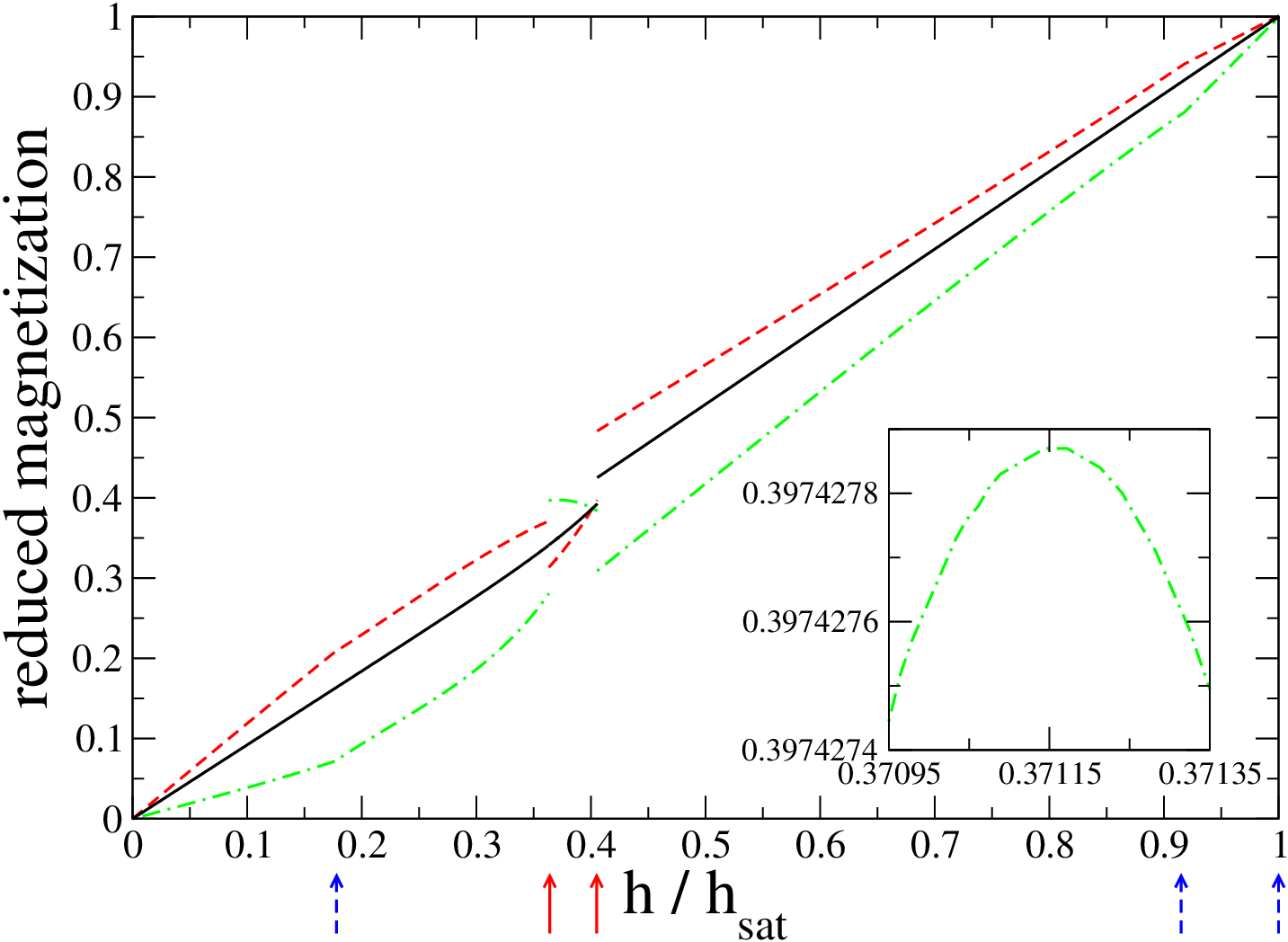}
\end{center}
\caption{Average contribution to the total magnetization along the field as a function of the magnetic field over its saturation value $\frac{h}{h_{sat}}$ in the ground state of Hamiltonian (\ref{eqn:model}) for a dimer of icosahedra ($N=2$) and $\omega=\frac{\pi}{1000}$ for connectivities IMC1 and IMC2. The (red) dashed line is the total magnetization per spin for spins not participating in intermolecular correlations $\frac{M_{np}^z}{8N}$, the (green) dot-dashed line is for ones that do $\frac{M_p^z}{4N}$, and the (black) solid line is the total reduced magnetization $\frac{M^z}{12N}$. The (red) solid arrows point at the locations of the magnetization discontinuities. The (blue) dashed arrows point at the locations of the susceptibility discontinuities.
%(~/basic/classical/icosahedronarraynew/N=2/omega=0.001PI)
}
\label{fig:magncontrN=2omega=piover1000}
\end{figure}

\begin{figure}[h]
\begin{center}
\includegraphics[width=3.5in,height=2.5in]{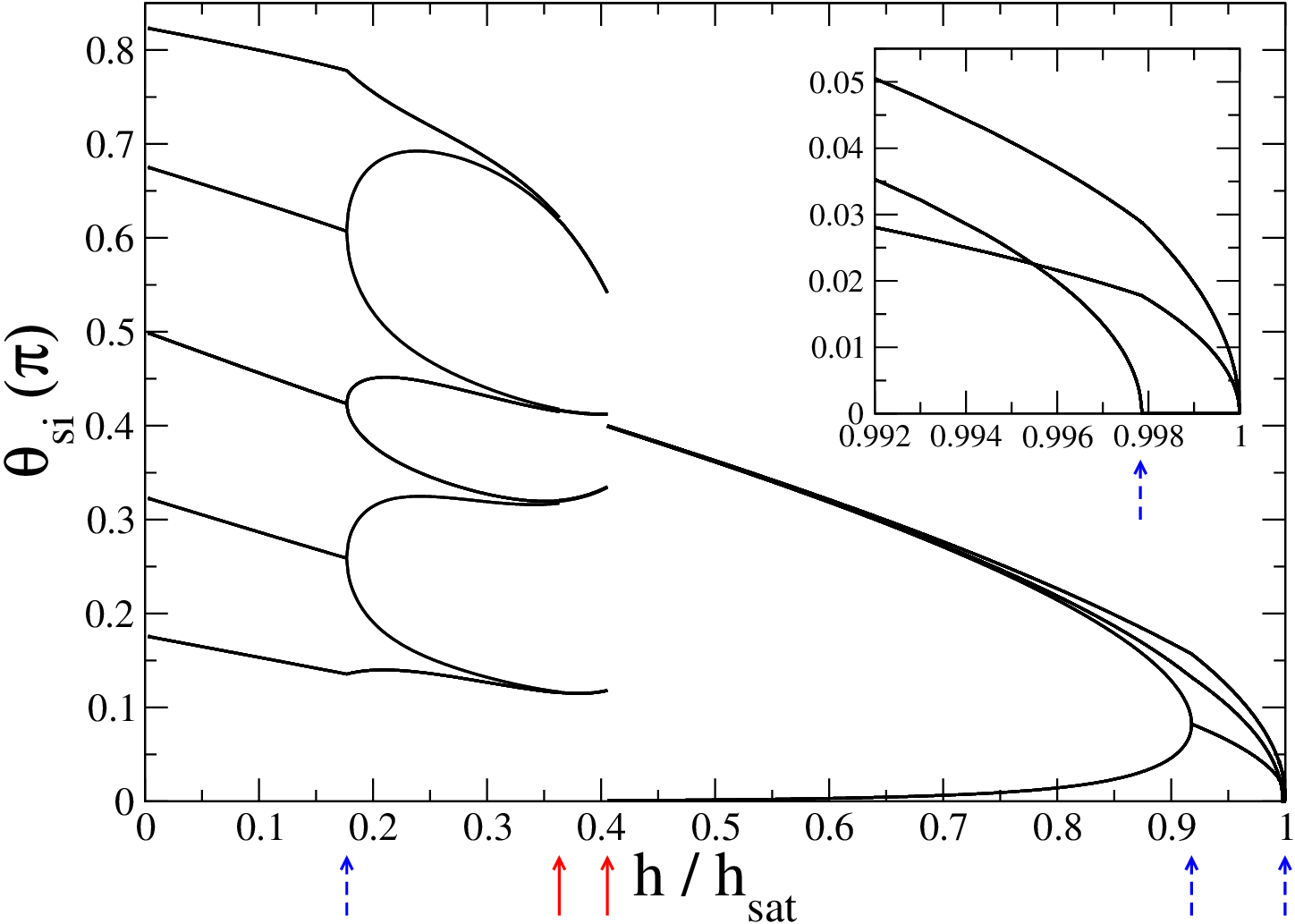}
\end{center}
\caption{Polar angles $\theta_{si}$ of the spins residing on the vertices of the icosahedra as a function of the magnetic field over its saturation value $\frac{h}{h_{sat}}$ in the ground state of Hamiltonian (\ref{eqn:model}) for a dimer of icosahedra ($N=2$) and $\omega=\frac{\pi}{1000}$ for connectivities IMC1 and IMC2. The (red) solid arrows point at the locations of the magnetization discontinuities. The (blue) dashed arrows point at the locations of the susceptibility discontinuities.
%(~/basic/classical/icosahedronarraynew/N=2/omega=0.001PI)
}
\label{fig:polaranglesN=2omega=piover1000}
\end{figure}

%Fig. \ref{fig:polaranglesN=2omega=piover1000} shows the ground-state spin polar angles as a function of the magnetic field, while
Fig. \ref{fig:energycontrN=2omega=piover1000} plots the average energy for the two different bond types and the average magnetic energy per spin. Fig. \ref{fig:magncontrN=2omega=piover1000} shows the reduced magnetization per spin along the field for the spins that do not participate in intermolecular bonds, $\frac{M_{np}^z}{8N}$, and for the ones that do, $\frac{M_p^z}{4N}$, as well as the one for all the spins, $\frac{M^z}{12N}$. The magnetic energy is decreasing with the field at the expense of the exchange energy, and the reduced magnetizations are mostly increasing functions of the field. At $\frac{h}{h_{sat}}=0.17698$ a susceptibility discontinuity occurs, with three of the five distinct spin polar angles $\theta_{si}$ splitting in two in a continuous fashion (Fig. \ref{fig:polaranglesN=2omega=piover1000}). Then a magnetization discontinuity appears at $\frac{h}{h_{sat}}=0.363290756$ bringing the lowest-energy configuration to the form before the isolated-icosahedron jump \cite{NPK15}, by reducing the number of distinct spin polar angles by a factor of two in a noncontinuous way.

Only between the two magnetization jumps does $\frac{M_p^z}{4N}$ become bigger than $\frac{M_{np}^z}{8N}$ (Fig. \ref{fig:magncontrN=2omega=piover1000}). The contributing spins in the former case are antiparallel in zero field and expected to be more resistant to the influence of the magnetic energy. $\frac{M_p^z}{4N}$ even decreases for $\frac{h}{h_{sat}}>0.3712$ and up to the next magnetization jump. It again becomes lower than $\frac{M_{np}^z}{8N}$ for $\frac{h}{h_{sat}}>0.4013793$. Again contrary to what is expected, the intermolecular correlations decrease for fields higher than $\frac{h}{h_{sat}}=0.3695$ and up to the second magnetization discontinuity (Fig. \ref{fig:energycontrN=2omega=piover1000}). This jump, which corresponds to the one of the isolated icosahedron, reduces $\frac{M_p^z}{4N}$, as well as the intermolecular bond exchange energy. As in the isolated-icosahedron case it makes the magnetic energies of all individual spins negative, with two spins for each molecule almost and not perfectly aligned with the field, as in an isolated icosahedron, due to the finite value of $\omega$ (Fig. \ref{fig:polaranglesN=2omega=piover1000}).

Two susceptibility discontinuities appear for higher fields. They occur when two distinct polar angles $\theta_{si}$ become equal at $\frac{h}{h_{sat}}=0.91790$ (Fig. \ref{fig:polaranglesN=2omega=piover1000}), and when one of the distinct polar angles becomes zero at $\frac{h}{h_{sat}}=0.99785$. This is the unique polar angle that corresponds to the interacting spins belonging to different icosahedra. Since these are aligned with the field, the two interacting icosahedra are equivalent to two noninteracting eight-site clusters in a magnetic field.

\begin{figure}[h]
\begin{center}
\includegraphics[width=3.5in,height=2.5in]{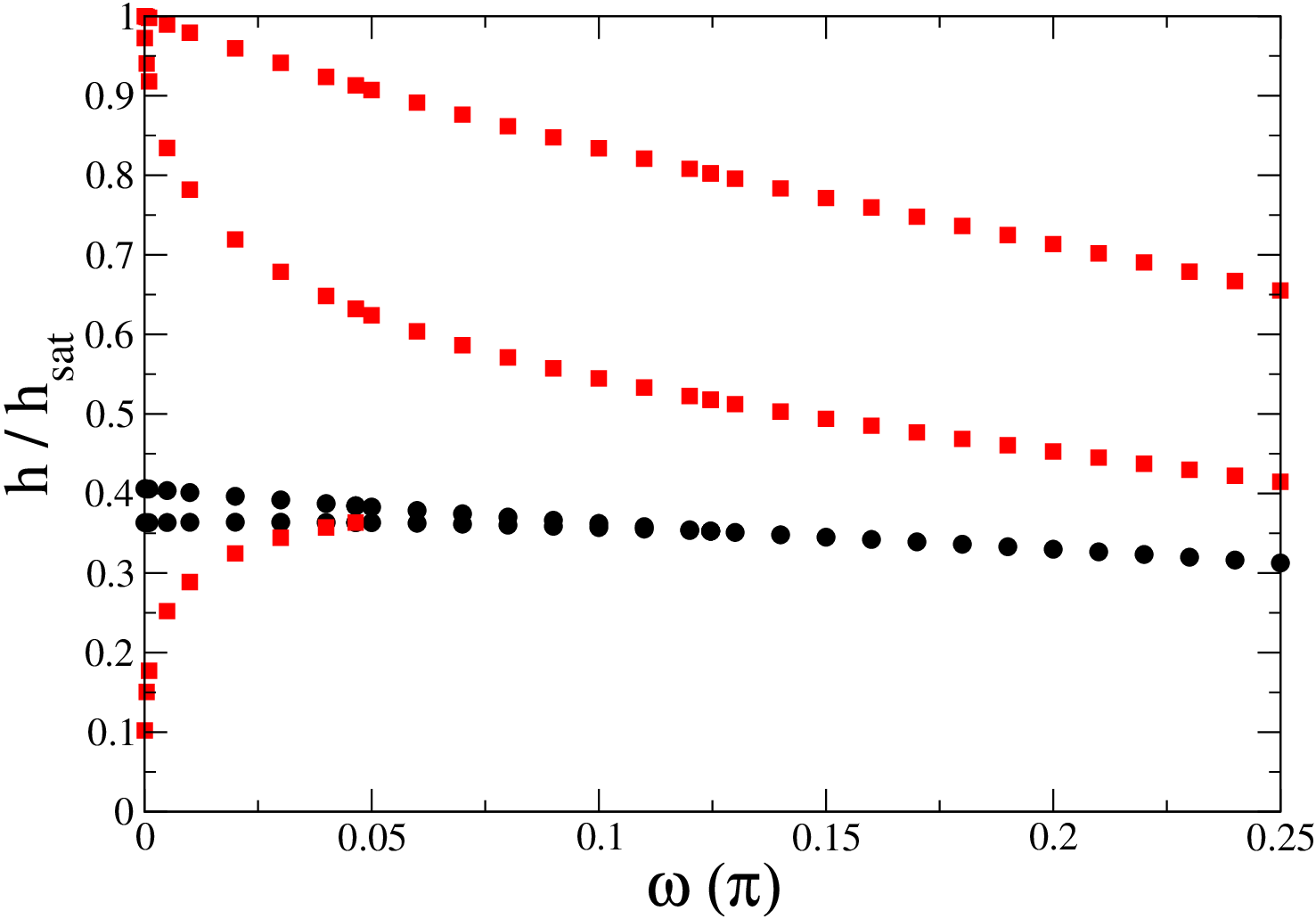}
\end{center}
\caption{Magnetic field over its saturation value $\frac{h}{h_{sat}}$ for the magnetization (black circles) and susceptibility (red squares) discontinuities in the ground state of Hamiltonian (\ref{eqn:model}) for a dimer of icosahedra ($N=2$) as a function of $\omega$ for connectivities IMC1 and IMC2. The saturation field $h_{sat}=5cos\omega+sin\omega+\sqrt{4cos^2\omega+sin(2\omega)+1}$ (App. \ref{app:saturationfield}).
%(~/basic/classical/icosahedronarraynew/N=2)
}
\label{fig:discN=2}
\end{figure}

\begin{figure}[h]
\begin{center}
\includegraphics[width=2.5in,height=3.5in,angle=270]{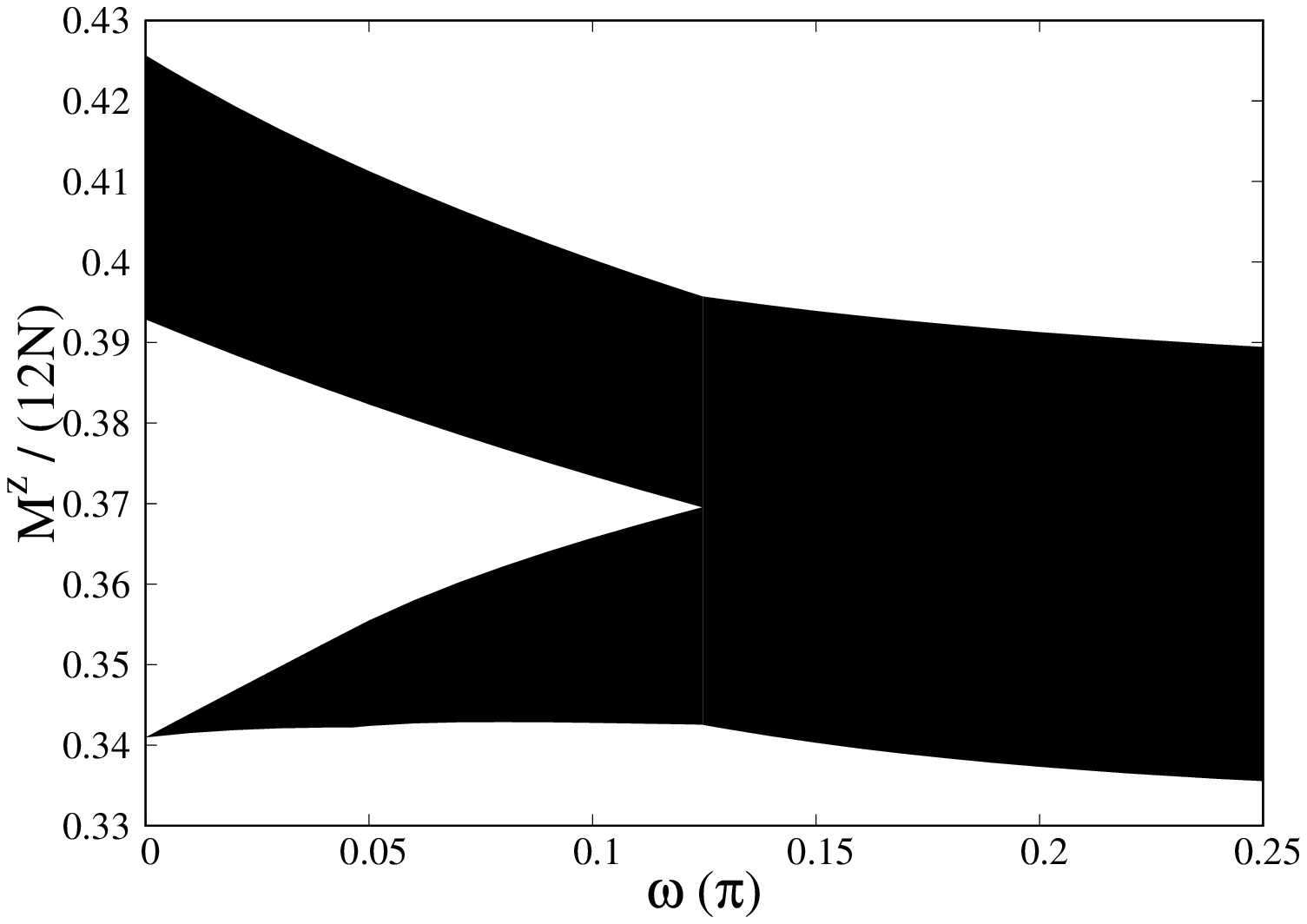}
\end{center}
\caption{Magnetization per spin $\frac{M^z}{12N}$ values that never correspond to the lowest-energy configuration due to the discontinuities in the ground state of Hamiltonian (\ref{eqn:model}) for a dimer of icosahedra ($N=2$) as a function of $\omega$ for connectivities IMC1 and IMC2.
%(~/basic/classical/icosahedronarraynew/N=2)
}
\label{fig:icosahedronarraydiscontinuitiesmagnetization}
\end{figure}

Fig. \ref{fig:discN=2} plots the evolution of the locations of the magnetization and susceptibility discontinuities as a function of $\omega$. The susceptibility jumps emerge at zero and saturation field as $\omega \to 0$. The low-field susceptibility jump disappears for $\omega>0.046490\pi$, while the two magnetization discontinuities merge into one for $\omega>0.12461\pi$, resulting in a single but wider gap. This is shown in Fig. \ref{fig:icosahedronarraydiscontinuitiesmagnetization} that plots the reduced magnetization values that never correspond to the lowest-energy configuration due to the discontinuities. The lower-field jump widens as the coupling of the two icosahedra strengthens. This results in a wider range of inaccessible magnetizations in comparison with the isolated icosahedron. The isolated-icosahedron discontinuity shifts to lower magnetization values for higher $\omega$.
%, either in two or in one part.
The high-field ground state, where the interacting spins belonging to different icosahedra are pointing along the $z$ axis, occupies a wider field range with increasing $\omega$ and allows for the calculation of the saturation field $h_{sat}=5cos\omega+sin\omega+\sqrt{4cos^2\omega+sin(2\omega)+1}$ (App. \ref{app:saturationfield}).

\begin{figure}[h]
\begin{center}
\includegraphics[width=3.5in,height=2.5in]{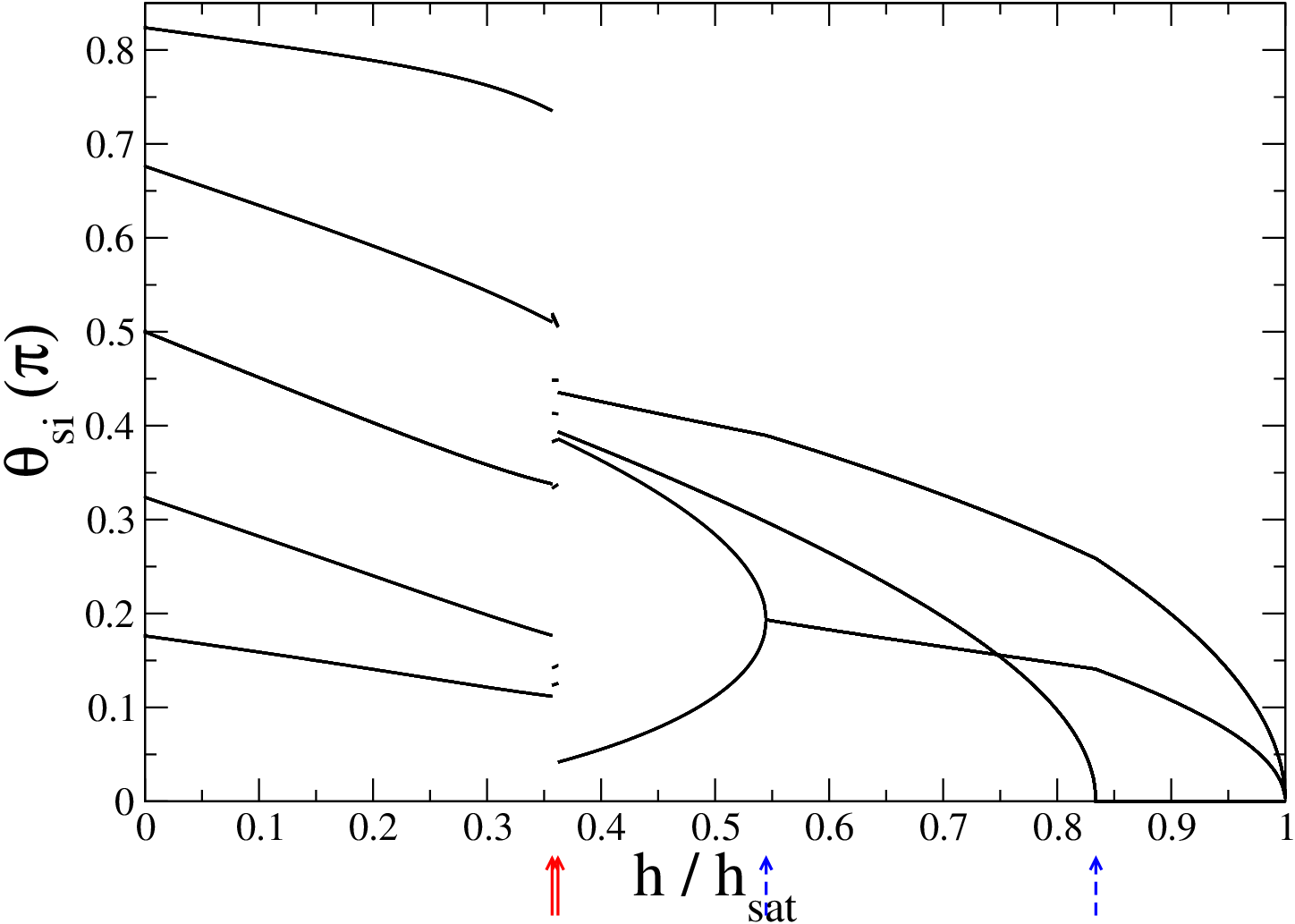}
\end{center}
\caption{Polar angles $\theta_{si}$ of the spins residing on the vertices of the icosahedra as a function of the magnetic field over its saturation value $\frac{h}{h_{sat}}$ in the ground state of Hamiltonian (\ref{eqn:model}) for a dimer of icosahedra ($N=2$) and $\omega=\frac{\pi}{10}$ for connectivities IMC1 and IMC2. The (red) solid arrows point at the locations of the magnetization discontinuities. The (blue) dashed arrows point at the locations of the susceptibility discontinuities.
%(~/basic/classical/icosahedronarraynew/N=2/omega=0.1PI)
}
\label{fig:polaranglesN=2omega=piover10}
\end{figure}

The value $\omega=\frac{\pi}{10}$ corresponds to a stronger intermolecular coupling for which the  two magnetization jumps are comparable in size (Table \ref{table:N=2omega=pi/10discontinuities} and Fig. \ref{fig:icosahedronarraydiscontinuitiesmagnetization}). Fig. \ref{fig:polaranglesN=2omega=piover10} shows the evolution of the polar angles as a function of the field for $\omega=\frac{\pi}{10}$. The low-field susceptibility jump has vanished. The ground state between the two magnetization discontinuities is now confined to a narrow window of the field. The spins that used to be practically aligned with the field after the second magnetization jump now significantly deviate from it due to the stronger intermolecular interaction, unlike the case of the isolated icosahedron.

%The ground-state polar angles and nearest-neighbor correlations are shown in Figs \ref{fig:polaranglesN=2omega=piover10} and \ref{fig:dotproductsN=2omega=piover10} as a function of the magnetic field. For low fields the spins form five unique polar angles with the $z$ axis and the antiferromagnetic correlations prevail over the magnetic energy, similarly to what happens for a dimer of dodecahedra \cite{NPK16}. A magnetization discontinuity at $\frac{h}{h_{sat}}=0.3570803484285$ (Table \ref{table:N=2omega=pi/10discontinuities}) leads to a less symmetric lowest-energy configuration with eight unique polar angles. Another jump occurs at $\frac{h}{h_{sat}}=0.3622518336845$, with the number of unique polar angles reduced to four. A susceptibility discontinuity follows at $\frac{h}{h_{sat}}= 0.54466$ after which two unique polar angles become equal and four pairs of spins become parallel. Finally at $\frac{h}{h_{sat}}=0.83397$ one of the unique polar angles becomes zero, resulting in another susceptibility jump and four nearest-neighbor pairs of spins aligning with the field. The saturation field $h_{sat}=5cos\omega+sin\omega+\sqrt{4cos^2\omega+sin(2\omega)+1}$ (App. \ref{app:saturationfield}).

%\begin{figure}[h]
%\begin{center}
%\includegraphics[width=3.5in,height=2.5in]{dotproductsN=2phi=piover10}
%\end{center}
%\caption{Nearest-neighbor correlations $\vec{s}_i \cdot \vec{s}_j$ of the spins residing on the vertices of the icosahedra as a function of the magnetic field over its saturation value $\frac{h}{h_{sat}}$ in the ground state of Hamiltonian (\ref{eqn:model}) for a dimer of $N=2$ icosahedra for $\omega=\frac{\pi}{10}$.
%%(~/basic/classical/icosahedronarraynew/N=2/omega=0.1PI)
%}
%\label{fig:dotproductsN=2omega=piover10}
%\end{figure}

\begin{table}
\begin{center}
\caption{Ground-state magnetization discontinuities of Hamiltonian (\ref{eqn:model}) for a dimer of icosahedra ($N=2$) and $\omega=\frac{\pi}{10}$ for connectivities IMC1 and IMC2. The columns give the magnetic field over the saturation field $\frac{h}{h_{sat}}$ for which the discontinuities occur, the total magnetization along the field just below and just above them over the number of sites $12N$, and the corresponding strength of the magnetization jumps. The two susceptibility discontinuities occur at $\frac{h}{h_{sat}}=0.54466$ and 0.83397, at the corresponding values of the magnetization 0.58487 and 0.86389.}
\begin{tabular}{c|c|c|c}
$\frac{h}{h_{sat}}$ & $(\frac{M^z}{12N})_-$ & $(\frac{M^z}{12N})_+$ & $\frac{\Delta M^z}{12N}$ ($\times 10^{-3}$) \\ % & \\
\hline
 0.3570803484285 & 0.34278164 & 0.36574414 & 22.96250 \\ % & run1 \\
\hline
 0.3622518336845 & 0.37343617 & 0.40032108 & 26.88491 \\ % & run2 \\
%\hline
%$\chi$ & 0.54466 & 0.58487 & 0.58487 & 0 \\ % & run3 \\
%\hline
%$\chi$ & 0.83397 & 0.86389 & 0.86389 & 0 \\ % & run4
\end{tabular}
\label{table:N=2omega=pi/10discontinuities}
\end{center}
\end{table}

\section{IMC3}
\label{sec:imc3}

\subsection{Triangle of Icosahedra ($N=3$)}
\label{subsec:N=3}

%For a triangle of icosahedra the lowest possible energy in zero field is achieved with IMC2, in which case the magnetic response is identical to the one of Sec. \ref{sec:imc1andimc2}. Minimum frustration in the absence of a field is also achieved with IMC3, and this case is considered below.
The smallest ring that achieves minimum frustration in the ground state in zero magnetic field with connectivity IMC3 is the triangle.
%
%The zero-field ground state of Hamiltonian (\ref{eqn:model}) on an isolated icosahedron has all nearest-neighbor correlations equal to $-\frac{\sqrt{5}}{5}$ \cite{Schmidt03}. The introduction of an external field leads to a magnetization discontinuity as the field is ramped up \cite{Schroeder05}. Such a discontinuity is not expected in the absence of magnetic anisotropy and originates from the frustrated connectivity of the molecule.
%
%The minimum number of icosahedra forming a ring with no extra frustration from the one of an isolated molecule is $N=3$.
%The minimum number of interacting icosahedra arranged in a ring with a zero-field ground state with nearest-neighbor intramolecular correlations equal to the ones of an isolated icosahedron is three. In this triangle interacting spins belonging to neighboring icosahedra are antiparallel in the ground state.
The magnetic units are now taken to be the total magnetizations of the individual icosahedra $\vec{M}_i$, $i=1,2,3$, with each forming a polar angle $\theta_i$ with the field. This is not necessarily zero, as it is for an isolated icosahedron. Unlike the intermolecular connectivities IMC1 and IMC2 for the dimer of icosahedra (Sec. \ref{sec:imc1andimc2}), here the intermolecular connectivity IMC3 allows the $\vec{M}_i$ to differ in magnitude, breaking translational symmetry, and also to develop finite components perpendicular to the field in the ground state, leading to finite values for the $\theta_i$. The triangle of icosahedra is contrasted with the one of single spins, perhaps the simplest frustrated unit. In the ground state of the isotropic nearest-neighbor AHM in a magnetic field on the triangle the field turns the spins toward its direction, with each spin forming the same angle with it. %while the difference in the spin azimuthal angles equals $\frac{2\pi}{3}$.

\begin{figure}[h]
\begin{center}
\includegraphics[width=3.5in,height=2.5in]{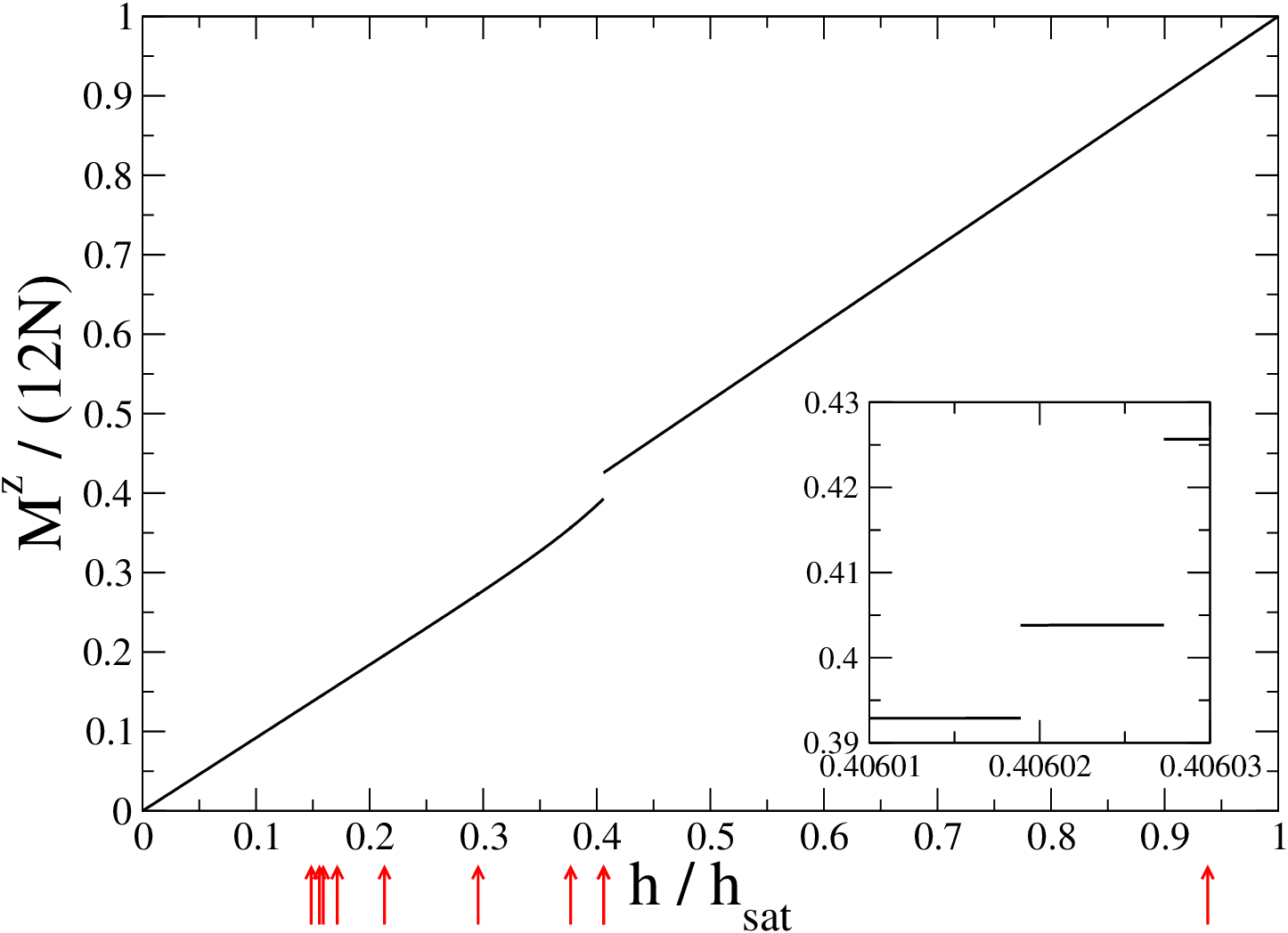}
\end{center}
\caption{Magnetization per spin $\frac{M^z}{12N}$ along the field as a function of the magnetic field over its saturation value $\frac{h}{h_{sat}}$ in the ground state of Hamiltonian (\ref{eqn:model}) for a triangle of icosahedra ($N=3$) and $\omega=\frac{9\pi}{50000}$ for connectivity IMC3. The (red) solid arrows point at the locations of the magnetization discontinuities.
%(~/discoverer/basic/classical/icosahedronarraynew/N=3/third/parallel/omega=0.00018PI)
}
\label{fig:magnetizationN=3omega=9piover50000}
\end{figure}

To establish the connection between the isolated icosahedron response and the one of the triangle of icosahedra, the case of a very small $\omega=\frac{9\pi}{50000}$ is again considered. The intermolecular coupling results in a total of ten ground-state magnetization discontinuities (Table \ref{table:N=3omega=9pi/50000discontinuities}). Figure \ref{fig:magnetizationN=3omega=9piover50000} plots the total reduced magnetization along the field $\frac{M^z}{12N}$. Comparing with Ref. \cite{Schroeder05} and unlike the dimer of icosahedra, the isolated-icosahedron discontinuity is split in two parts, at $\frac{h}{h_{sat}}=0.406018880$ and 0.406027285, with the second twice as wide as the first. Across the first part one of the molecules undergoes the isolated-icosahedron jump, while across the second the other two icosahedra follow (inset of Fig. \ref{fig:magnetizationN=3omega=9piover50000}). The total field magnetizations of the individual icosahedra are again slightly renormalized with respect to their isolated counterpart.
%The polar angles the three molecular magnetizations make with the field axis are shown in Fig. \ref{fig:polaranglesN=3omega=9piover50000}.
It is concluded that the magnetization of each individual molecule behaves differently, in contrast to an isolated triangle of spins.

\begin{table}
\begin{center}
\caption{Ground-state magnetization discontinuities of Hamiltonian (\ref{eqn:model}) for a triangle of icosahedra ($N=3$) and $\omega=\frac{9\pi}{50000}$ for connectivity IMC3. The columns give the magnetic field over the saturation field $\frac{h}{h_{sat}}$ for which the discontinuities occur, the total magnetization along the field just below and just above them over the number of sites $12N$, and the corresponding strength of the magnetization jumps.}
\begin{tabular}{c|c|c|c}
 $\frac{h}{h_{sat}}$ & $(\frac{M^z}{12N})_-$ & $(\frac{M^z}{12N})_+$ & $\frac{\Delta M^z}{12N}$ ($\times 10^{-6}$) \\
\hline
 0.14864731 & 0.13681455 & 0.13681510 & 0.55 \\ % & run11 \\
\hline
 0.15575485 & 0.14334708 & 0.14334848 & 1.40 \\ % & run15 \\
\hline
 0.159096207 & 0.146419311 & 0.146423429 & 4.118 \\ % & run16 \\
\hline
 0.171512062 & 0.15783451 & 0.15784125 & 6.74 \\ % & run13 \\
\hline
 0.21302403 & 0.196029476 & 0.196033501 & 4.025 \\ % & run7 \\
\hline
 0.295472492 & 0.272855137 & 0.272857645 & 2.508 \\ % & run9 \\
\hline
 0.376957522 & 0.356427723 & 0.356431658 & 3.935 \\ % & run5 \\
\hline
 0.406018880 & 0.39290446 & 0.40381798 & 10913.52 \\ % & run8 \\
\hline
 0.406027285 & 0.40382834 & 0.42566260 & 21834.26 \\ % & run10 \\
\hline
 0.93787418 & 0.93994592 & 0.93994706 & 1.14 \\ % & run
\end{tabular}
\label{table:N=3omega=9pi/50000discontinuities}
\end{center}
\end{table}

%\begin{figure}[h]
%\begin{center}
%\includegraphics[width=3.5in,height=2.5in]{magnetizationsmagnN=3omega=9piover50000}
%\end{center}
%\caption{Magnitude of the total magnetization per spin $\frac{M_i}{12}$ of each icosahedron $i=1,2,3$ as a function of the magnetic field over its saturation value $\frac{h}{h_{sat}}$ in the ground state of Hamiltonian (\ref{eqn:model}) for a triangle of icosahedra ($N=3$) and $\omega=\frac{9\pi}{50000}$ for connectivity IMC3. The (red) solid arrows point at the locations of the magnetization discontinuities.
%%(~/discoverer/basic/classical/icosahedronarraynew/N=3/third/parallel/omega=0.00018PI)
%}
%\label{fig:magnetizationsmagnN=3omega=9piover50000}
%\end{figure}

\begin{figure}[h]
\begin{center}
\includegraphics[width=3.5in,height=2.5in]{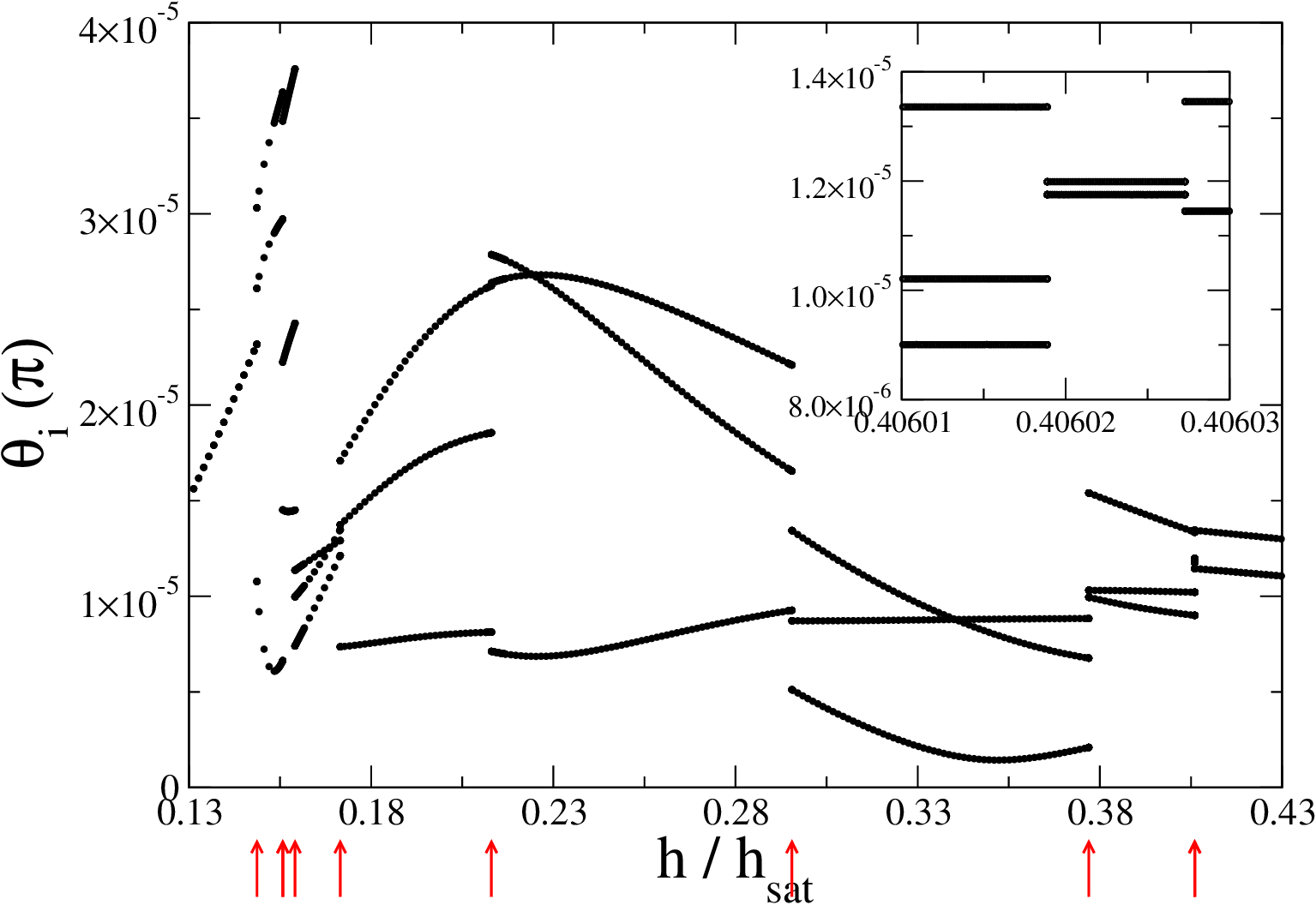}
\end{center}
\caption{Polar angles $\theta_i$ of the magnetizations of the icosahedra $\vec{M}_i$ as a function of the magnetic field over its saturation value $\frac{h}{h_{sat}}$ in the ground state of Hamiltonian (\ref{eqn:model}) for a triangle of icosahedra ($N=3$) and $\omega=\frac{9\pi}{50000}$ for connectivity IMC3. The (red) solid arrows point at the locations of the magnetization discontinuities.
%(~/discoverer/basic/classical/icosahedronarraynew/N=3/third/parallel/omega=0.00018PI)
}
\label{fig:polaranglesN=3omega=9piover50000}
\end{figure}

\begin{figure}[h]
\begin{center}
\includegraphics[width=3.5in,height=2.5in]{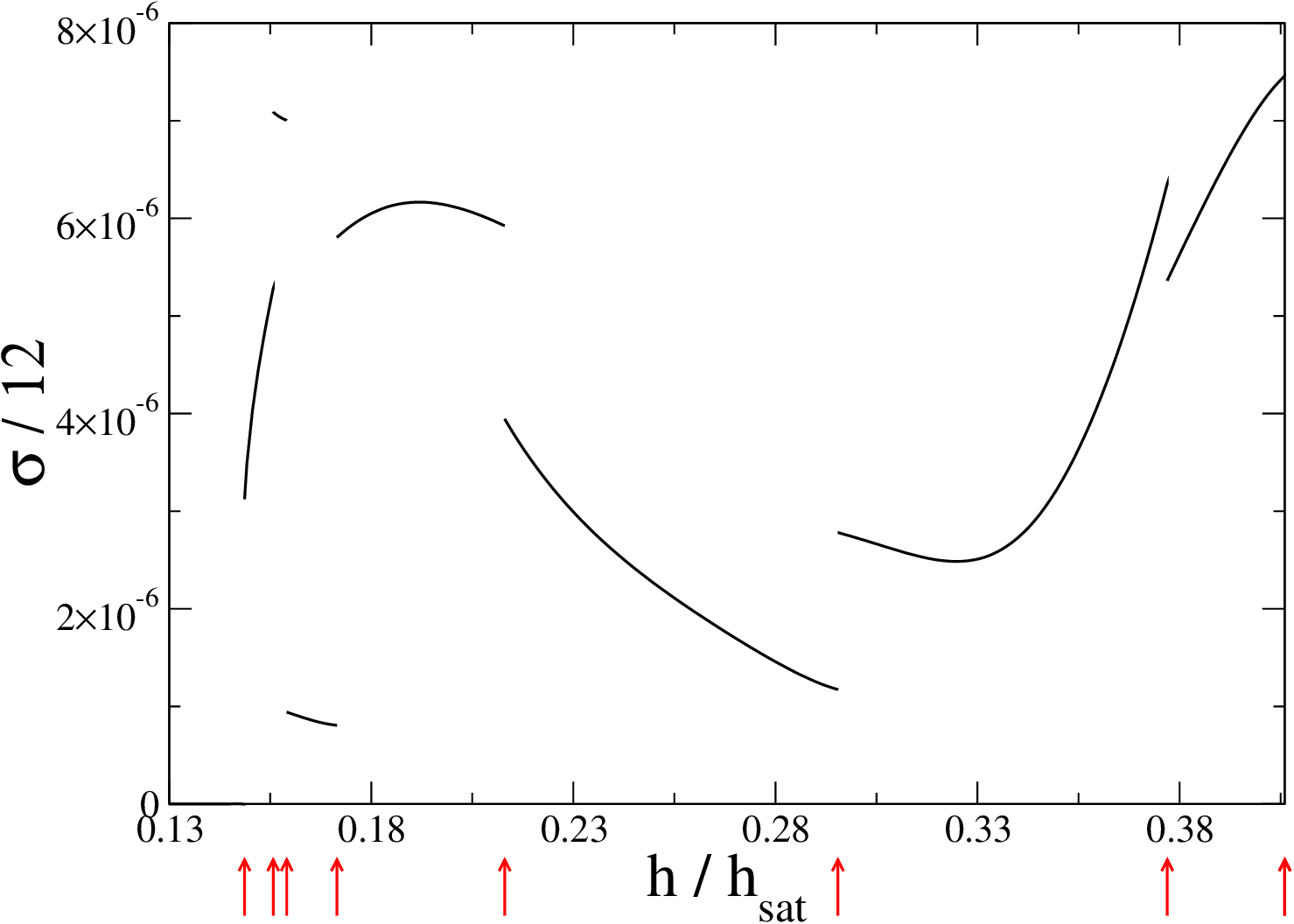}
\end{center}
\caption{Standard deviation $\sigma$ of the magnitude of the molecular magnetization per spin $\frac{M_i}{12}$ as a function of the magnetic field over its saturation value $\frac{h}{h_{sat}}$ in the ground state of Hamiltonian (\ref{eqn:model}) for a triangle of icosahedra ($N=3$) and $\omega=\frac{9\pi}{50000}$ for connectivity IMC3. The (red) solid arrows point at the locations of the magnetization discontinuities.
%(~/discoverer/basic/classical/icosahedronarraynew/N=3/third/parallel/omega=0.00018PI)
}
\label{fig:magnetizationsmagnstandarddeviationN=3omega=9piover50000}
\end{figure}

The nonequal projections of the magnetization of the three molecules originate from their different total magnetizations. Not only do the individual $\vec{M}_i$ differ in magnitude, unlike the case of the dimer of icosahedra, but they can also be directed away from the field axis, albeit at a very small angle (Fig. \ref{fig:polaranglesN=3omega=9piover50000}). At zero field each icosahedron has no residual magnetization. When the field is switched on the icosahedra acquire a common finite $M_i$, emerging for each one at the finite polar angle $\theta_i=4 \times 10^{-6}\pi$ as $h \to 0$. For low magnetic fields the $\vec{M}_i$ form the same polar angle with the $z$ axis,
%. Their azimuthal angles differ by $\frac{2\pi}{3}$, further
resembling a triangle of spins. The common polar angle at first decreases with the field, achieving a local minimum at $\frac{h}{h_{sat}} \sim 0.04$. However, in contrast to what happens for a triangle of spins, for higher fields this angle increases. This is also in contrast with the polar angles of the individual spins of an isolated icosahedron, where not all of them can be increasing functions of $h$, even for a short field range \cite{NPK15}. Still, the projections of the molecular magnetizations along the $z$ axis $M_i^z$ are getting bigger with the field as expected and are equal to $\frac{M^z}{N}$, since the $M_i$ increase with $h$ for each icosahedron, compensating for the bigger polar angles.

The first seven magnetization discontinuities (Fig. \ref{fig:magnetizationN=3omega=9piover50000} and Table \ref{table:N=3omega=9pi/50000discontinuities}) are weak in strength and generated from the intermolecular coupling. They lead to ground states where each icosahedron has its own $M_i$, and Fig. \ref{fig:magnetizationsmagnstandarddeviationN=3omega=9piover50000} plots their standard deviation from their average value. The angles the $M_i$ form with the field are different (Fig. \ref{fig:polaranglesN=3omega=9piover50000}),
%, while their azimuthal angles do not differ now by $\frac{2\pi}{3}$.
and the $M_i^z$ are also different. This lowest-energy configuration is again in contrast with the umbrella state of a triangle of spins. Furthermore, polar angles can still grow bigger with the field, even all of them together.
%Fig. xxx shows the total magnetic energy of the spins that participate in both intramolecular and intermolecular bonds. At the transition it becomes less negative, contrary to the magnetic energy of the spins that only participate in intramolecular bonds (Fig. xxx).

\begin{figure}[h]
\begin{center}
\includegraphics[width=3.5in,height=2.5in]{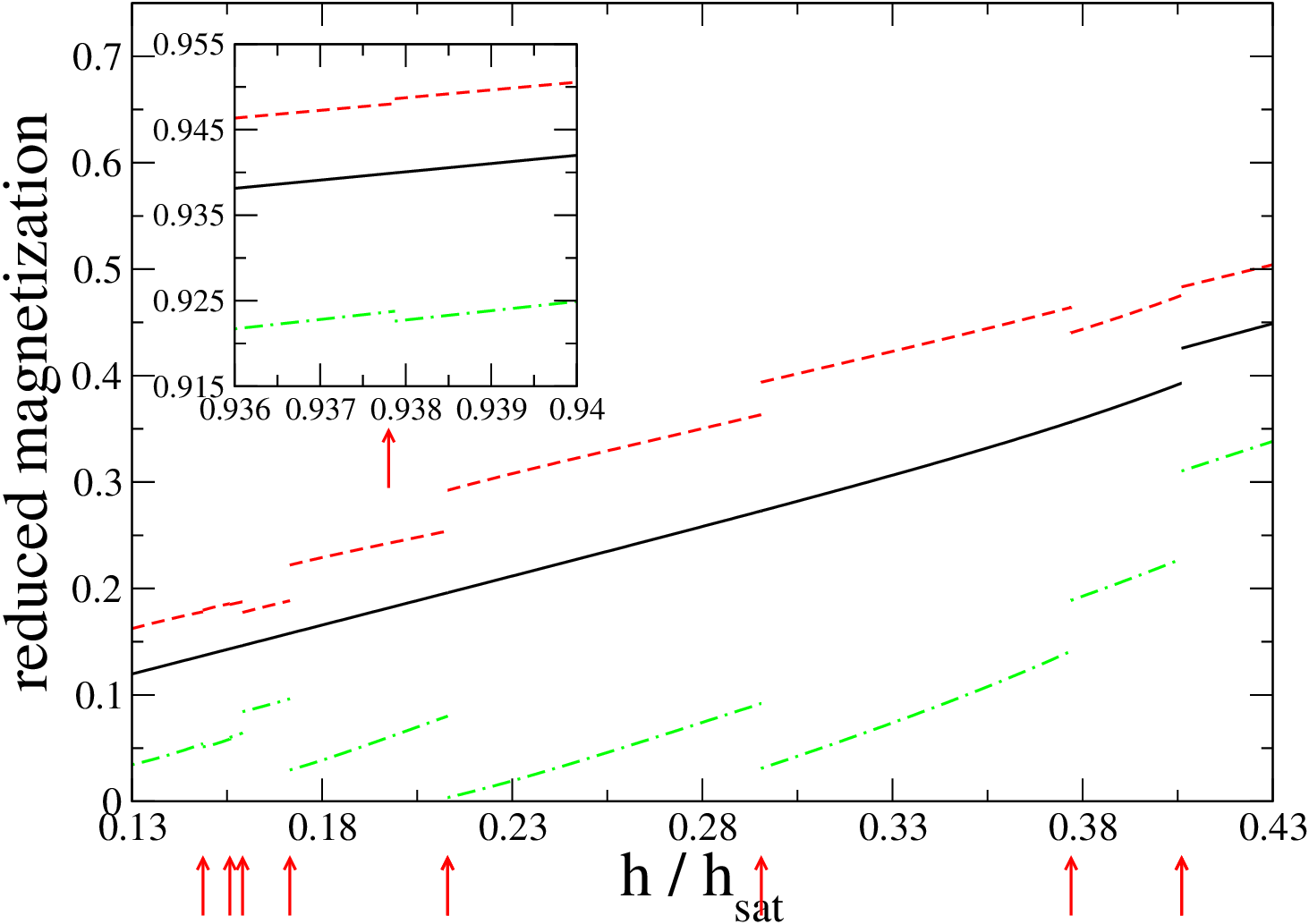}
\end{center}
\caption{Average contribution to the total magnetization along the field as a function of the magnetic field over its saturation value $\frac{h}{h_{sat}}$ in the ground state of Hamiltonian (\ref{eqn:model}) for a triangle of icosahedra ($N=3$) and $\omega=\frac{9\pi}{50000}$ for connectivity IMC3. The (red) dashed line is the total magnetization per spin for spins not participating in intermolecular correlations $\frac{M_{np}^z}{8N}$, the (green) dot-dashed line is for ones that do $\frac{M_p^z}{4N}$, and the (black) solid line is the total reduced magnetization $\frac{M^z}{12N}$. The (red) solid arrows point at the locations of the magnetization discontinuities.
%(~/discoverer/basic/classical/icosahedronarraynew/N=3/third/parallel/omega=0.00018PI)
}
\label{fig:magnetizationlocationN=3omega=piover1000}
\end{figure}

\begin{figure}[h]
\begin{center}
\includegraphics[width=3.5in,height=2.5in]{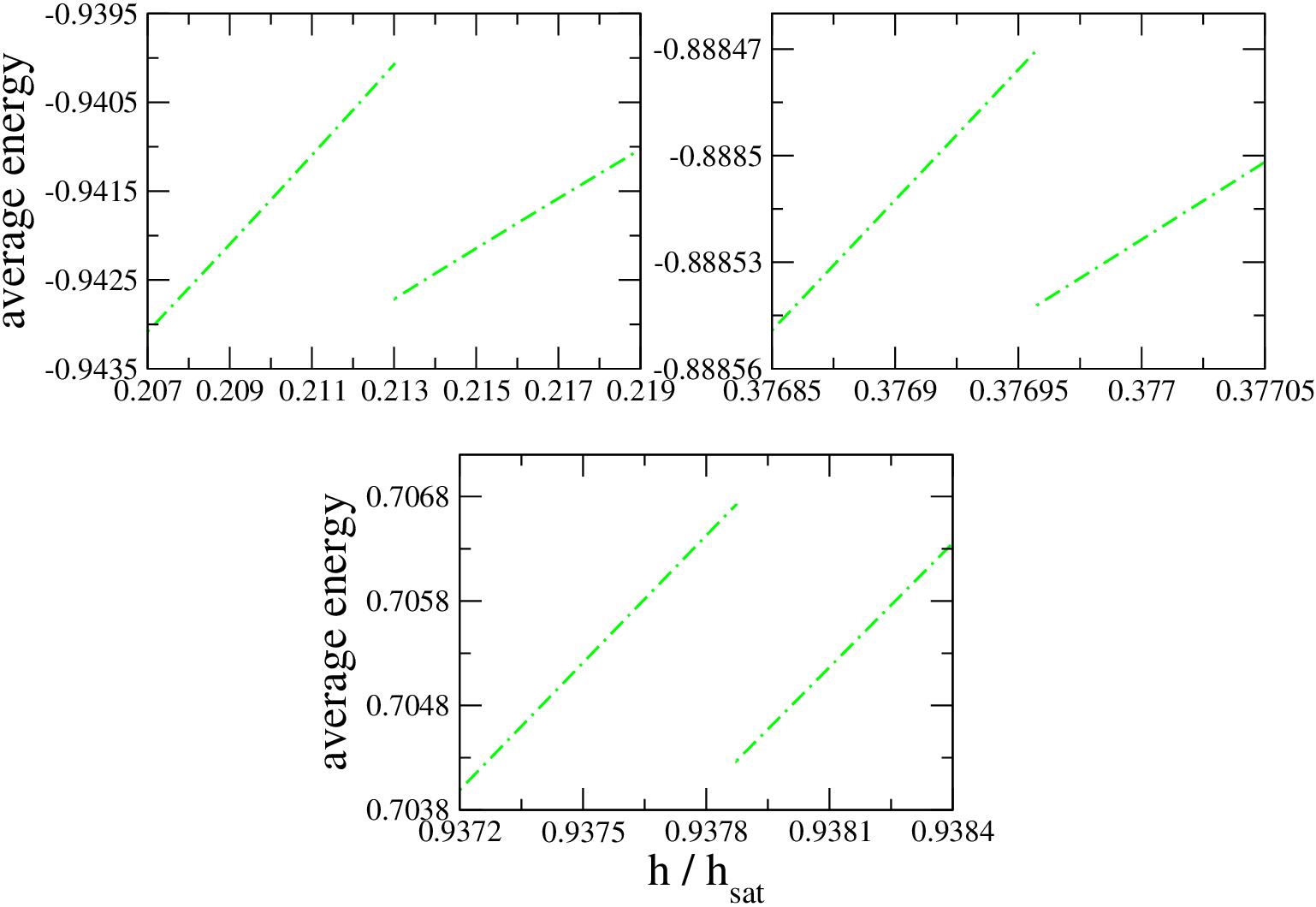}
\end{center}
\vspace{-0.5cm}
\caption{Average contribution to the total energy of bonds in neighboring icosahedra as a function of the magnetic field over its saturation value $\frac{h}{h_{sat}}$ in the ground state of Hamiltonian (\ref{eqn:model}) for a triangle of icosahedra ($N=3$) and $\omega=\frac{9\pi}{50000}$ for connectivity IMC3.
%(~/discoverer/basic/classical/icosahedronarraynew/N=3/third/parallel/omega=0.00018PI)
}
\label{fig:energycontrN=3omega=9piover50000}
\end{figure}

Fig. \ref{fig:magnetizationlocationN=3omega=piover1000} shows that across the first seven jumps the change of the total magnetization along the field of spins not participating in intermolecular bonds is opposite in sign to the total magnetization of spins that do. This was also the case in Sec. \ref{sec:imc1andimc2} for the discontinuity that did not relate to the isolated-icosahedron discontinuity. Furthermore, following the fifth discontinuity the total magnetization field projection of the latter becomes small. It is also noted that across the fifth, the seventh, and the tenth discontinuity the contribution of the intermolecular correlations to the energy becomes more negative (Fig. \ref{fig:energycontrN=3omega=9piover50000}).

Following the eighth jump two icosahedra have the same $M_i$, as already mentioned. Above the ninth discontinuity the three $\vec{M}_i$ have the same magnitude and the polar angles of two of them are equal. Finally, crossing the tenth discontinuity all three $\vec{M}_i$ have the same magnitude and make the same angle with the field axis, resembling a triangle of spins.

\begin{figure}[h]
\begin{center}
\includegraphics[width=3.5in,height=2.5in]{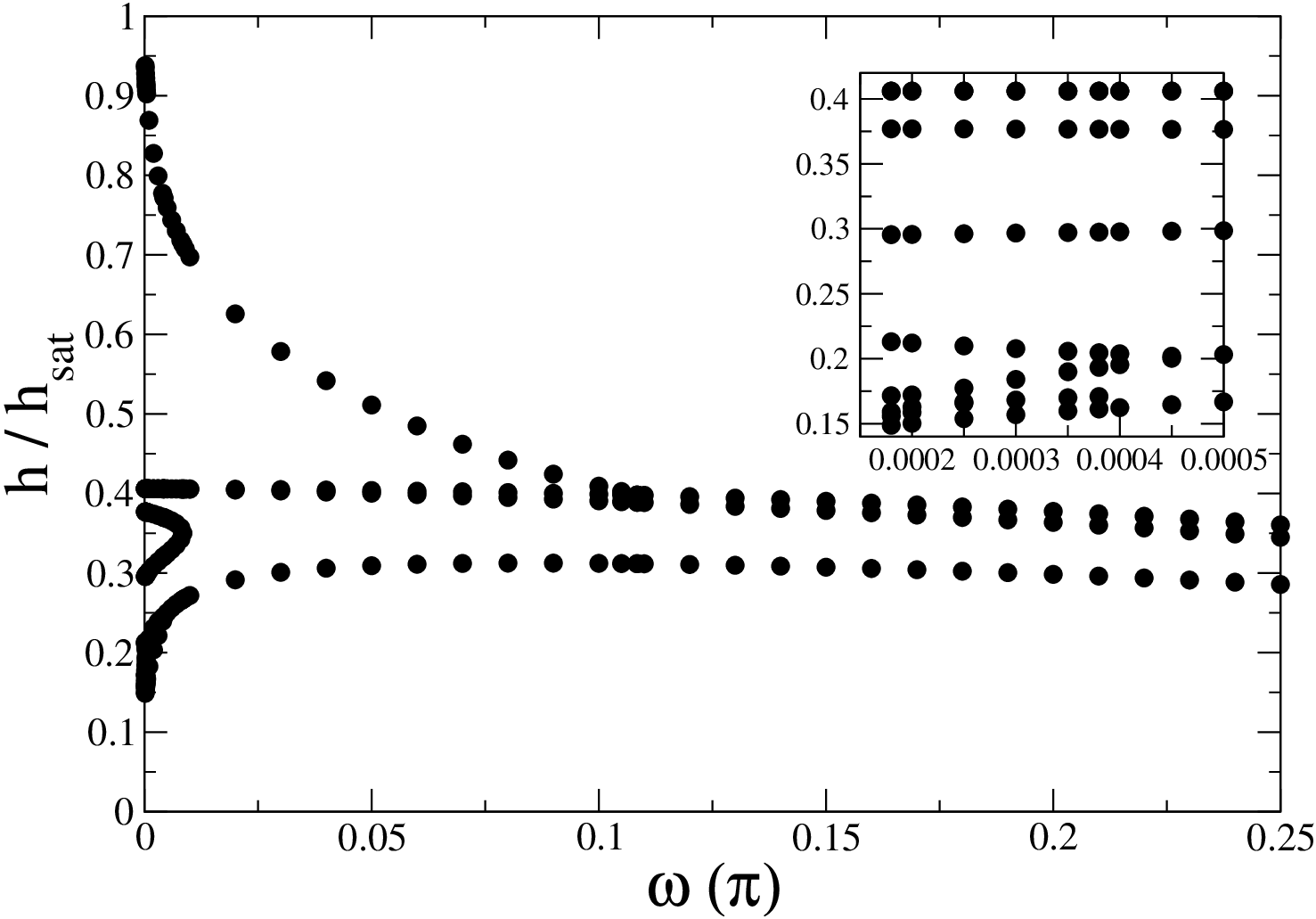}
\end{center}
\caption{Magnetic field over its saturation value $\frac{h}{h_{sat}}$ for the magnetization discontinuities in the ground state of Hamiltonian (\ref{eqn:model}) for a triangle of icosahedra ($N=3$) as a function of $\omega$ for connectivity IMC3.%The saturation field $h_{sat}=5cos\omega+sin\omega+\sqrt{4cos^2\omega+sin(2\omega)+1}$ (App. \ref{app:saturationfield}).
%(~/basic/classical/icosahedronarraynew/N=3/third)
}
\label{fig:discN=3}
\end{figure}

\begin{figure}[h]
\begin{center}
\includegraphics[width=2.5in,height=3.5in,angle=270]{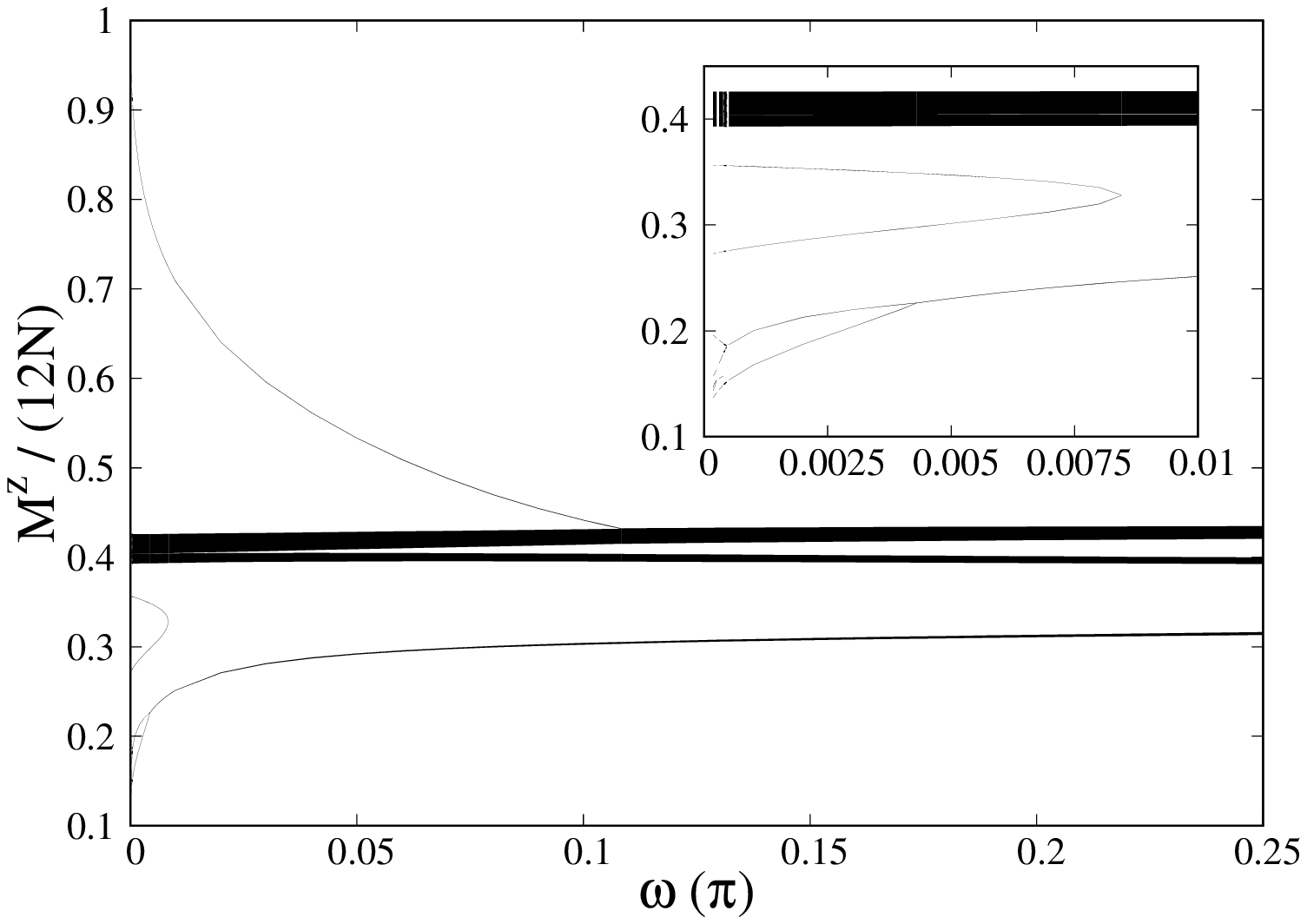}
\end{center}
\caption{Magnetization per spin $\frac{M^z}{12N}$ values that never correspond to the lowest-energy configuration due to the discontinuities in the ground state of Hamiltonian (\ref{eqn:model}) for a triangle of icosahedra ($N=3$) as a function of $\omega$ for connectivity IMC3.
%(~/basic/classical/icosahedronarraynew/N=3/third)
}
\label{fig:icosahedronarraydiscontinuitiesmagnetizationN=3}
\end{figure}

The evolution of the locations of the magnetization discontinuities as a function of $\omega$ is shown in Fig. \ref{fig:discN=3}. The widths of the corresponding magnetization jumps are plotted in Fig. \ref{fig:icosahedronarraydiscontinuitiesmagnetizationN=3}. Three discontinuities eventually appear for higher $\omega$. The two highest-field discontinuities of small $\omega$ eventually merge. For low fields only one jump exists. As in the case of $N=2$, the isolated-icosahedron jump can be traced up to $\omega=\frac{\pi}{4}$.

\begin{figure}[h]
\begin{center}
\includegraphics[width=3.5in,height=2.5in]{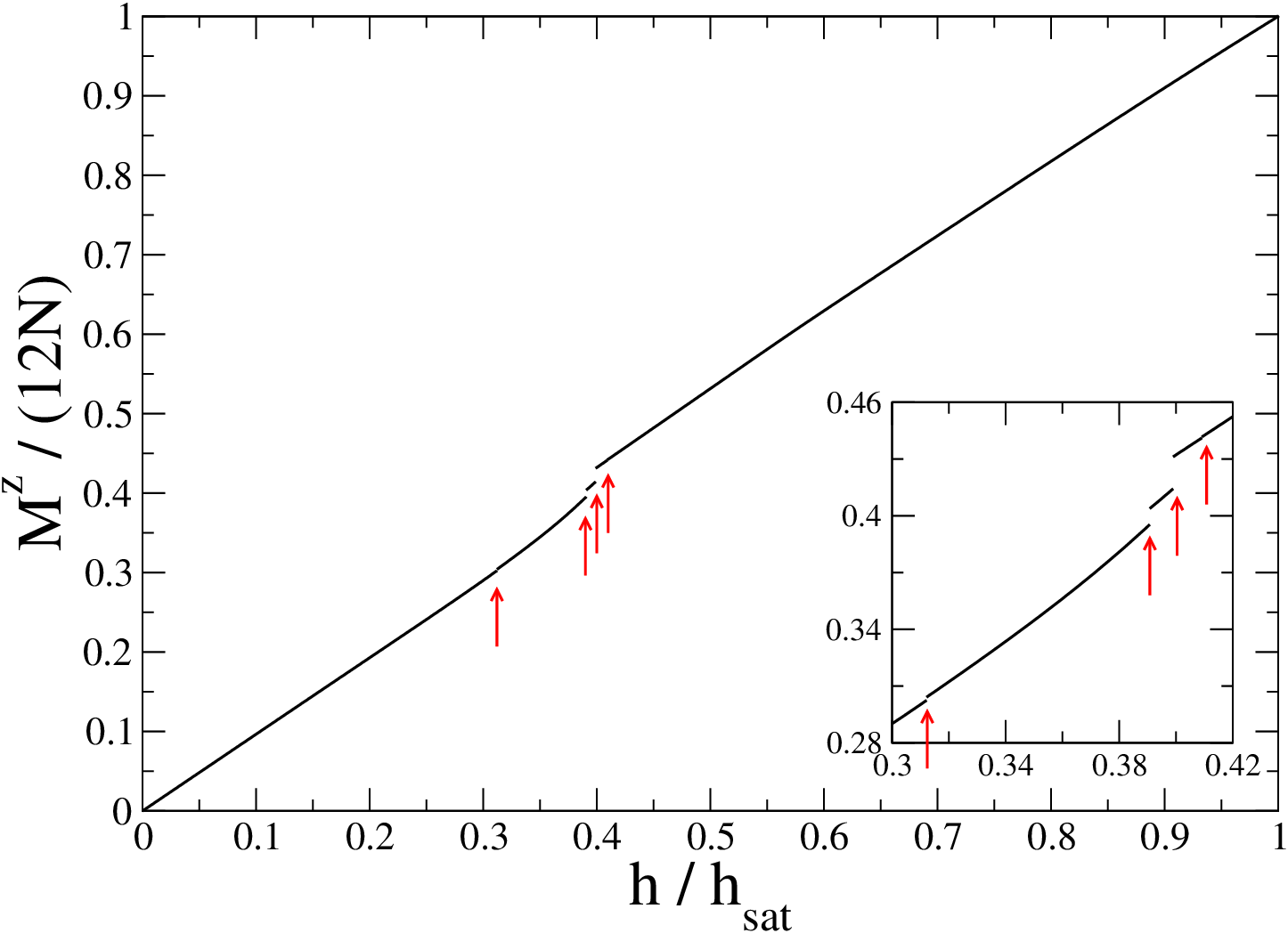}
\end{center}
\caption{Magnetization per spin $\frac{M^z}{12N}$ along the field as a function of the magnetic field over its saturation value $\frac{h}{h_{sat}}$ in the ground state of Hamiltonian (\ref{eqn:model}) for a triangle of icosahedra ($N=3$) and $\omega=\frac{\pi}{10}$ for connectivity IMC3. The (red) solid arrows point at the locations of the magnetization discontinuities.
%(~/basic/classical/icosahedronarray/N=3/omega=0.1PI)
}
\label{fig:magnetizationN=3omega=piover10}
\end{figure}

\begin{figure}[h]
\begin{center}
\includegraphics[width=3.5in,height=2.5in]{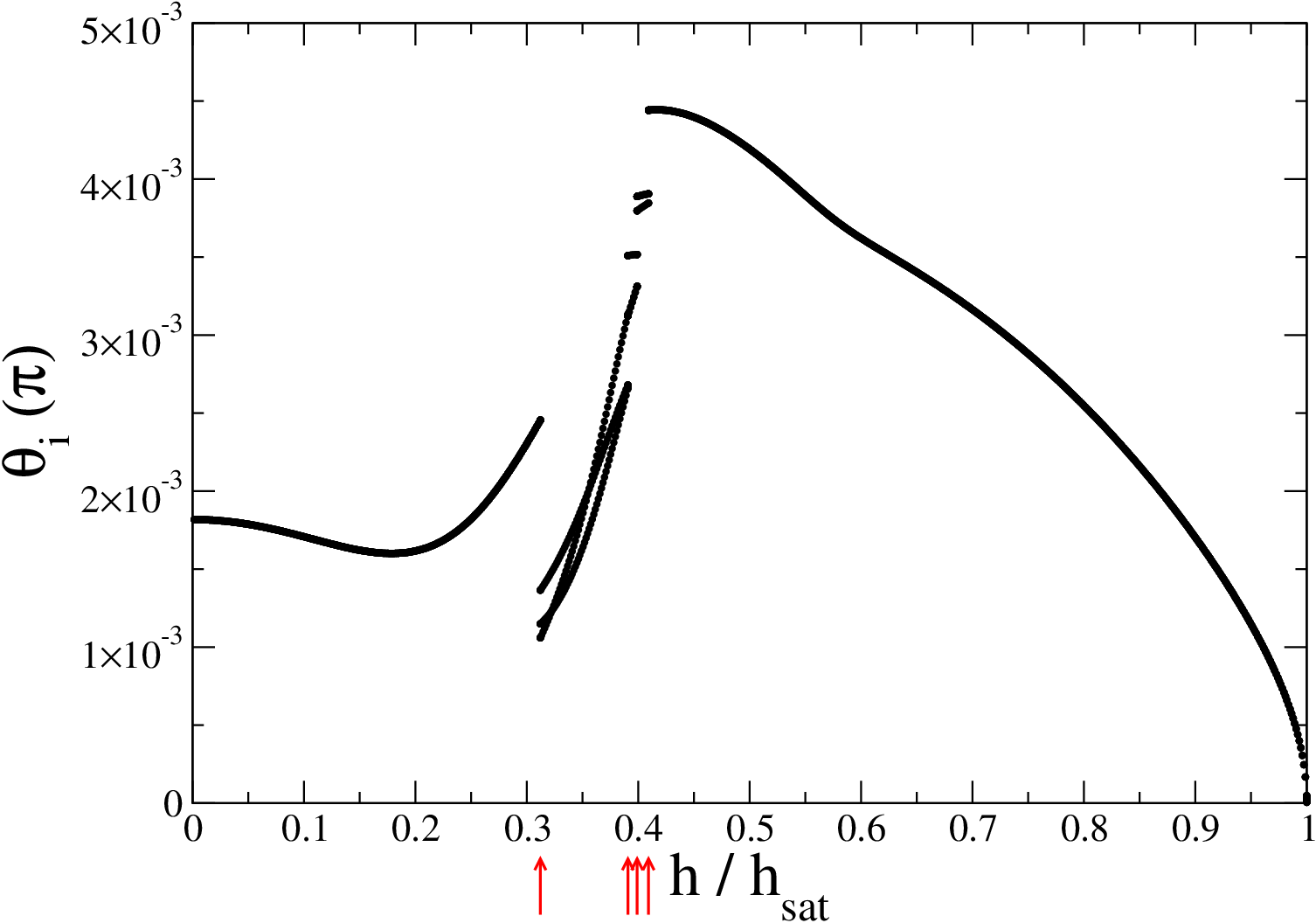}
\end{center}
\caption{Polar angles $\theta_i$ of the magnetizations of the icosahedra $\vec{M}_i$ as a function of the magnetic field over its saturation value $\frac{h}{h_{sat}}$ in the ground state of Hamiltonian (\ref{eqn:model}) for a triangle of icosahedra ($N=3$) and $\omega=\frac{\pi}{10}$ for connectivity IMC3. The (red) solid arrows point at the locations of the magnetization discontinuities.
%(~/basic/classical/icosahedronarray/N=3/omega=0.1PI)
}
\label{fig:polaranglesN=3omega=piover10}
\end{figure}

Further insight in the regime of stronger intermolecular coupling can be gained by again considering the value $\omega=\frac{\pi}{10}$. Figure \ref{fig:magnetizationN=3omega=piover10} plots the total reduced magnetization along the field $\frac{M^z}{12N}$. The total number of discontinuities is smaller and equal to four (Table \ref{table:N=3omega=pi/10discontinuities} and Fig. \ref{fig:icosahedronarraydiscontinuitiesmagnetizationN=3}). The finite common $\vec{M}_i$ of the icosahedra for an infinitesimal field emerges now at a bigger polar angle $\theta_i=1.818 \times 10^{-3}\pi$ (Fig. \ref{fig:polaranglesN=3omega=piover10}), and the common polar angle of the $\vec{M}_i$ achieves its local minimum
%. Their azimuthal angles differ by $\frac{2\pi}{3}$, further
at a higher $\frac{h}{h_{sat}}=0.178$.
%This is also in contrast with the polar angles of an isolated icosahedron, where not all of them can be increasing functions of $h$, even for a short field range \cite{NPK15}. Still the projections of the molecular magnetizations along the $z$ axis $M_i^z$ are getting bigger with the field as expected and are equal to $\frac{M^z}{N}$, since the $M_i$ increase with $h$ for each icosahedron, compensating for the bigger polar angles.

\begin{table}
\begin{center}
\caption{Ground-state magnetization discontinuities of Hamiltonian (\ref{eqn:model}) for a triangle of icosahedra ($N=3$) and $\omega=\frac{\pi}{10}$ for connectivity IMC3. The columns give the magnetic field over the saturation field $\frac{h}{h_{sat}}$ for which the discontinuities occur, the total magnetization along the field just below and just above them over the number of sites $12N$, and the corresponding strength of the magnetization jumps.}
\begin{tabular}{c|c|c|c}
 $\frac{h}{h_{sat}}$ & $(\frac{M^z}{12N})_-$ & $(\frac{M^z}{12N})_+$ & $\frac{\Delta M^z}{12N}$ ($\times 10^{-3}$) \\
\hline
 0.31214447 & 0.30243210 & 0.30412410 & 1.69200 \\ % & run41 \\
\hline
 0.390745820 & 0.39535993 & 0.40383713 & 8.47720 \\ % & run42 \\
\hline
 0.398995468 & 0.41428942 & 0.43122713 & 16.93771 \\ % & run43 \\
\hline
 0.40916029 & 0.44123730 & 0.44174704 & 0.50974 % & run44
\end{tabular}
\label{table:N=3omega=pi/10discontinuities}
\end{center}
\end{table}

%Even though the polar angle becomes bigger with $h$, simultaneously the magnetization of each icosahedron increases, and the combined effect is a rise of the total magnetization of the ring along the field even though the individual icosahedra magnetizations move further away from it.

\begin{figure}[h]
\begin{center}
\includegraphics[width=3.5in,height=2.5in]{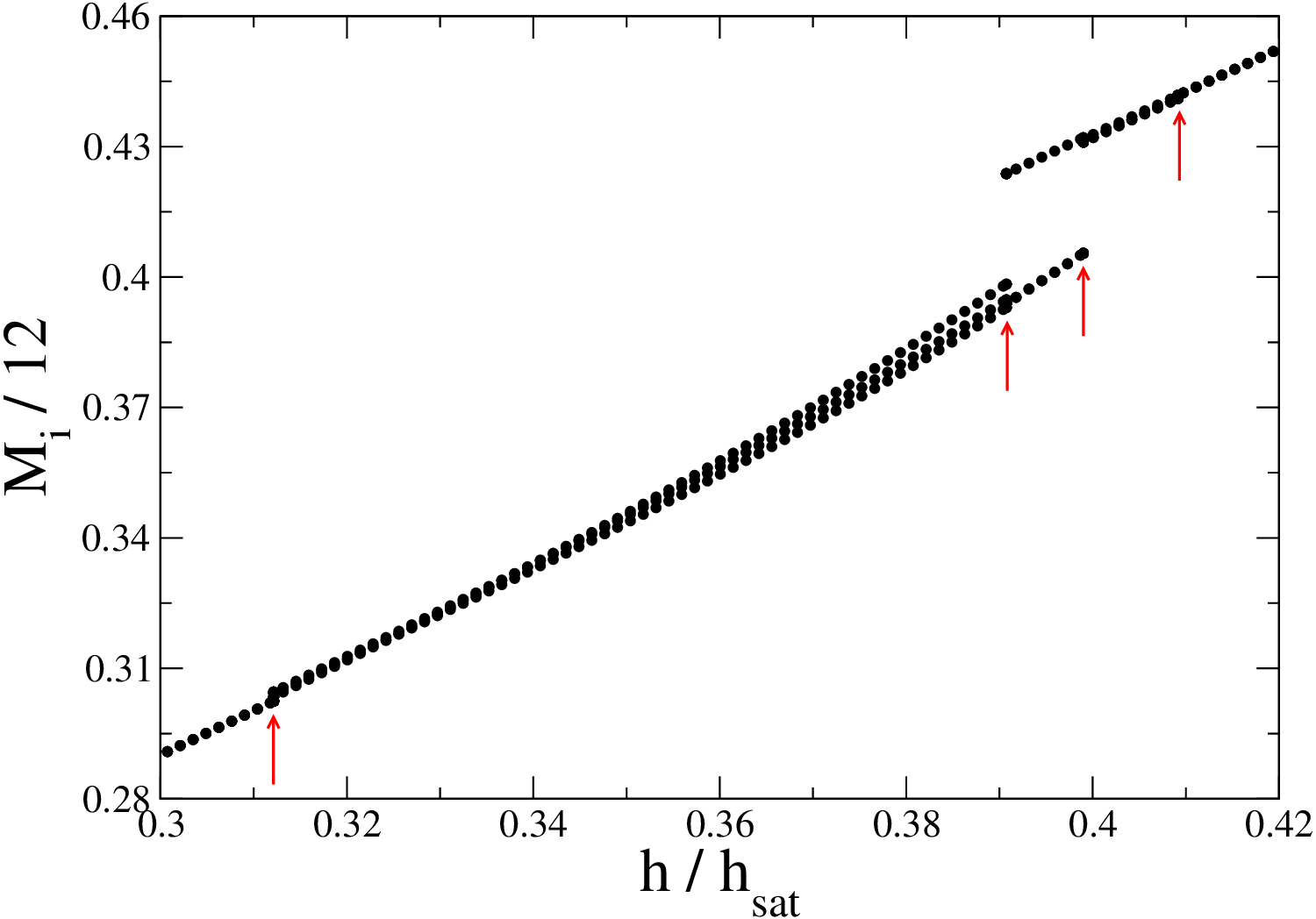}
\end{center}
\caption{Magnitude of the total magnetization per spin $\frac{M_i}{12}$ of each icosahedron $i=1,2,3$ as a function of the magnetic field over its saturation value $\frac{h}{h_{sat}}$ in the ground state of Hamiltonian (\ref{eqn:model}) for a triangle of icosahedra ($N=3$) and $\omega=\frac{\pi}{10}$ for connectivity IMC3. The (red) solid arrows point at the locations of the magnetization discontinuities.
%(~/basic/classical/icosahedronarray/N=3/omega=0.1PI)
}
\label{fig:magnetizationsmagnN=3omega=piover10}
\end{figure}

\begin{figure}[h]
\begin{center}
\includegraphics[width=3.5in,height=2.5in]{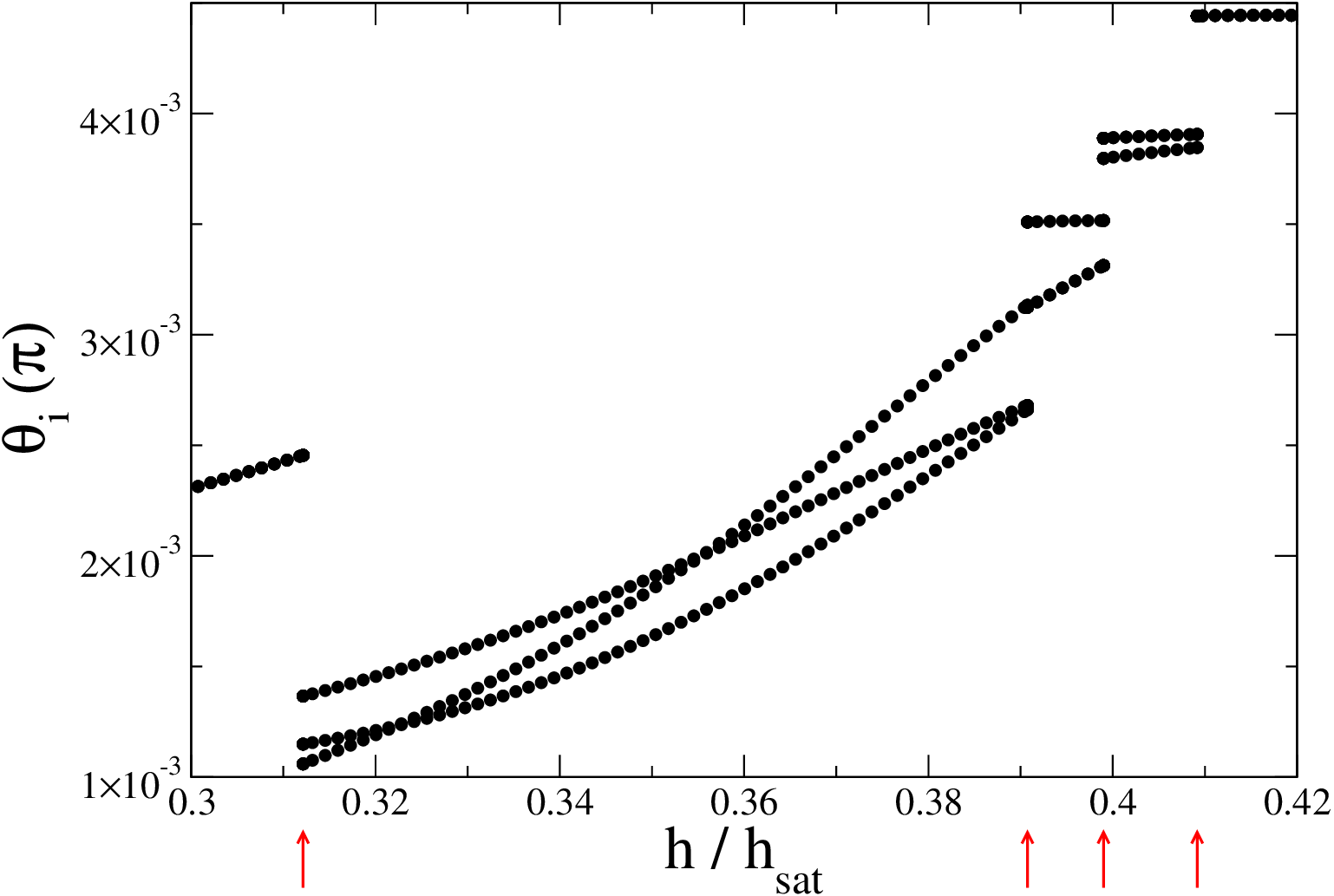}
\end{center}
\caption{Part of Fig. \ref{fig:polaranglesN=3omega=piover10} in greater detail.
%(~/basic/classical/icosahedronarray/N=3/omega=0.1PI)
}
\label{fig:polaranglesN=3omega=piover10a}
\end{figure}

\begin{figure}[h]
\begin{center}
\includegraphics[width=3.5in,height=2.5in]{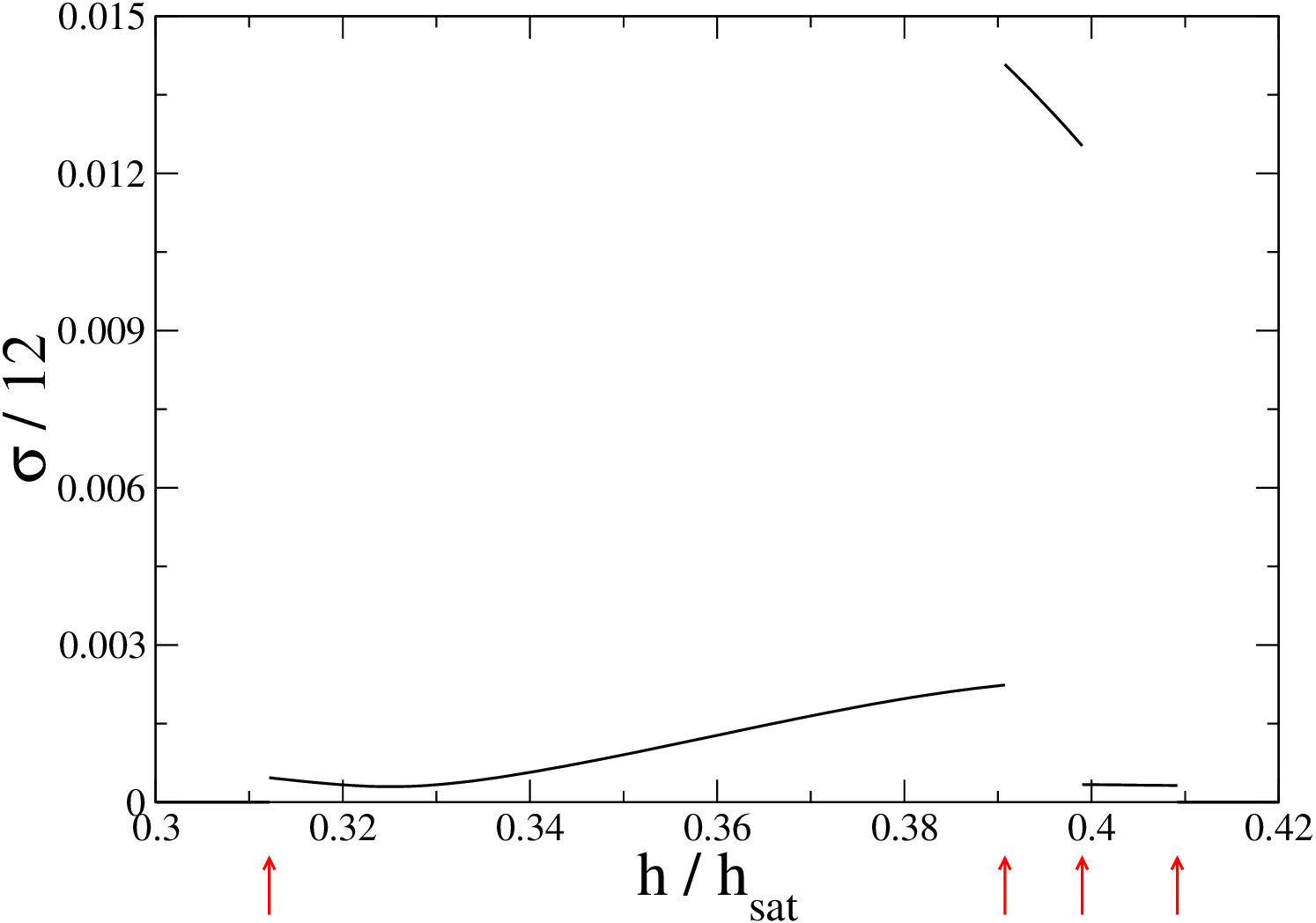}
\end{center}
\caption{Standard deviation $\sigma$ of the magnitude of the molecular magnetization per spin $\frac{M_i}{12}$ as a function of the magnetic field over its saturation value $\frac{h}{h_{sat}}$ in the ground state of Hamiltonian (\ref{eqn:model}) for a triangle of icosahedra ($N=3$) and $\omega=\frac{\pi}{10}$ for connectivity IMC3. The (red) solid arrows point at the locations of the magnetization discontinuities.
%(~/basic/classical/icosahedronarray/N=3/omega=0.1PI)
}
\label{fig:magnetizationsmagnstandarddeviationN=3omega=piover10}
\end{figure}

\begin{figure}[h]
\begin{center}
\includegraphics[width=3.5in,height=2.5in]{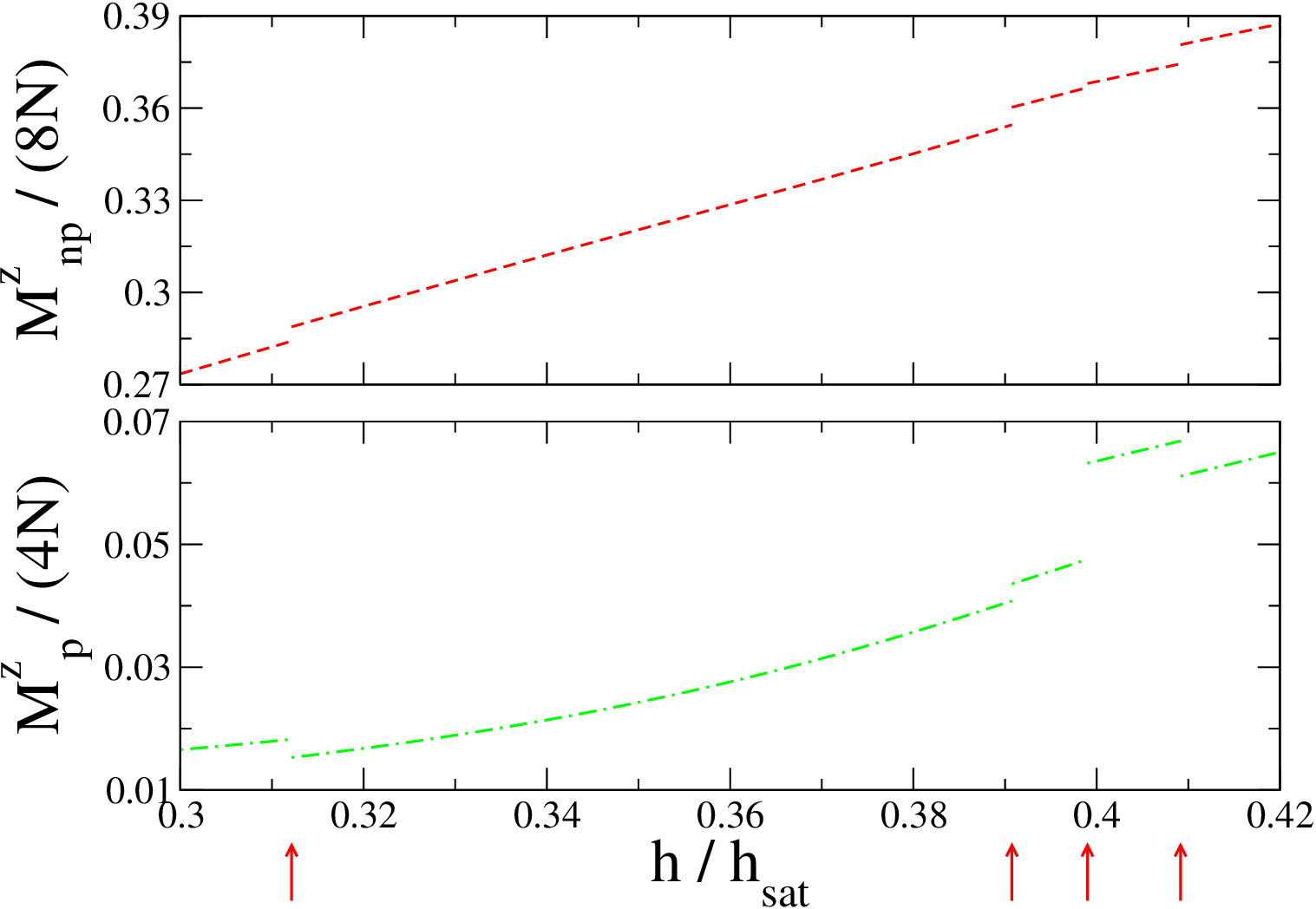}
\end{center}
\caption{Total magnetization per spin along the field of the spins that participate in intericosahedral bonds $\frac{M_p^z}{4N}$ and the ones that do not $\frac{M_{np}^z}{8N}$ as a function of the magnetic field over its saturation value $\frac{h}{h_{sat}}$ in the ground state of Hamiltonian (\ref{eqn:model}) for a triangle of icosahedra ($N=3$) and $\omega=\frac{\pi}{10}$ for connectivity IMC3. The (red) solid arrows point at the locations of the magnetization discontinuities.
%(~/basic/classical/icosahedronarray/N=3/omega=0.1PI)
}
\label{fig:magnetizationlocationN=3omega=piover10}
\end{figure}

The first magnetization discontinuity at $\frac{h}{h_{sat}}=0.31214447$
%(Fig. \ref{fig:magnetizationN=3omega=piover10} and Table \ref{table:N=3omega=pi/10discontinuities})
leads to a ground state where again each icosahedron has its own $M_i$ (Fig. \ref{fig:magnetizationsmagnN=3omega=piover10}) and polar angle (Fig. \ref{fig:polaranglesN=3omega=piover10a}). Fig. \ref{fig:magnetizationsmagnstandarddeviationN=3omega=piover10} plots the standard deviation of the $M_i$. Their projections along the field axis
%, while their azimuthal angles do not differ now by $\frac{2\pi}{3}$.
$M_i^z$ are different.
%This lowest-energy configuration is again in contrast with the umbrella state of a triangle of spins.
All the polar angles keep getting bigger with the field.
%Fig. xxx shows the total magnetic energy of the spins that participate in both intramolecular and intermolecular bonds. At the transition it becomes less negative, contrary to the magnetic energy of the spins that only participate in intramolecular bonds (Fig. xxx).
Fig. \ref{fig:magnetizationlocationN=3omega=piover10} shows the projection of the total magnetization per spin along the field for the spins that do not participate in intermolecular bonds, and for the ones that do. The value for the latter decreases at the jump.

The next discontinuity occurs at $\frac{h}{h_{sat}}=0.390745820$, with the change in $M^z$ significantly bigger than the one of the first jump (Table \ref{table:N=3omega=pi/10discontinuities}). Again across it only one of the molecules undergoes a jump and ends up with a magnetization with bigger magnitude and polar angle, while the other two have the same magnitude and polar angle for their $\vec{M}_i$ (Figs \ref{fig:magnetizationsmagnN=3omega=piover10} and \ref{fig:polaranglesN=3omega=piover10a}). The icosahedron with the bigger $M_i$ has undergone a jump similar to the one of the isolated molecule. This is attested by the strength of the jump and the discontinuity field, both still close in value to the ones of the isolated icosahedron \cite{Schroeder05}. Even though $\omega$ is now bigger the triangle of icosahedra has locally two distinct $M_i$, with the individual icosahedra being on different sides of the magnetization jump of an isolated icosahedron. The standard deviation of the $M_i$ acquires now maximum values (Fig. \ref{fig:magnetizationsmagnstandarddeviationN=3omega=piover10}). The other two icosahedra undergo the jump that corresponds to the one of the isolated icosahedron at the next magnetization discontinuity, which occurs at $\frac{h}{h_{sat}}=0.398995468$. As was the case for $\omega=\frac{9\pi}{50000}$, this discontinuity is almost twice as wide as the previous one (Table \ref{table:N=3omega=pi/10discontinuities}). A stronger $\omega$ does not prohibit the molecules from effecting the isolated icosahedron discontinuity in two steps through local variations of the magnetization. The number of distinct magnetizations is still two, but the standard deviation of the $M_i$ becomes smaller (Fig. \ref{fig:magnetizationsmagnstandarddeviationN=3omega=piover10}). The icosahedron with the bigger magnetization forms the smallest polar angle with the field.

Following a discontinuity weak in strength at $\frac{h}{h_{sat}}=0.40916029$ all the icosahedra acquire the same $M_i$, with each one forming the same angle with the field all the way to saturation, again resembling a triangle of spins. The spins that participate in intermolecular bonds again decrease their total magnetization along the field at this jump (Fig. \ref{fig:magnetizationlocationN=3omega=piover10}). The polar angle with the field keeps increasing up to $\frac{h}{h_{sat}}=0.417$, but this is again compensated by the rise of the $M_i$ to result in $M_i^z$ that increase with the field. %The azimuthal angles of the $M_i$ again form in pairs a relative angle of $\frac{2\pi}{3}$.

The above results show that an external field generates locally different $\vec{M}_i$ for specific field ranges. The triangle of icosahedra is magnetically acting as being made of more than one parts. Each part reacts differently to the field, hosting distinctly different magnetic configurations that locally change by ramping up the field. Furthermore, the magnetization jump of an isolated icosahedron is distributed in two clear-cut jumps in the triangle, with the change in the total magnetization (Figs  \ref{fig:magnetizationN=3omega=9piover50000} and \ref{fig:magnetizationN=3omega=piover10}) localized to one or two individual molecules in each jump. This effect is robust and present even when the intermolecular coupling becomes stronger.

\subsection{Hexagon of Icosahedra ($N=6$)}
\label{subsec:N=6}

\begin{figure}[h]
\begin{center}
\includegraphics[width=3.5in,height=2.5in]{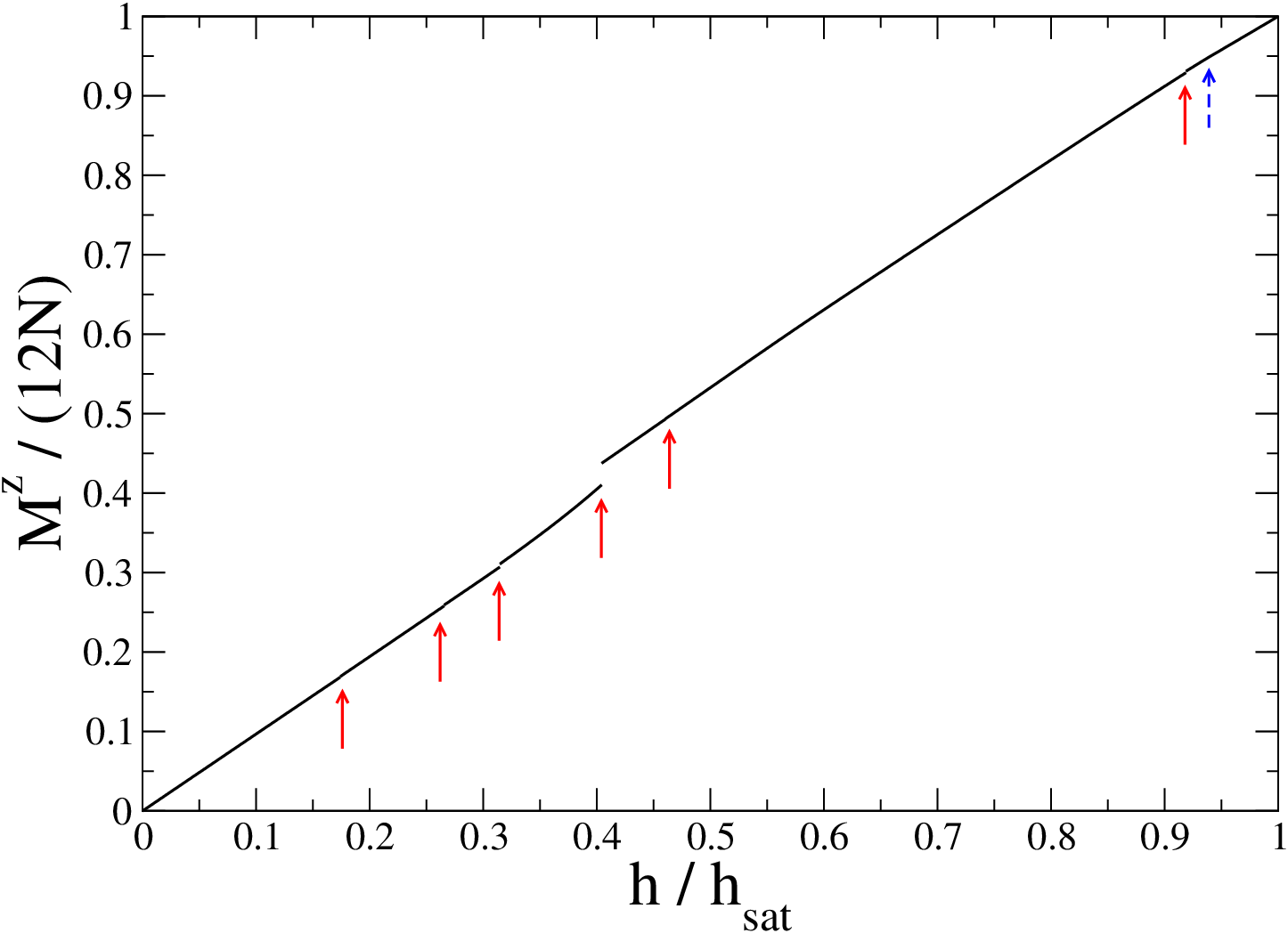}
\end{center}
\caption{Reduced magnetization $\frac{M^z}{12N}$ along the field as a function of the magnetic field over its saturation value $\frac{h}{h_{sat}}$ in the ground state of Hamiltonian (\ref{eqn:model}) for a hexagon of icosahedra ($N=6$) and $\omega=\frac{\pi}{10}$ for connectivity IMC3. The (red) solid arrows point at the locations of the magnetization discontinuities. The (blue) dashed arrow points at the location of the susceptibility discontinuity.
%(~/basic/classical/icosahedronarray/N=6/omega=0.1PI)
}
\label{fig:magnetizationN=6omega=piover10}
\end{figure}

The next-bigger ring of icosahedra achieving minimum frustration in zero field with connectivity IMC3 is the hexagon. Due to computational requirements the magnetization response is calculated only for $\omega=\frac{\pi}{10}$. Figure \ref{fig:magnetizationN=6omega=piover10} shows the total ground-state magnetization along the field $M^z$ of Hamiltonian (\ref{eqn:model}). The total number of magnetization discontinuities in a field is six, and close to saturation the susceptibility is also discontinuous (Table \ref{table:N=6omega=pi/10discontinuities}).

\begin{table}
\begin{center}
\caption{Ground-state magnetization discontinuities of Hamiltonian (\ref{eqn:model}) for a hexagon of icosahedra ($N=6$) and $\omega=\frac{\pi}{10}$ for connectivity IMC3. The columns give the magnetic field over the saturation field $\frac{h}{h_{sat}}$ for which the discontinuities occur, the total magnetization along the field just below and just above them over the number of sites $12N$, and the corresponding strength of the magnetization jumps. The susceptibility discontinuity occurs at $\frac{h}{h_{sat}}=0.93948$, at the corresponding value of the magnetization 0.94913.
  }
\begin{tabular}{c|c|c|c}
 $\frac{h}{h_{sat}}$ & $(\frac{M}{12N})_-$ & $(\frac{M}{12N})_+$ & $\frac{\Delta M}{12N}$ ($\times 10^{-3}$) \\
\hline
 0.17405567 & 0.16827873 & 0.16907497 & 0.79624 \\ % & run50 \\
\hline
 0.2655524 & 0.2579651 & 0.2591112 & 1.1461 \\ % & run51 \\
\hline
 0.3146333 & 0.3070018 & 0.3108143 & 3.8125 \\ % & run52 \\
\hline
 0.404313853660 & 0.4103450 & 0.4375794 & 27.2344 \\ % & run73 - new run22 \\
\hline
 0.4604222 & 0.4930904 & 0.4934528 & 0.3624 \\ % & run54 \\
\hline
 0.918823548 & 0.92918918 & 0.93102521 & 1.83603 \\ % & run55 \\
%\hline
%$\chi$ & 0.93948 & 0.94913 & 0.94913 & 0 % & run56
\end{tabular}
\label{table:N=6omega=pi/10discontinuities}
\end{center}
\end{table}

\begin{figure}[h]
\begin{center}
\includegraphics[width=3.5in,height=2.5in]{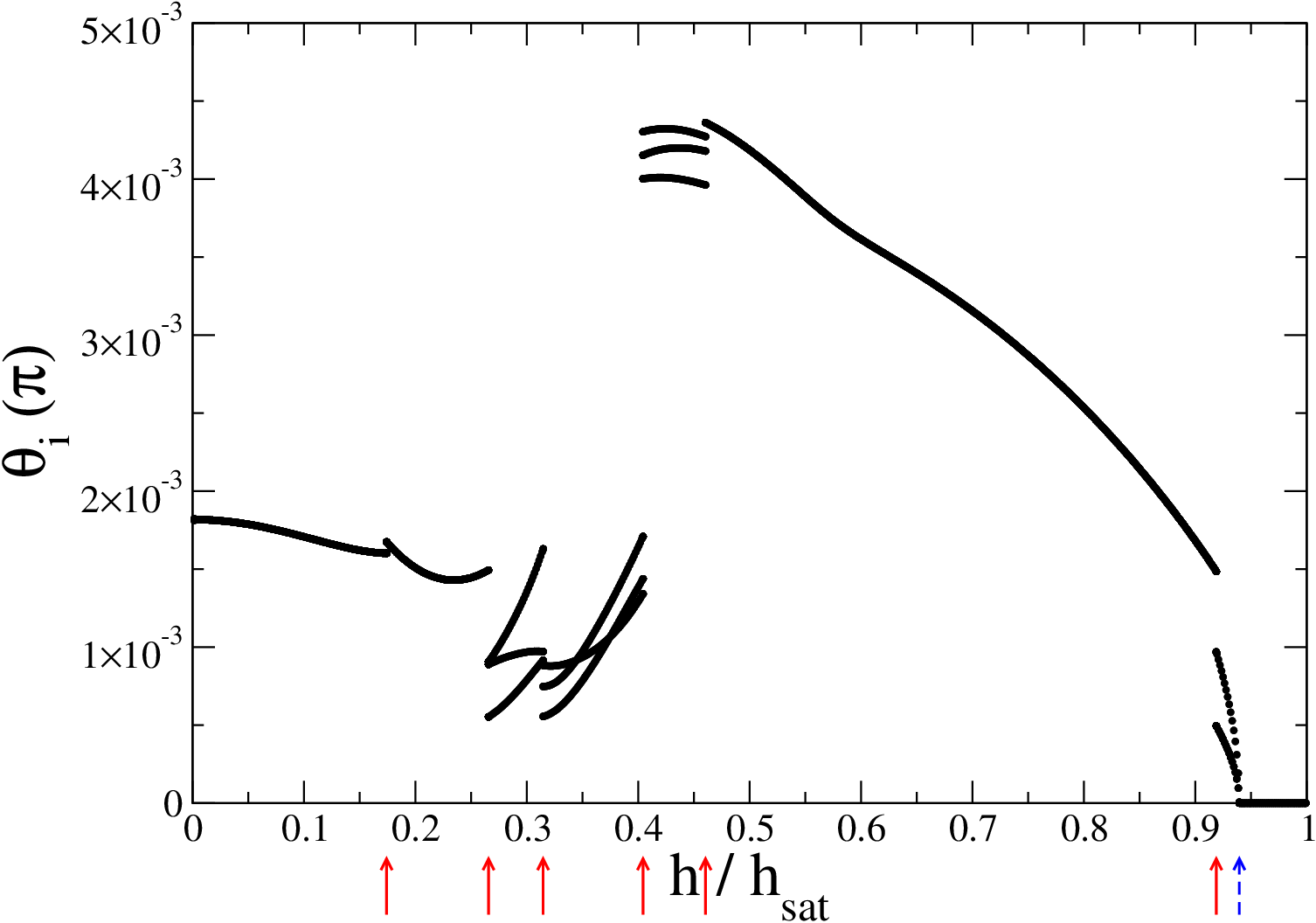}
\end{center}
\caption{Polar angles $\theta_i$ of the magnetizations of the icosahedra $\vec{M}_i$ as a function of the magnetic field over its saturation value $\frac{h}{h_{sat}}$ in the ground state of Hamiltonian (\ref{eqn:model}) for a hexagon of icosahedra ($N=6$) and $\omega=\frac{\pi}{10}$ for connectivity IMC3. The (red) solid arrows point at the locations of the magnetization discontinuities. The (blue) dashed arrow points at the location of the susceptibility discontinuity.
%(~/basic/classical/icosahedronarray/N=6/omega=0.1PI)
}
\label{fig:polaranglesN=6omega=piover10}
\end{figure}

An infinitesimal field generates a ground-state magnetization $\vec{M}_i$ for each icosahedron at a polar angle with the field equal to $1.818 \times 10^{-3}\pi$ (Fig. \ref{fig:polaranglesN=6omega=piover10}), the same with the one of $N=3$ (Sec. \ref{subsec:N=3}). %The difference in azimuthal angles between nearest neighbors is $\frac{2\pi}{3}$, while
The common polar angle again decreases with the field. At $\frac{h}{h_{sat}}=0.17405567$ a magnetization jump occurs, leading to a lowest-energy configuration
%with a jump in the two-spin vector chirality $\kappa = \vec{s}_i \times \vec{s}_j$ (Fig. xxx). The chirality of same-icosahedron spins now assumes two distinct values.
where the $\vec{M}_i$ still share the polar angle, which acquires a higher value. Then it decreases and reaches a minimum at $\frac{h}{h_{sat}}=0.234$, following which the $\vec{M}_i$ turn away from the field. %The nearest-neighbor azimuthal angles are not equal to $\frac{2\pi}{3}$ any more.

\begin{figure}[h]
\begin{center}
\includegraphics[width=3.5in,height=2.5in]{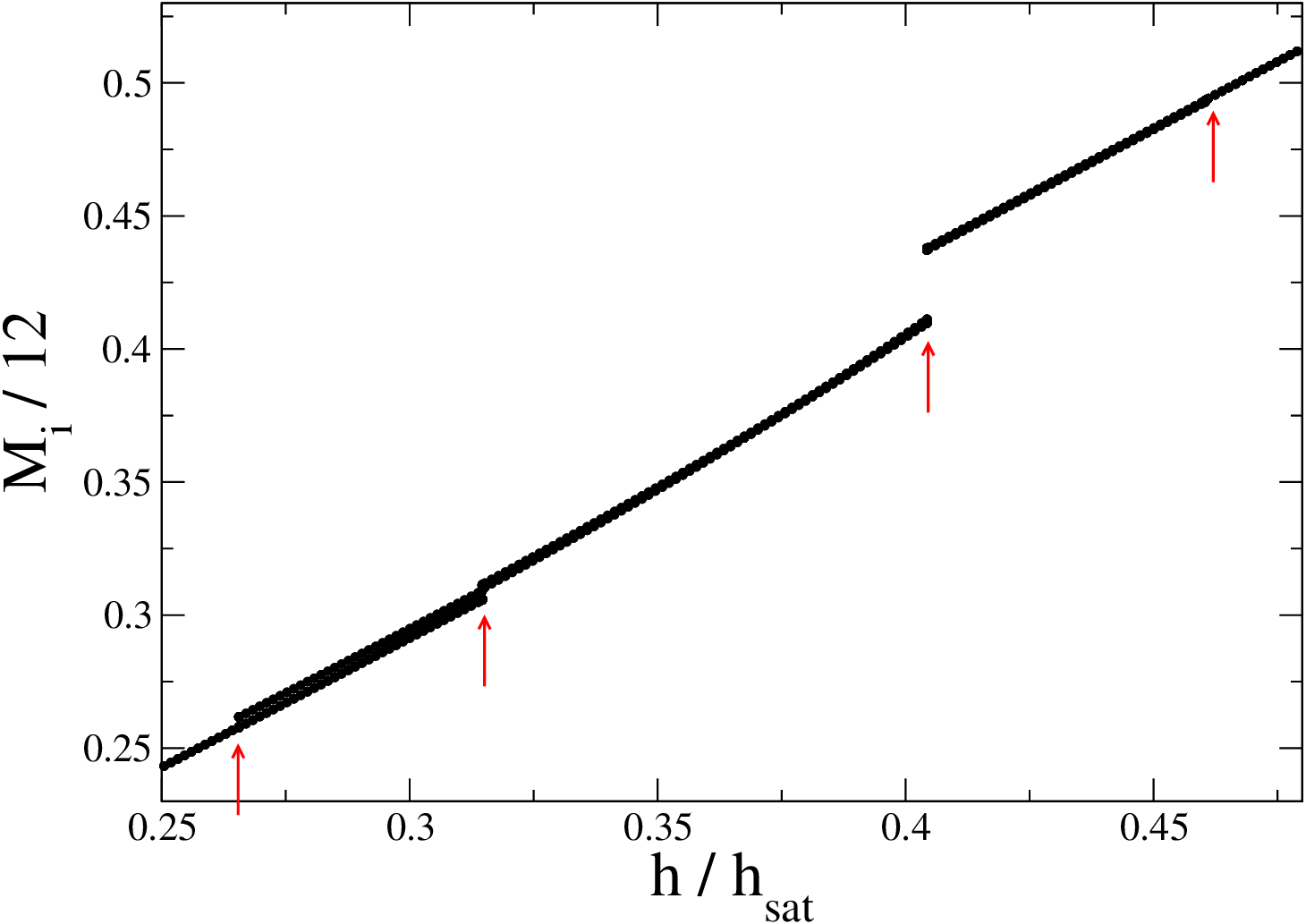}
\end{center}
\caption{Magnitude of the total magnetization per spin $\frac{M_i}{12}$ of each icosahedron $i=1,2,\dots,6$ as a function of the magnetic field over its saturation value $\frac{h}{h_{sat}}$ in the ground state of Hamiltonian (\ref{eqn:model}) for a hexagon of icosahedra ($N=6$) and $\omega=\frac{\pi}{10}$ for connectivity IMC3. The (red) solid arrows point at the locations of the magnetization discontinuities.
%(~/basic/classical/icosahedronarray/N=6/omega=0.1PI)
}
\label{fig:magnetizationsmagnN=6omega=piover10}
\end{figure}

\begin{figure}[h]
\begin{center}
\includegraphics[width=3.5in,height=2.5in]{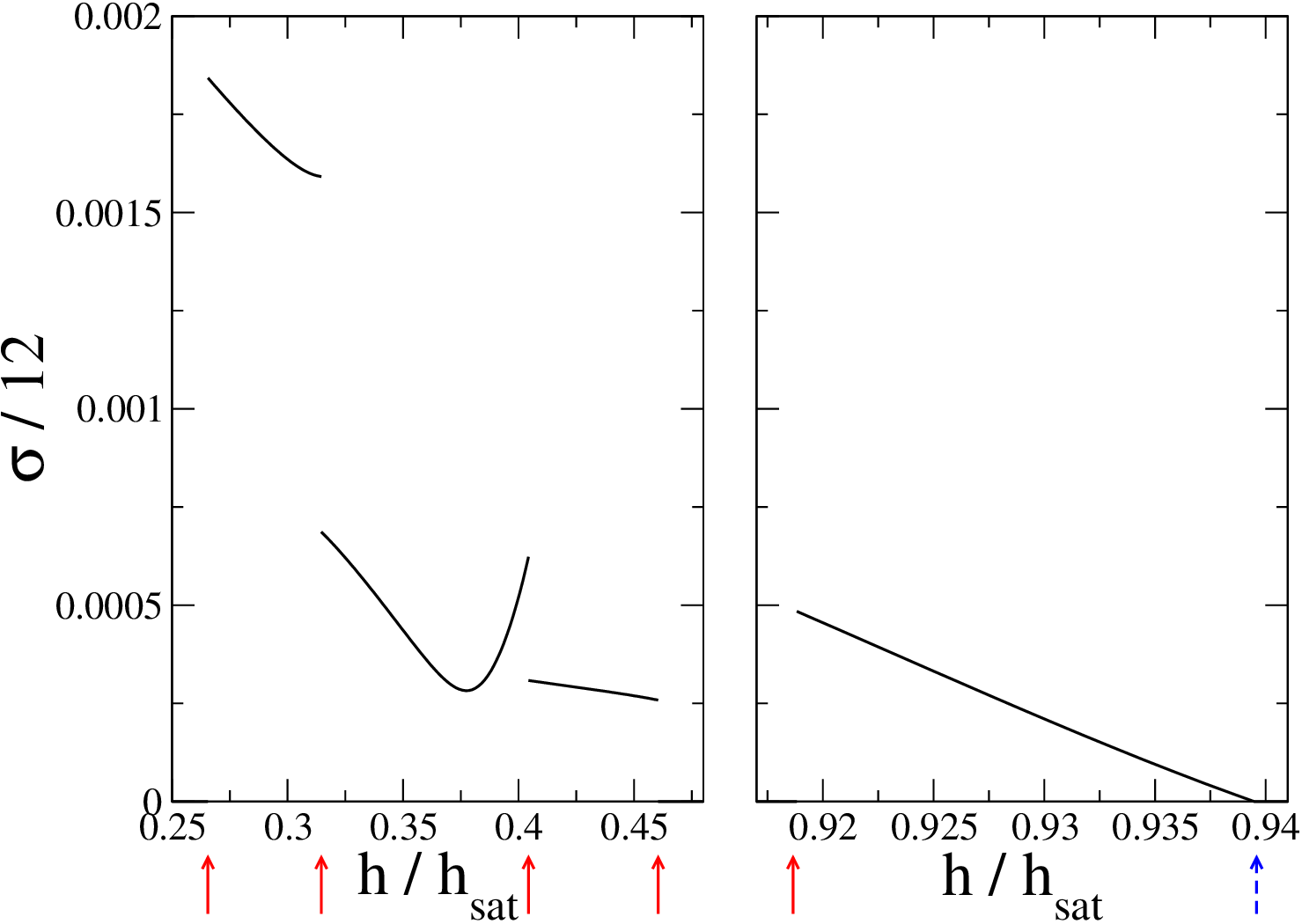}
\end{center}
\caption{Standard deviation $\sigma$ of the magnitude of the molecular magnetization per spin $\frac{M_i}{12}$ as a function of the magnetic field over its saturation value $\frac{h}{h_{sat}}$ in the ground state of Hamiltonian (\ref{eqn:model}) for a hexagon of icosahedra ($N=6$) and $\omega=\frac{\pi}{10}$ for connectivity IMC3. The (red) solid arrows point at the locations of the magnetization discontinuities. The (blue) dashed arrow points at the location of the susceptibility discontinuity.
%(~/basic/classical/icosahedronarray/N=6/omega=0.1PI)
}
\label{fig:magnetizationsmagnstandarddeviationN=6omega=piover10}
\end{figure}

A new magnetization jump occurs at $\frac{h}{h_{sat}}=0.2655524$. Following it translational symmetry is broken, with the icosahedra sharing $M_i$ (Fig. \ref{fig:magnetizationsmagnN=6omega=piover10}) and polar angle values (Fig. \ref{fig:polaranglesN=6omega=piover10}) in pairs. Two of the pairs are adjacent to each other. Fig. \ref{fig:magnetizationsmagnstandarddeviationN=6omega=piover10} shows the standard deviation of the $M_i$. The polar angles mostly increase with the field. The projections of the molecular magnetizations along the field $M_i^z$ follow the pattern of the $M_i$ and their polar angles.

%\begin{figure}[h]
%\begin{center}
%\includegraphics[width=3.5in,height=2.5in]{polaranglesN=6phi=piover10a}
%\end{center}
%\caption{Part of Fig. \ref{fig:polaranglesN=6omega=piover10} in greater detail.
%%(~/basic/classical/icosahedronarray/N=6/omega=0.1PI)
%}
%\label{fig:polaranglesN=6omega=piover10a}
%\end{figure}

The next magnetization jump occurs at $\frac{h}{h_{sat}}=0.3146333$. The lowest-energy configuration does not change. The polar angles increase with the field and the standard deviation of the $M_i$ is now smaller. This discontinuity occurs at a field very close to the one of the first discontinuity of $N=3$ and has similar characteristics with it (Sec. \ref{subsec:N=3}).
%The two-spin vector chirality around the transition field is plotted in Fig. xxx.

The next discontinuity occurs at $\frac{h}{h_{sat}}=0.404313853660$ and the lowest-energy configuration is still the same. Each $M_i^z$ undergoes the magnetization jump of the isolated icosahedron, with the corresponding field value very close to the one of the isolated molecule \cite{Schroeder05}. This is in contrast with what happens for $N=3$, where the isolated-icosahedron jump occurs at different fields for different molecules. At this jump the rise in the magnetization allows the polar angles to increase.

\begin{figure}[h]
\begin{center}
\includegraphics[width=3.5in,height=2.5in]{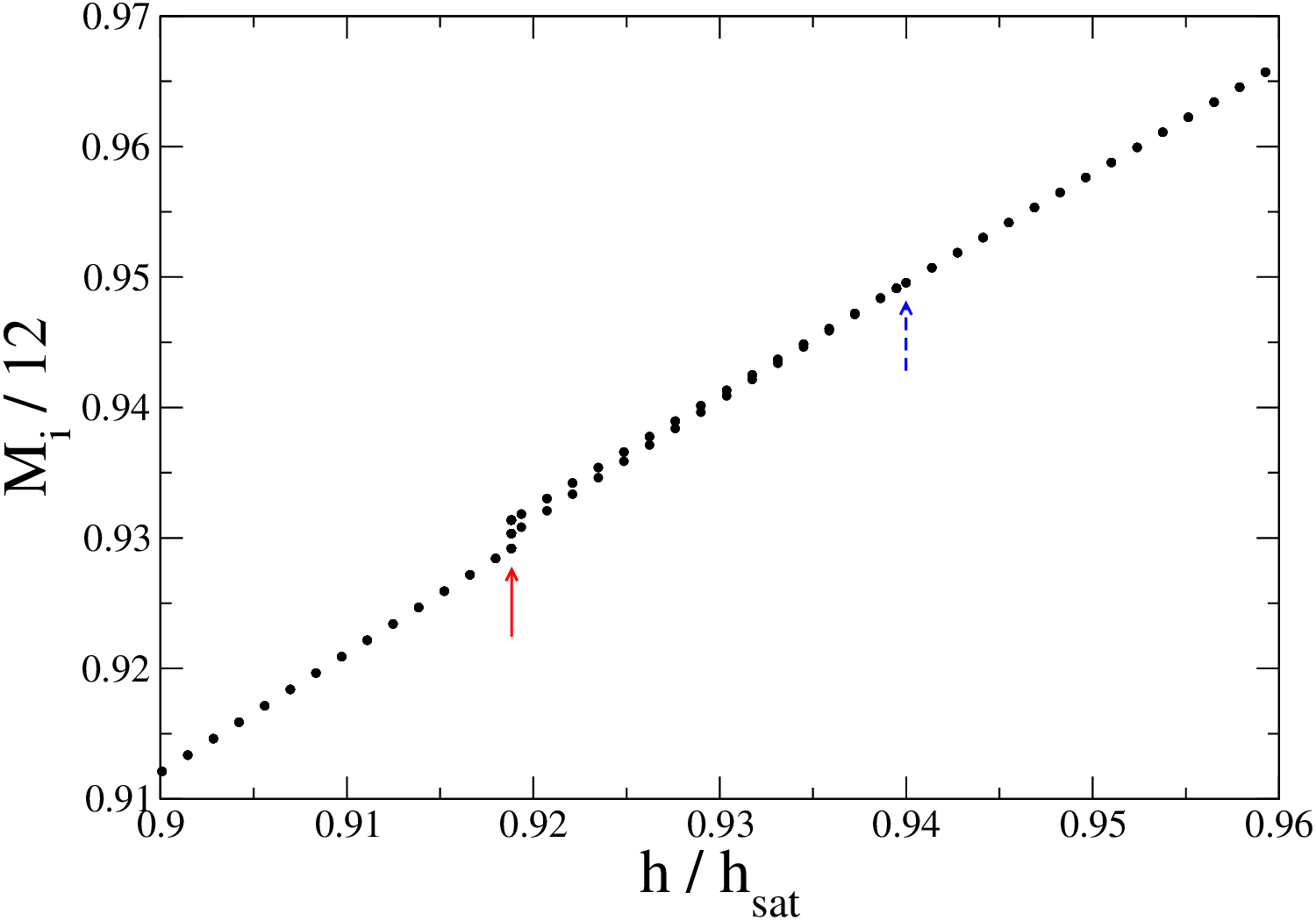}
\end{center}
\caption{Same as Fig. \ref{fig:magnetizationsmagnN=6omega=piover10} but close to saturation. The (blue) dashed arrow points at the location of the susceptibility discontinuity.
%(~/basic/classical/icosahedronarray/N=6/omega=0.1PI)
}
\label{fig:magnetizationsmagnN=6omega=piover10a}
\end{figure}

Above a magnetization jump at $\frac{h}{h_{sat}}=0.4604222$ all $M_i$ and polar angles become equal, as in a triangle of spins. %and nearest-neighbor azimuthal angles differ by $\frac{2\pi}{3}$, as in the case of a hexagon of single spins.
The $\vec{M}_i$ turn toward the field but at the last discontinuity at $\frac{h}{h_{sat}}=0.918823548$ translational symmetry is again lost. Two distinct $M_i$ occur now in the ground state, with groups of four and two icosahedra sharing the $M_i$ value (Fig. \ref{fig:magnetizationsmagnN=6omega=piover10a}) as well as the polar angle (Fig. \ref{fig:polaranglesN=6omega=piover10}), providing another way to break the translational symmetry. The two icosahedra that share one of the $M_i$ values are maximally distanced from one another. Following a susceptibility discontinuity at $\frac{h}{h_{sat}}=0.93948$ all $\vec{M}_i$ acquire the same magnitude and align with the field, even though saturation has not been reached yet. This is similar to what happens for the dimer of icosahedra, where there are no components of the molecules' magnetizations away from the field axis (Sec. \ref{subsec:twoicosahedra}). Further increasing the field results in higher $M_i$'s up to saturation.i

%\begin{figure}[h]
%\begin{center}
%\includegraphics[width=3.5in,height=2.5in]{polaranglesN=6phi=piover10b}
%\end{center}
%\caption{Part of Fig. \ref{fig:polaranglesN=6omega=piover10} in greater detail.
%%(~/basic/classical/icosahedronarray/N=6/omega=0.1PI)
%}
%\label{fig:polaranglesN=6omega=piover10b}
%\end{figure}

\begin{figure}[h]
\begin{center}
\includegraphics[width=3.5in,height=2.5in]{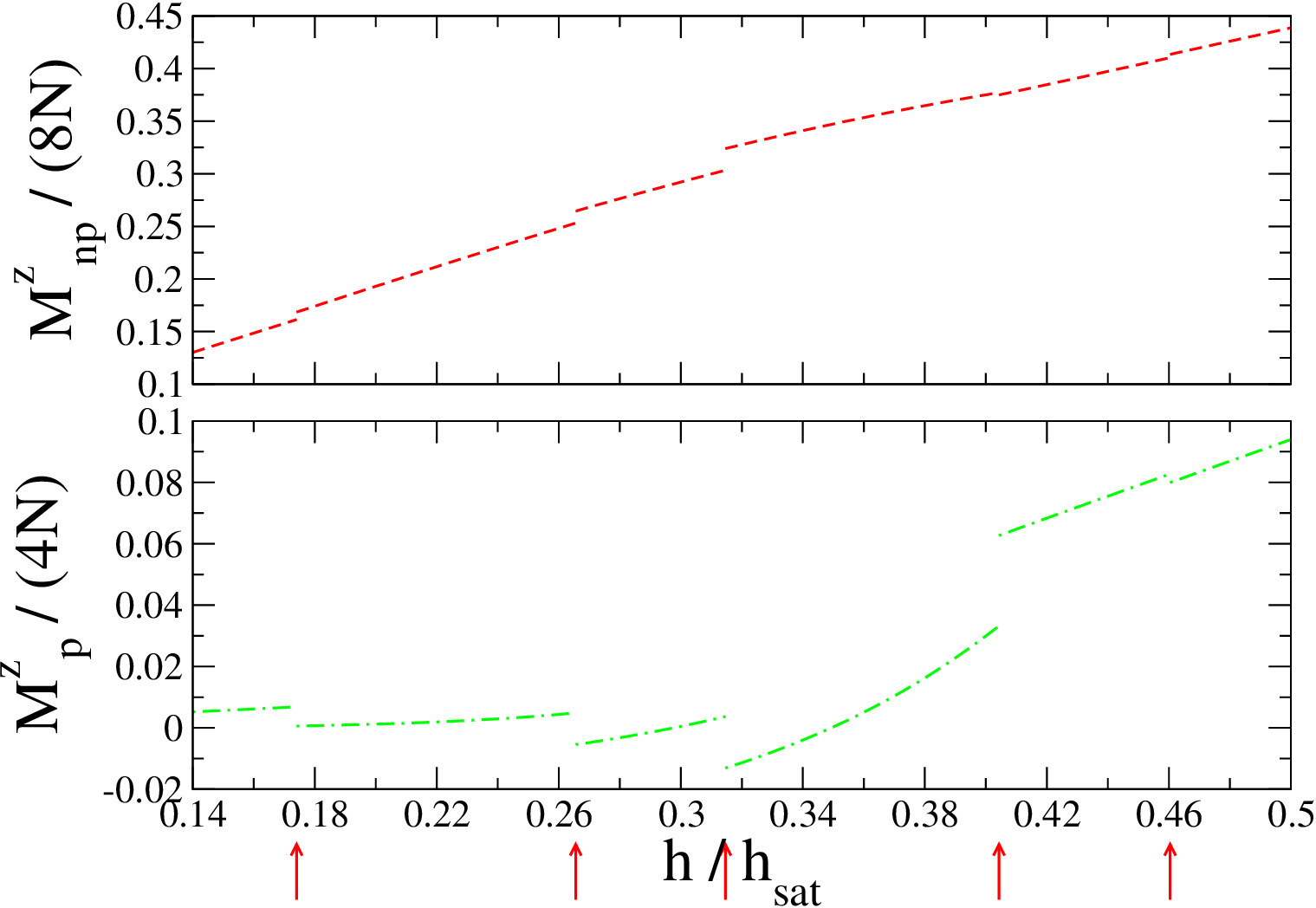}
\end{center}
\caption{Total magnetization per spin along the field of the spins that participate in intericosahedral bonds $\frac{M_p^z}{4N}$ and the ones that do not $\frac{M_{np}^z}{8N}$ as a function of the magnetic field over its saturation value $\frac{h}{h_{sat}}$ in the ground state of Hamiltonian (\ref{eqn:model}) for a hexagon of icosahedra ($N=6$) and $\omega=\frac{\pi}{10}$ for connectivity IMC3. The (red) solid arrows point at the locations of the magnetization discontinuities.
%(~/basic/classical/icosahedronarray/N=6/omega=0.1PI)
}
\label{fig:magnetizationlocationN=6omega=piover10}
\end{figure}

\begin{figure}[h]
\begin{center}
\includegraphics[width=3.5in,height=2.5in]{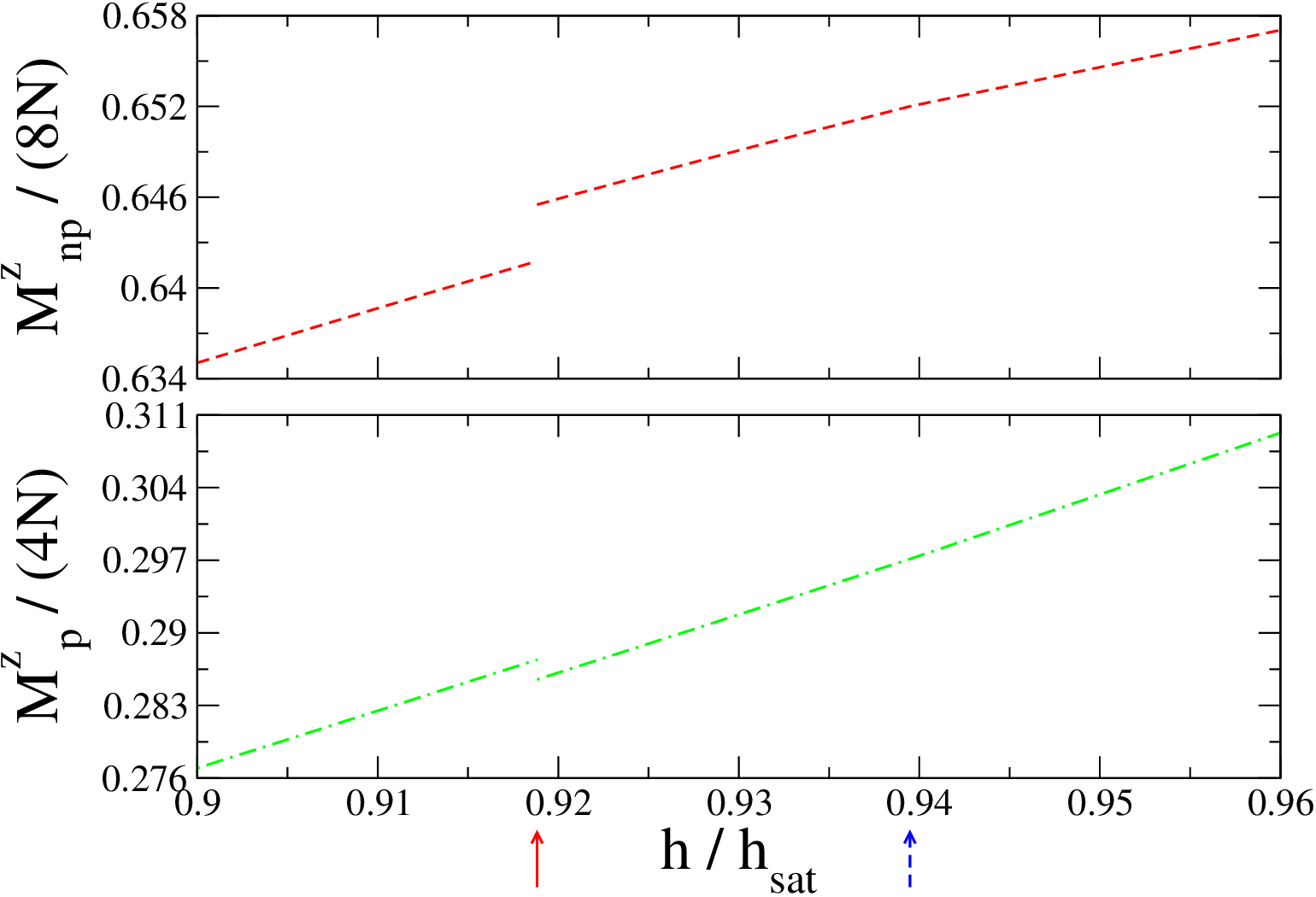}
\end{center}
\caption{Same as Fig. \ref{fig:magnetizationlocationN=6omega=piover10} but close to saturation. The (blue) dashed arrow points at the location of the susceptibility discontinuity.
%(~/basic/classical/icosahedronarray/N=3/omega=0.1PI)
}
\label{fig:magnetizationlocationN=6omega=piover10a}
\end{figure}

The projection of the total magnetization per spin along the field for the spins that participate in intermolecular bonds decreases across the discontinuities, with the exception of the jump that corresponds to the one of the isolated icosahedron (Figs \ref{fig:magnetizationlocationN=6omega=piover10} and \ref{fig:magnetizationlocationN=6omega=piover10a}). On the other hand, for the spins that only participate in intramolecular bonds this is the only jump that their corresponding projection gets smaller.

\section{IMC4}
\label{sec:IMC4}

\subsection{Pentagon of Icosahedra ($N=5$)}
\label{subsec:fiveicosahedraIMC4}

The discontinuities in the ground-state magnetization of Hamiltonian (\ref{eqn:model}) for a pentagon of icosahedra and $\omega=\frac{\pi}{10}$ are listed in Table \ref{table:N=5omega=pi/10discontinuities}. The isolated-icosahedron jump is divided in two parts, as was the case for the triangle of icosahedra with connectivity IMC3 (Sec. \ref{subsec:N=3}). In both cases the number of molecules is odd. Two further discontinuities of the magnetization and two of the susceptibility appear for smaller fields.

\begin{table}
\begin{center}
\caption{Ground-state magnetization discontinuities of Hamiltonian (\ref{eqn:model}) for a pentagon of icosahedra ($N=5$) and $\omega=\frac{\pi}{10}$ for connectivity IMC4. The columns give the magnetic field over the saturation field $\frac{h}{h_{sat}}$ for which the discontinuities occur, the total magnetization along the field just below and just above them over the number of sites $12N$, and the corresponding strength of the magnetization jumps. The two susceptibility discontinuities occur at $\frac{h}{h_{sat}}=0.08104$ and 0.32756, at the corresponding values of the magnetization 0.07725 and 0.31905.}
\begin{tabular}{c|c|c|c}
 $\frac{h}{h_{sat}}$ & $(\frac{M^z}{12N})_-$ & $(\frac{M^z}{12N})_+$ & $\frac{\Delta M^z}{12N}$ ($\times 10^{-3}$) \\
%\hline
%$\chi$ & 0.08104 & 0.07725 & 0.07725 & 0 \\ %& run3 \\
\hline
 0.116150063 & 0.11074345 & 0.11329057 & 2.54712 \\ % & run21 \\
\hline
 0.255183070 & 0.24545915 & 0.24759449 & 2.13534 \\ % & run20 \\
%\hline
%$\chi$ & 0.32756 & 0.31905 & 0.31905 & 0 \\ % & run7 \\
\hline
 0.375470596 & 0.37328982 & 0.39707656 & 23.78674 \\ % & run22 \\
\hline
 0.37747069482 & 0.39932649 & 0.41212926 & 12.80277 \\ % & run17
\end{tabular}
\label{table:N=5omega=pi/10discontinuities}
\end{center}
\end{table}

\begin{figure}[h]
\begin{center}
\includegraphics[width=3.5in,height=2.5in]{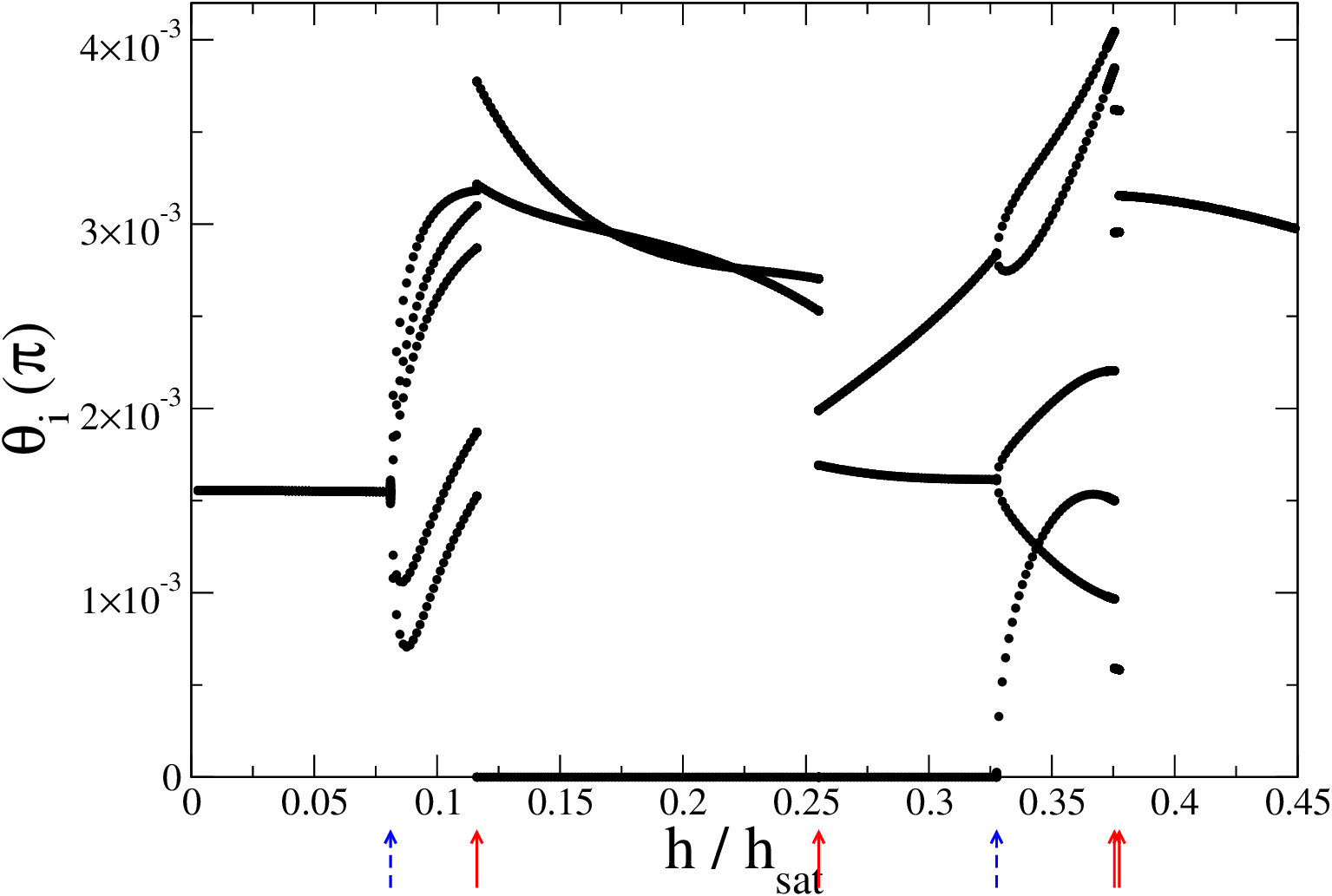}
\end{center}
\caption{Polar angles $\theta_i$ of the magnetizations of the icosahedra $\vec{M}_i$ as a function of the magnetic field over its saturation value $\frac{h}{h_{sat}}$ in the ground state of Hamiltonian (\ref{eqn:model}) for a pentagon of icosahedra ($N=5$) and $\omega=\frac{\pi}{10}$ for connectivity IMC4. The (red) solid arrows point at the locations of the magnetization discontinuities. The (blue) dashed arrows point at the locations of the susceptibility discontinuities.
%(~/discoverer/basic/classical/icosahedronarraynew/N=5/fourth/parallel/omega=0.1PI)
}
\label{fig:polaranglesN=5fourthomega=piover10}
\end{figure}

The magnetizations of the icosahedra emerge at an infinitesimal field at a polar angle of  $1.555 \times 10^{-3}\pi$ (Fig. \ref{fig:polaranglesN=5fourthomega=piover10}). Their polar angles can also here increase with the field. However in the present case, between the first magnetization and second susceptibility discontinuity, one of the icosahedra has its magnetization aligned with the $z$ axis. The remaining $\vec{M}_i$ distribute themselves symmetrically with respect to this molecule, with its neighboring icosahedra having their own magnetization magnitude and polar angle, while the same is true for the remaining two icosahedra.

\begin{figure}[h]
\begin{center}
\includegraphics[width=3.5in,height=2.5in]{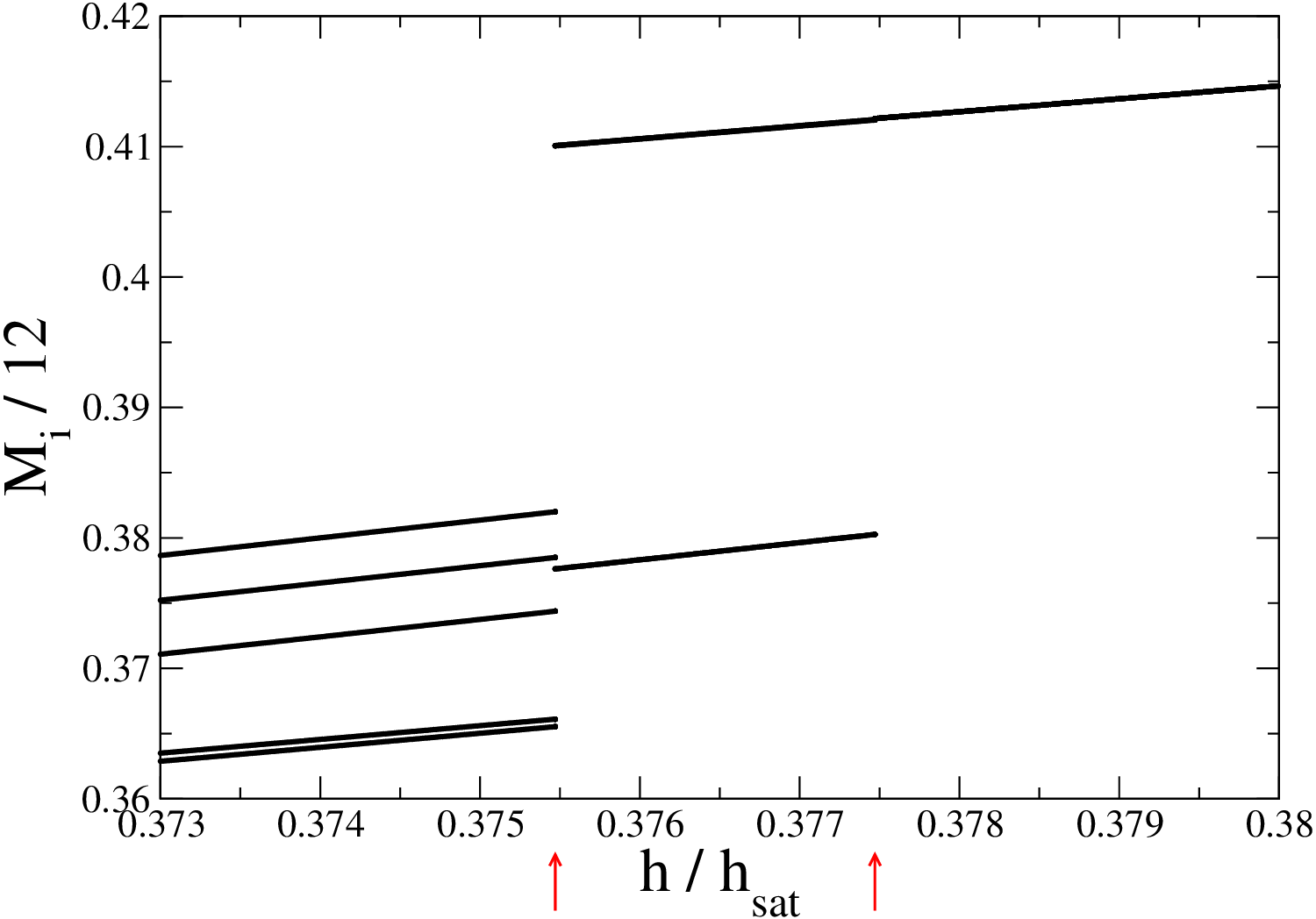}
\end{center}
\caption{Magnitude of the total magnetization per spin $\frac{M_i}{12}$ of each icosahedron $i=1,2,...,5$ as a function of the magnetic field over its saturation value $\frac{h}{h_{sat}}$ in the ground state of Hamiltonian (\ref{eqn:model}) for a pentagon of icosahedra ($N=5$) and $\omega=\frac{\pi}{10}$ for connectivity IMC4. The (red) solid arrows point at the locations of the magnetization discontinuities.
%(~/discoverer/basic/classical/icosahedronarraynew/N=5/fourth/parallel/omega=0.1PI)
}
\label{fig:magnetizationsN=5fourthomega=piover10}
\end{figure}

\begin{figure}[h]
\begin{center}
\includegraphics[width=3.5in,height=2.5in]{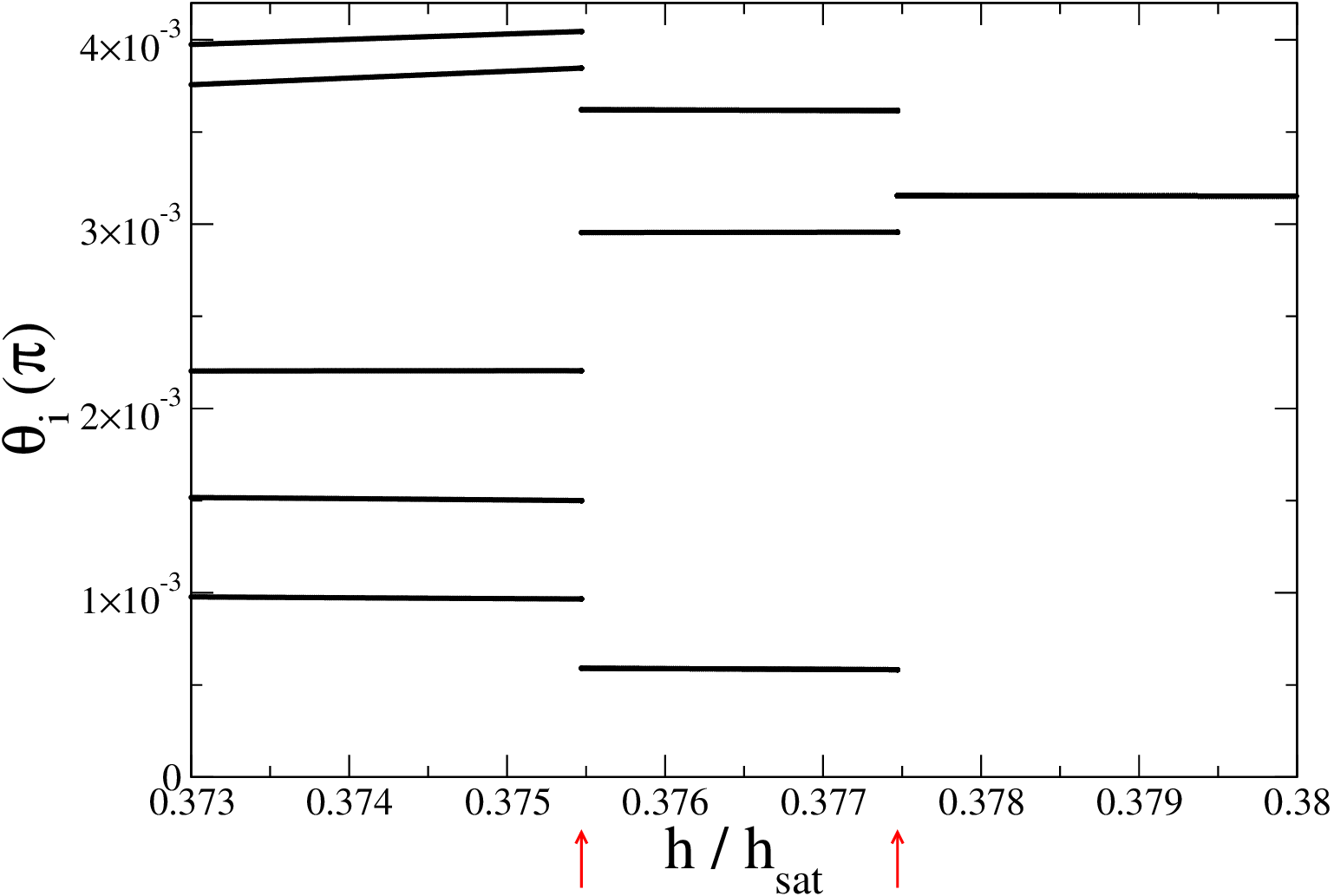}
\end{center}
\caption{Part of Fig. \ref{fig:polaranglesN=5fourthomega=piover10} in greater detail.
%(~/discoverer/basic/classical/icosahedronarraynew/N=5/fourth/parallel/omega=0.1PI)
}
\label{fig:polaranglesN=5fourthomega=piover10a}
\end{figure}

The isolated-icosahedron jump occurs in two steps. Below the magnetization discontinuity at $\frac{h}{h_{sat}}=0.375470596$ each icosahedron has its own $\vec{M}_i$ in magnitude and polar angle (Figs \ref{fig:magnetizationsN=5fourthomega=piover10} and \ref{fig:polaranglesN=5fourthomega=piover10a}). Above this discontinuity three adjacent icosahedra have undergone the jump of the isolated icosahedron. The other two molecules follow at the next magnetization discontinuity at $\frac{h}{h_{sat}}=0.37747069482$. For even higher fields the $\vec{M}_i$ share the magnitude and the polar angle up to saturation, resembling a pentagon of single spins.

\subsection{Hexagon of Icosahedra ($N=6$)}
\label{subsec:sixicosahedraIMC4}

Table \ref{table:N=6omega=pi/10discontinuities-1} lists the discontinuities in the ground-state magnetic response of Hamiltonian (\ref{eqn:model}) for a hexagon of icosahedra and $\omega=\frac{\pi}{10}$. In this case the isolated-icosahedron jump occurs at the same field value for all six molecules. This was also the case for the hexagon of icosahedra with connectivity IMC3 (Sec. \ref{subsec:N=6}). A total of six magnetization and three susceptibility discontinuities appear as the field is ramped up.

\begin{table}
\begin{center}
\caption{Ground-state magnetization discontinuities of Hamiltonian (\ref{eqn:model}) for a hexagon of icosahedra ($N=6$) and $\omega=\frac{\pi}{10}$ for connectivity IMC4. The columns give the magnetic field over the saturation field $\frac{h}{h_{sat}}$ for which the discontinuities occur, the total magnetization along the field just below and just above them over the number of sites $12N$, and the corresponding strength of the magnetization jumps. The three susceptibility discontinuities occur at $\frac{h}{h_{sat}}=0.26483$, 0.62436, and 0.829189, at the corresponding values of the magnetization 0.25739, 0.66009, and 0.861122.}
\begin{tabular}{c|c|c|c}
 $\frac{h}{h_{sat}}$ & $(\frac{M}{12N})_-$ & $(\frac{M}{12N})_+$ & $\frac{\Delta M}{12N}$ ($\times 10^{-3}$) \\
\hline
 0.139941791311 & 0.136820296 & 0.137680554 & 0.860258 \\ % run4 \\
\hline
 0.17567368441 & 0.172006678 & 0.173339331 & 1.332653 \\ % run5 \\
%\hline
%$\chi$ & 0.26483 & 0.25739 & 0.25739 & 0 \\ % & discoverer run \\
\hline
 0.393429952289 & 0.404109925 & 0.429666037 & 25.556112 \\ % & discoverer run1 \\
\hline
 0.4413982693 & 0.477475836 & 0.477492362 & 0.016526 \\ % & run7 \\
\hline
 0.581242500758 & 0.616716376 & 0.617248771 & 0.532395 \\ % & run12 \\
%\hline
%$\chi$ & 0.62436 & 0.66009 & 0.66009 & 0 \\ % & discoverer run2 \\
\hline
 0.64883569 & 0.68432545 & 0.68433829 & 0.01284 \\ % & run14 \\
%\hline
%$\chi$ & 0.829189 & 0.861122 & 0.861122 & 0 \\ % & run15
\end{tabular}
\label{table:N=6omega=pi/10discontinuities-1}
\end{center}
\end{table}

\begin{figure}[h]
\begin{center}
\includegraphics[width=3.5in,height=2.5in]{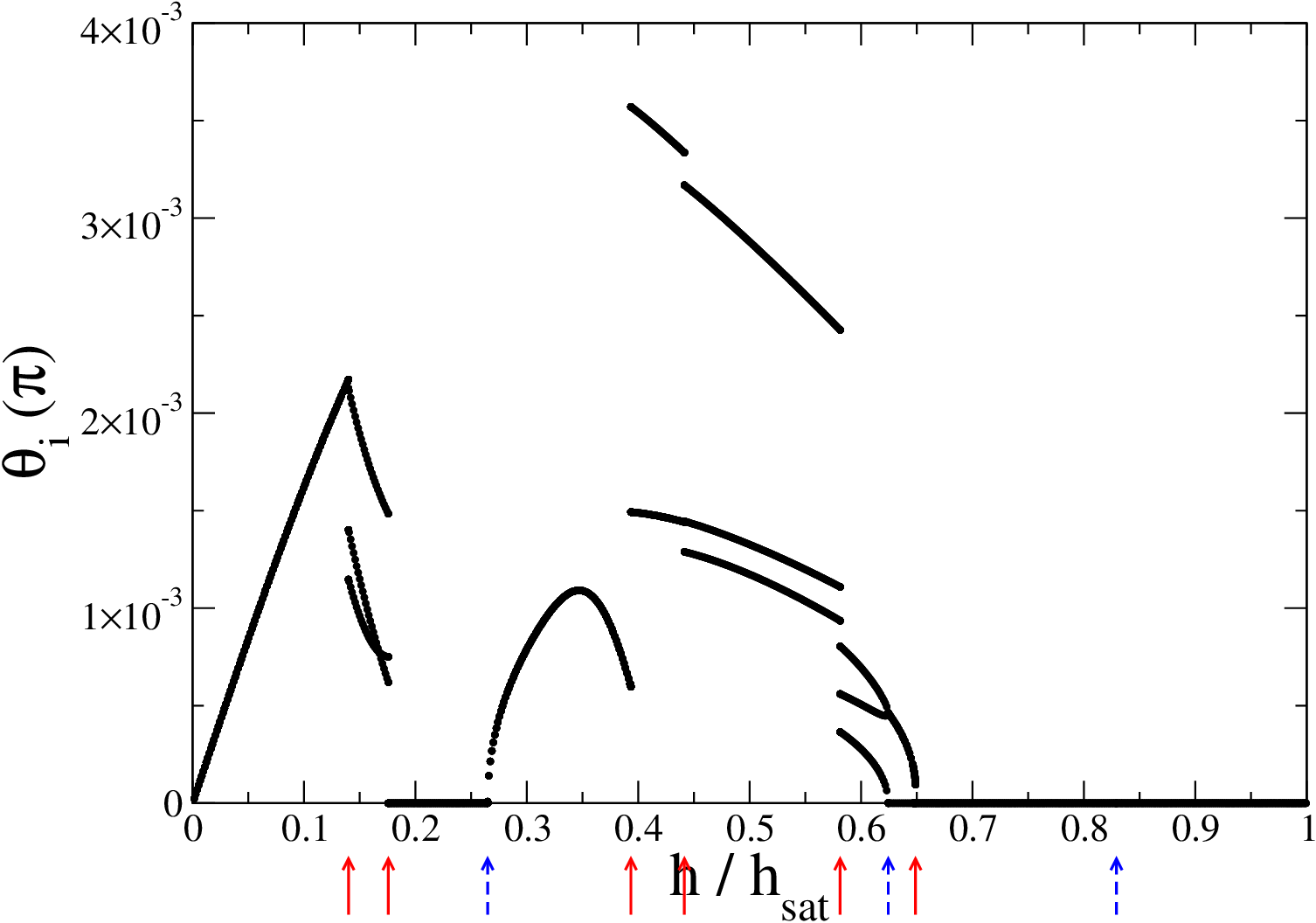}
\end{center}
\caption{Polar angles $\theta_i$ of the magnetizations of the icosahedra $\vec{M}_i$ as a function of the magnetic field over its saturation value $\frac{h}{h_{sat}}$ in the ground state of Hamiltonian (\ref{eqn:model}) for a hexagon of icosahedra ($N=6$) and $\omega=\frac{\pi}{10}$ for connectivity IMC4. The (red) solid arrows point at the locations of the magnetization discontinuities. The (blue) dashed arrows point at the locations of the susceptibility discontinuities.
%(~/discoverer/basic/classical/icosahedronarraynew/N=6/fourth/parallel/omega=0.1PI)
}
\label{fig:polaranglesN=6fourthomega=piover10}
\end{figure}

\begin{figure}[h]
\begin{center}
\includegraphics[width=3.5in,height=2.5in]{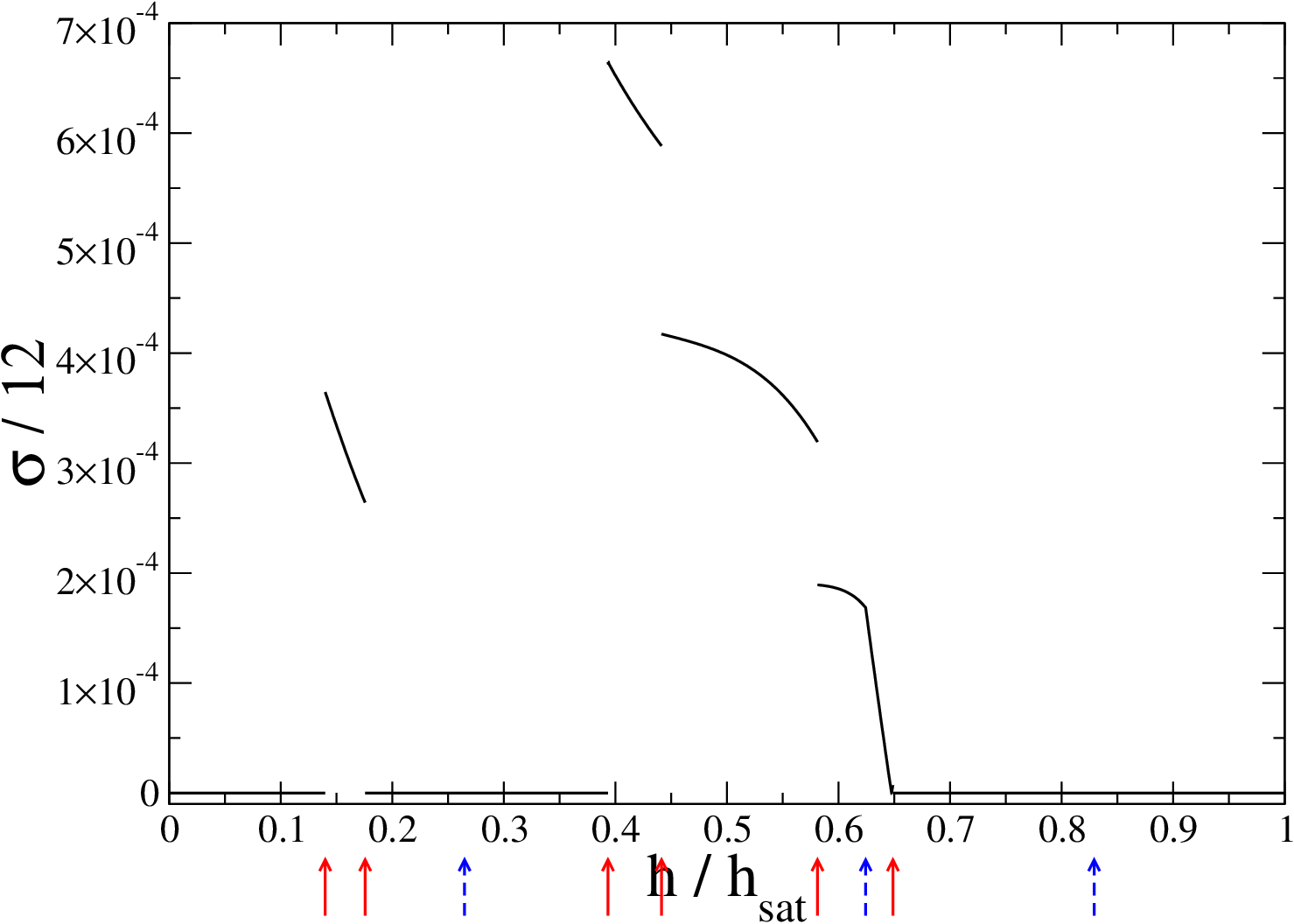}
\end{center}
\caption{Standard deviation $\sigma$ of the magnitude of the molecular magnetization per spin $\frac{M_i}{12}$ as a function of the magnetic field over its saturation value $\frac{h}{h_{sat}}$ in the ground state of Hamiltonian (\ref{eqn:model}) for a hexagon of icosahedra ($N=6$) and $\omega=\frac{\pi}{10}$ for connectivity IMC4. The (red) solid arrows point at the locations of the magnetization discontinuities. The (blue) dashed arrows point at the locations of the susceptibility discontinuities.
%(~/discoverer/basic/classical/icosahedronarraynew/N=6/fourth/parallel/omega=0.1PI)
}
\label{fig:magnetizationsmagnstandarddeviationN=6fourthomega=piover10}
\end{figure}

\begin{figure}[h]
\begin{center}
\includegraphics[width=3.5in,height=2.5in]{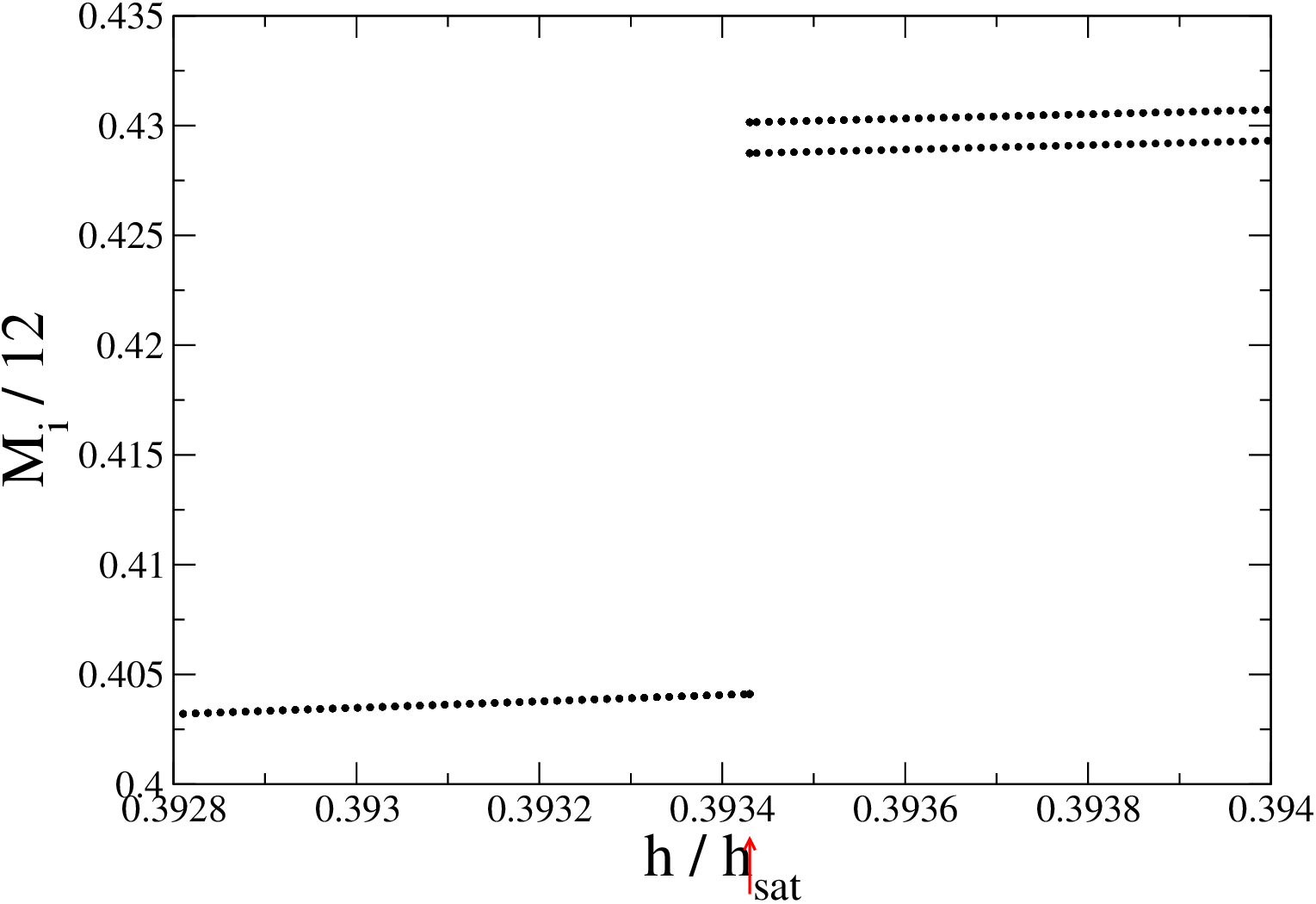}
\end{center}
\caption{Magnitude of the magnetizations $\vec{M}_i$ of the icosahedra as a function of the magnetic field over its saturation value $\frac{h}{h_{sat}}$ in the ground state of Hamiltonian (\ref{eqn:model}) for a hexagon of icosahedra ($N=6$) and $\omega=\frac{\pi}{10}$ for connectivity IMC4. The (red) solid arrow points at the location of the magnetization discontinuity.
%(~/discoverer/basic/classical/icosahedronarraynew/N=6/fourth/parallel/omega=0.1PI)
}
\label{fig:magnetizationsN=6fourthomega=piover10}
\end{figure}

For small fields the $\vec{M}_i$ share again the magnitude, but now they emerge from zero and not a finite polar angle for an infinitesimal field (Fig. \ref{fig:polaranglesN=6fourthomega=piover10}). Following the first magnetization discontinuity at $\frac{h}{h_{sat}}=0.139941791311$ the polar angles decrease and the $M_i$ differ (Fig. \ref{fig:magnetizationsmagnstandarddeviationN=6fourthomega=piover10}), only to become equal and point along the field direction after the second magnetization jump at $\frac{h}{h_{sat}}=0.17567368441$. The isolated-icosahedron jump occurs at $\frac{h}{h_{sat}}=0.393429952289$ (Fig. \ref{fig:magnetizationsN=6fourthomega=piover10}), forcing the icosahedra to assume two different pairs of values for $M_i$ and the polar angle, one for four molecules and the other for two. The former appear in two pairs separated by a molecule of the latter type.

For fields higher than the one of the last magnetization discontinuity the $\vec{M}_i$ align with the field. The last susceptibility discontinuity follows, and has a signature only on the magnitudes of the molecular magnetizations and not their polar angles.

%\section{N=9}
%\label{sec:N=9}

%1.23, 1.86, 2.15, 2.2

%$\frac{h}{h_{sat}}=0.17$.

%Transition from configuration with same magnetization and polar angle that decreases with the field to nine different magnetizations and polar angles.

%The other transitions also lead to magnetizations and polar angles unique to every icosahedron.

%\begin{table}
%\begin{center}
%\caption{Ground-state magnetization discontinuities of Hamiltonian (\ref{eqn:model}) for a triangle of icosahedra ($N=9$) and $\omega=\frac{\pi}{10}$. The columns give the magnetic field over the saturation field $\frac{h}{h_{sat}}$ for which the discontinuities occur, the total magnetization along the field just below and just above the discontinuity over the number of sites $12N$, and the corresponding strength of the magnetization jumps.}
%\begin{tabular}{c|c|c|c|c}
%$\frac{h}{h_{sat}}$ & $(\frac{M}{12N})_-$ & $(\frac{M}{12N})_+$ & $\frac{\Delta M}{12N}$ ($\times 10^{-3}$) & \\
%\hline
%0.168676 & 0.1630661 & 0.1634936 & 0.4275 & run12 \\
%\hline
%0.2566765 & 0.2488398 & 0.2494214 & 0.5816 & run13 \\
%\hline
%0.296432 & 0.288655 & 0.289201 & 0.546 & run15 \\
%\hline
%0.300566 & 0.293307 & 0.296485 & 3.178 & run16 \\
%\hline
% &  &  &  & run?? \\
%\hline
% &  &  &  & run?? \\
%\hline
%0.412 & 0.440 & 0.445 & 5 & run?? \\
%\hline
%0.48(5-7) & 0.52 & 0.52 & 0 & run20 \\
%\hline
%0.90779927 & 0.91900613 & 0.92104005 & 2.03392 & run21 \\
%\hline
% &  &  &  & run??
%\end{tabular}
%\label{table:N=9omega=pi/10discontinuities}
%\end{center}
%\end{table}

\section{Conclusions}
\label{sec:conclusions}

In this paper the classical magnetization response of rings of icosahedra was calculated with the spin interactions described by the AHM, which lacks any anisotropy.  The magnetic response originates from the properties of an isolated icosahedron and the interactions between neighboring icosahedra. It is characterized by multiple magnetization discontinuities and lowest-energy configurations that have locally varying magnetization. For the three and the five-membered ring the magnetization jump associated with the one of an isolated icosahedron appears at different magnetic fields for the different molecules of the ring. This results to an even more pronounced local variation of the magnetization and points to a difference in the magnetic behavior between even and odd-membered rings of molecules.

It is of interest to consider larger rings, which will reveal how the magnetic response evolves when going closer to the thermodynamic limit and what ground-state configurations develop in this case. However this is more demanding computationally. Another area of interest are structures of icosahedra that are more than one dimensional, going also in the direction of icosahedral quasicrystals and their approximants \cite{Shechtman84,Goldman93,Suzuki21,Tsai00,Eto25,Novat26}.

\section{Acknowledgments}
\label{sec:acknowledgments}

The author is very thankful to one of the Referees for analytically calculating the zero-field ground state for different intermolecular couplings and numbers of molecules. He also thanks Samir Lounis for a discussion. Part of the calculations have been carried out on Discoverer at Sofia Tech Park in Bulgaria, under project BG-PHYS-05 ''Investigation of strong correlations in spin systems''.

\begin{appendix}

\section{Saturation Field for a Dimer of Icosahedra ($N=2$) for IMC1 and IMC2}
\label{app:saturationfield}

%Close to saturation spins 3, 5, 7, 8, 15, 17, 19, and 20 are aligned with the field axis. Spins 0, 1, 10, 11, 12, 13, 22, and 23 have polar angle $\theta_0$, while spins 2, 4, 6, 9, 14, 16, 18, and 21 have polar angle $\theta_1$. The energy functional is

Close to saturation interacting spins belonging to different icosahedra are aligned with the field axis. The rest belong to one of two unique polar angles $\theta_0$ and $\theta_1$. The energy functional is:

\begin{eqnarray}
%& & \frac{E}{4} = cos\omega[2+4cos(\theta_0+\theta_1)+4cos\theta_0+4cos\theta_1]+ \nonumber \\ & & (cos\omega+sin\omega)cos(2\theta_1) - 2h(1+cos\theta_0+cos\theta_1) \Rightarrow \nonumber \\ & & \frac{E}{4} - 2cos\omega+2h = 4cos\omega[cos(\theta_0+\theta_1)+cos\theta_0+ \nonumber \\ & & cos\theta_1]+(cos\omega+sin\omega)cos(2\theta_1) - 2h(cos\theta_0+cos\theta_1) \Rightarrow \nonumber \\
& & \frac{E}{4} - 2cos\omega+2h = 4cos\omega cos(\theta_0+\theta_1)+2(2cos\omega-h) \nonumber \\ & & (cos\theta_0+cos\theta_1) + (cos\omega+sin\omega)cos(2\theta_1)
\end{eqnarray}

Very close to saturation $\theta_0 \to 0$ and $\theta_1 \to 0$ and the energy functional becomes:

\begin{eqnarray}
%& & \frac{E}{4} - 2cos\omega+2h \approx 4cos\omega[1-\frac{(\theta_0+\theta_1)^2}{2}]+2(2cos\omega- \nonumber \\ & & h) (1-\frac{\theta_0^2}{2}+1-\frac{\theta_1^2}{2}) + (cos\omega+sin\omega)[1-\frac{(2\theta_1)^2}{2}] \Rightarrow \nonumber \\ & & \frac{E}{4} - 2cos\omega+2h \approx 4cos\omega-2cos\omega(\theta_0+\theta_1)^2+ \nonumber \\ & & 4(2cos\omega-h)-(2cos\omega-h)(\theta_0^2+\theta_1^2)+ \nonumber \\ & & (cos\omega+sin\omega)-2(cos\omega+sin\omega)\theta_1^2 \Rightarrow \nonumber \\ & & \frac{E}{4} - 15cos\omega-sin\omega+6h \approx -2cos\omega(\theta_0+\theta_1)^2- \nonumber \\ & & (2cos\omega-h) (\theta_0^2+\theta_1^2)-2(cos\omega+sin\omega)\theta_1^2 \Rightarrow \nonumber \\
& & \frac{E}{4} - 15cos\omega-sin\omega+6h \approx -2cos\omega(\theta_0+\theta_1)^2- \nonumber \\ & & (2cos\omega-h)\theta_0^2-(4cos\omega+2sin\omega-h)\theta_1^2
\end{eqnarray}

If $E'=\frac{E}{4}-15cos\omega-sin\omega+6h$, the partial derivatives with respect to $\theta_0$ and $\theta_1$ are:

%\begin{eqnarray}
%\frac{\partial E'}{\partial \theta_0} & = & -4cos\omega(\theta_0+\theta_1)-2(2cos\omega-h)\theta_0 \nonumber \\
%\frac{\partial E'}{\partial \theta_1} & = & -4cos\omega(\theta_0+\theta_1)-2(4cos\omega+2sin\omega-h)\theta_1
%\end{eqnarray}
%or
\begin{eqnarray}
\frac{\partial E'}{\partial \theta_0} & = & -2(4cos\omega-h)\theta_0-4cos\omega\theta_1 \nonumber \\
\frac{\partial E'}{\partial \theta_1} & = & -4cos\omega\theta_0-2(6cos\omega+2sin\omega-h)\theta_1
\end{eqnarray}

To find the minimum the derivatives are set equal to zero:
%\begin{eqnarray}
%0 & = & -2(4cos\omega-h)\theta_0-4cos\omega\theta_1 \nonumber \\
%0 & = & -4cos\omega\theta_0-2(6cos\omega+2sin\omega-h)\theta_1
%\end{eqnarray}
%or
\begin{eqnarray}
0 & = & (4cos\omega-h)\theta_0+2cos\omega\theta_1 \nonumber \\
0 & = & 2cos\omega\theta_0+(6cos\omega+2sin\omega-h)\theta_1
\end{eqnarray}

This system of equations has a solution if the determinant is zero.
%or $(4cos\omega-h)(6cos\omega+2sin\omega-h)-4cos^2\omega=0 \Rightarrow h^2-(10cos\omega+2sin\omega)h+20cos^2\omega+8cos\omega sin\omega=0 \Rightarrow h^2-2(5cos\omega+sin\omega)h+4(5cos^2\omega+2cos\omega sin\omega)=0$. Then $\Delta=4(5cos\omega+sin\omega)^2-16(5cos^2\omega+2cos\omega sin\omega)=4[(5cos\omega+sin\omega)^2-4(5cos^2\omega+2cos\omega sin\omega)]=4(25cos^2\omega+10cos\omega sin\omega+sin^2\omega-20cos^2\omega-8cos\omega sin\omega)=4(5cos^2\omega+2cos\omega sin\omega+sin^2\omega)=4[4cos^2\omega+sin(2\omega)+1] \Rightarrow \Delta=4[4cos^2\omega+sin(2\omega)+1]$.
%
%$\displaystyle h_{1,2}=\frac{2(5cos\omega+sin\omega) \pm 2\sqrt{4cos^2\omega+sin(2\omega)+1}}{2} \Rightarrow h_{1,2}=5cos\omega+sin\omega \pm \sqrt{4cos^2\omega+sin(2\omega)+1}$.
%
This leads to the saturation field $h_{sat}=5cos\omega+sin\omega+\sqrt{4cos^2\omega+sin(2\omega)+1}$.

\end{appendix}

\bibliography{icosahedronarray}

\end{document}